\documentclass[useAMS,usenatbib]{mn2e}
\usepackage{graphicx}
\usepackage{amsmath}
\usepackage{txfonts}
\usepackage{natbib}
\usepackage{longtable}
\usepackage[usenames]{color}
\usepackage{subfigure}

\def\cosmology{($\Omega_{M}$=0.3, $\Omega_{\Lambda}$=0.7, $w$=$-1$, $h$=0.7)}

\def\clusterfields{51}
\def\sdssfields{40}

\def\speczcluster{4735}
\def\speczfields{27}
\def\accuracy{0.03}

\def\codeurl{http://big-macs-calibrate.googlecode.com}
\def\totalsets{471}
\def\successfulsets{183}
\def\surveys{(e.g., PanSTARRS, Dark Energy Survey, SkyMapper, Large Synoptic Sky Survey)}

\bibpunct{(}{)}{;}{a}{}{,}

\DeclareMathAlphabet{\mymath}{U}{eus}{m}{n}

\def\la{\mathrel{\hbox{\rlap{\hbox{\lower4pt\hbox{$\sim$}}}\hbox{$<$}}}}
\def\ga{\mathrel{\hbox{\rlap{\hbox{\lower4pt\hbox{$\sim$}}}\hbox{$>$}}}}

\newcommand{\be}{\begin{equation}}
\newcommand{\ee}{\end{equation}}

\newcommand{\bi}{\begin{itemize}}
\newcommand{\ei}{\end{itemize}}

\newcommand{\ben}{\begin{enumerate}}
\newcommand{\een}{\end{enumerate}}

\newcommand{\bfig}{\begin{figure}\begin{minipage}{140mm}}
\newcommand{\efig}{\end{minipage}\end{figure}}

\newcommand{\btab}{\begin{table}\begin{minipage}{140mm}}
\newcommand{\etab}{\end{minipage}\end{table}}

\newcommand{\bfigMore}{\begin{figure}\begin{minipage}{160mm}}
\newcommand{\efigMore}{\end{minipage}\end{figure}}

\newcommand{\btabMore}{\begin{table}\begin{minipage}{160mm}}
\newcommand{\etabMore}{\end{minipage}\end{table}}

\newcommand{\bea}{\begin{eqnarray}}
\newcommand{\eea}{\end{eqnarray}}

\newcommand{\bega}{\begin{gather}}
\newcommand{\eega}{\end{gather}}

\newcommand{\bc}{\begin{center}}
\newcommand{\ec}{\end{center}}

\title
[Photometry \& Photometric Redshifts]
{Weighing the Giants II: Improved Calibration of Photometry from Stellar Colors and Accurate Photometric Redshifts}

\author[Patrick L. Kelly et al.]
{\parbox[t]{\textwidth}{
Patrick L. Kelly$^{1,2,3,4}$ \thanks{E-mail:pkelly@astro.berkeley.edu}, 
Anja von der Linden$^{1,2}$,
Douglas E. Applegate$^{1,2,3}$,
Mark T. Allen$^{1,2}$,
Steven W. Allen$^{1,2,3}$,
Patricia R. Burchat$^{1,2}$,
David L. Burke$^{1,3}$,
Harald Ebeling$^{5}$,
Peter Capak$^{6}$,
Oliver Czoske$^{7}$,
David Donovan$^{5}$,
Adam Mantz$^{8}$,
and
R. Glenn Morris$^{1,3}$
}
\\
        \vspace*{3pt}
\\
$^{1}$Kavli Institute for Particle Astrophysics and Cosmology,
Stanford University,
452 Lomita Mall,
Stanford, CA  94305-4085, USA\\
$^{2}$Department of Physics,
Stanford University,
382 Via Pueblo Mall,
Stanford, CA  94305-4060, USA\\
$^{3}$SLAC National Accelerator Laboratory,
2575 Sand Hill Road,
Menlo Park, CA 94025, USA\\
$^{4}$Department of Astronomy, 
University of California, 
Berkeley, CA 94720-3411, USA\\
$^{5}$Institute for Astronomy,
2680 Woodlawn Drive,
Honolulu, HI 96822, USA\\
$^{6}$California Institute of Technology, 
MC 249-17, 1200 East California Boulevard, 
Pasadena, CA 91125, USA\\
$^{7}$Universit\"at Wien, Institut f\"ur Astronomie,
T\"urkenschanzstra\"se 17, 1180 Wien, Austria\\
$^{8}$Kavli Institute for Cosmological Physics,
University of Chicago,
5640 South Ellis Avenue,
Chicago, IL 60637-1433, USA
}

\begin{document}

\date{}

\pagerange{\pageref{firstpage}--\pageref{lastpage}} \pubyear{2012}
\maketitle

\label{firstpage}

\begin{abstract}
We present improved methods for using stars found in astronomical exposures to calibrate both star and galaxy colors as well as to adjust the instrument flat field.
By developing a spectroscopic model for the SDSS stellar locus in color-color space, synthesizing an expected stellar locus, and 
simultaneously solving for all unknown zeropoints when fitting to the instrumental locus, 
we increase the calibration accuracy of stellar locus matching. 
We also use a new combined technique to estimate improved flat-field models for the Subaru SuprimeCam camera, 
forming `star-flats' based on the magnitudes of stars observed in multiple positions or through comparison with available measurements in the SDSS catalog. 
These techniques yield galaxy magnitudes with reliable color calibration ($\lesssim 0.01$ - 0.02 mag accuracy) that enable us to estimate photometric redshift probability distributions without spectroscopic training samples.
We test the accuracy of our photometric redshifts using spectroscopic redshifts $z_s$ for $\sim$5000 galaxies in \speczfields~cluster fields 
with at least five bands of photometry, as well as galaxies in the COSMOS field, finding
\mbox{$\sigma((z_p - z_s)/(1+z_s)) \approx $~\accuracy} for the most probable redshift $z_p$.
We show that the full posterior probability distributions for the redshifts of galaxies with five-band photometry
exhibit good agreement with redshifts estimated from thirty-band photometry in the COSMOS field. 
The growth of shear with increasing distance behind each galaxy cluster shows the expected redshift-distance relation for a flat $\Lambda$-CDM cosmology. 
Photometric redshifts and calibrated colors are used in subsequent papers to measure the masses of \clusterfields~galaxy clusters
from their weak gravitational shear and determine improved cosmological constraints.
We make our Python code for stellar locus matching publicly available at \codeurl; the code requires only input catalogs and filter transmission functions.

\end{abstract}


\section{Introduction}
\label{sect:intro}


A principal challenge for current and planned optical and near-IR wide-field surveys is to estimate accurate redshift probability distributions for millions of galaxies from broadband photometry.
Correct probability distributions are necessary, for example, for the weak lensing cosmological measurements
that current and upcoming surveys (e.g., Dark Energy Survey; Large Synoptic Survey Telescope) aim to extract from wide-field optical imaging. 
Photometric redshift algorithms, however, can show significant systematic biases if the input galaxy photometry has 
even modest ($\sim$0.03-0.04 mag) calibration error. 
To infer the weak lensing masses of galaxy clusters using photometric redshifts estimated from Subaru and CFHT photometry, we have developed and applied several techniques to calibrate broadband galaxy colors
to an accuracy of $\sim$0.01-0.02 magnitudes, without requiring specific standard star observations.
The relative distribution of counts recorded during flat-field exposures of an illuminated screen or of the sky can 
differ from the actual instrument sensitivity by up to $\sim$10\% across the focal plane (\citealt{manfroid01}; \citealt{koc03}; \citealt{magnier04}; \citealt{caa07}; \citealt{rcg09}) because of a combination of geometric distortion, superposed reflections, and imperfect flat-field sources. 
We combine two methods to measure the Subaru SuprimeCam `star flat,' the map of the spatially dependent zeropoint error that remains after traditional flat-field correction, across 
a decade of observations and camera upgrades. 
Our star-flat model is informed by the magnitudes of the same stars observed in multiple locations on the focal plane, as well as by comparisons with available SDSS catalog magnitudes. 



A powerful color calibration strategy suitable for use with medium-to-wide field data uses the fact that almost all of the stars observed in any field lie along a well-understood one-dimensional locus in color-color space (e.g., \citealt{high09}; \citealt{macdonald04}).  
According to this technique, the zeropoints of the filters are shifted until the position of the observed stellar locus matches the expected locus.
The resulting calibration automatically corrects for Milky Way dust extinction. 
We improve the accuracy of this technique by constructing a spectroscopic model for the SDSS stellar locus, and
by developing a numerical algorithm that fits for all unknown zeropoints simultaneously.

 
Comparison of photometric redshifts estimated from our calibrated galaxy magnitudes against spectroscopic redshifts show that these bootstrapped calibration techniques are effective. 
We find excellent agreement between the most probable photometric redshift $z_p$ and the spectroscopic redshift $z_s$, with a measurement error of \mbox{$\sigma((z_p - z_s)/(1+z_s)) \approx $~\accuracy}.
Five-band $p(z)$ distributions 
summed over different sets of galaxies
($\sum_{\mathrm{gal}} p(z)$)  
show congruence with the \citet{ics09} 30-band photometric redshift distributions for the same sets of galaxies.
The growth of lensing shear with increasing redshift of galaxies behind each cluster, sensitive to the photometric redshifts of galaxies too faint to be represented in spectroscopic samples, exhibits the shape expected for a flat $\Lambda$-CDM cosmology.

The algorithms and techniques described in this paper may be useful as a primary means of calibration or as a demanding test of zeropoint accuracy. 
This paper is the second in a series (``Weighing the Giants") addressing the specific task
of measuring accurate galaxy cluster masses using shear-based weak lensing methods. 
Paper I \citep{vaa12} in this series describes the overall project strategy, the cluster sample and the data reduction methods. 
Paper III \citep{avk12} presents a Bayesian approach to measuring galaxy-cluster masses, that uses the full photometric redshift probability distributions reported here; 
these masses are compared to those measured with a standard `color-cut' method based on 
three-filter photometry for each field.

Section \ref{sec:sample} of this paper summarizes the wide-field imaging data used here.
In Section~\ref{sec:flat},
we describe how we determine the SuprimeCam star flats,
which we use to extract consistent magnitudes across the CCD array.
Section~\ref{sec:slr} describes the stellar locus calibration algorithm and the spectroscopic model we have developed for the stellar locus.
In Section~\ref{sec:photoz}, we discuss the algorithms and the templates for galaxy spectra that we use
to estimate photometric redshift probability distributions $p(z)$.
A method for finding the zeropoints of  $u^*$- and $B_J$- band photometry is presented in Section~\ref{sec:extrapolate}.
In Section~\ref{sec:confirm}, we use  the galaxy cluster red sequence and spectroscopic redshift measurements in the cluster fields
to evaluate the accuracy of our photometric calibrations and redshift estimates. 
In Section~\ref{sec:prob}, we compare the redshift probability distributions determined from 
calibrated photometry in five bands
($B_JV_Jr^+i^+z^+$)
against both the zCOSMOS spectroscopic redshift sample and the most probable redshift inferred from thirty imaging bands in the COSMOS field.
General agreement between the observed growth of weak lensing shear with distance behind the massive clusters and the $\Lambda$-CDM expectation is found in
Section~\ref{sec:shearratio}. 
In Section~\ref{sec:conclusions}, we summarize the calibration techniques and the quality of the photometric redshift estimates.

\section{Subaru and CFHT Imaging Data}
\label{sec:sample}

Our imaging data set consists of wide-field optical exposures of \clusterfields~X-ray--luminous galaxy clusters that span the  redshift interval 
$0.15 < z < 0.7$.
The data were acquired between 2000 and 2009 with SuprimeCam mounted on the 8.3-meter Subaru telescope and with MegaPrime on the 3.6-meter Canada-France-Hawaii telescope (CFHT). 
The field of view of SuprimeCam is $34\arcmin \times 27\arcmin$ (0.2$''$ pixel$^{-1}$), while the MegaPrime has a $1^{\circ} \times 1^{\circ}$ field of view (0.187$''$ pixel$^{-1}$). 
Each cluster field was imaged in at least three separate broadband filters, and 27 fields were imaged with five or more SuprimeCam ($B_JV_JR_CI_Cz^+$) or MegaPrime ($u^*g^*r^*i^*z^*$) bandpasses. 
Paper I describes the cluster sample and observations, the overscan, bias, and dark corrections, as well as the flat-field and superflat processing steps.

\section{Measuring the Star Flat}   
\label{sec:flat}

The intensity of a pixelated image recorded by a CCD camera mounted on a telescope depends, in part, on the properties of the CCD sensors and the readout electronics, as well as the illumination of the
focal plane by the telescope optics. 
Correcting images for spatial sensitivity variations is necessary to be able
to extract accurate fluxes for galaxies and stars.

Calibration images of illuminated dome screens, the twilight sky, and the night sky vary smoothly on scales of tens of arcseconds, and these exposures of flat-field sources are useful for measuring local pixel-to-pixel sensitivity variations.   
The variation of `flat-field' images taken with a wide-field camera on several arcminute scales, however, 
can disagree with the camera's actual sensitivity to point sources. 
The discrepancy can be up to $\sim$10\%  across the field of view of wide-field instruments (\citealt{manfroid01}; \citealt{koc03}; \citealt{magnier04}; \citealt{caa07}; \citealt{rcg09}).
Dividing wide-field science images by flat-field exposures leads to
objects near the center of the focal plane that are systematically  fainter than sources
at the periphery.
 
Geometric distortion along the optical paths of many wide-field telescopes, such as 
Subaru/SuprimeCam and CFHT/MegaPrime, results 
in a decrease in pixel scale with increasing 
distance from the center of the focal plane. 
In SuprimeCam, the pixel scale decreases by  $\sim$1.5\% between the center and 15$'$ from the center,
with a corresponding decrease of $\sim$3\% in the solid angle subtended by a pixel.
(See Fig.~3 in Paper I.)
This effect means that pixels near the center of the field will receive proportionally greater 
flux than they otherwise would when illuminated by a hypothetical calibration source with constant flux per unit solid angle. 
The effect of variation in pixel solid angle due to geometric distortion can be explicitly corrected using the Jacobian of the astrometric distortion (e.g., \citealt{caa07}), or left for correction by a `star flat' (e.g., \citealt{rcg09}), which is the approach we take.

Light that scatters off the surfaces of reflective and refractive optical elements 
also contributes to the spatial distribution of counts in flat-field images.
On average, photons scattering from filters and the CCD sensors are redirected toward the field center (e.g., \citealt{rcg09}).
This can be seen, for example, in the halos that surround bright stars. Stellar halos, which are due to extra reflections, are each centered on a point that is offset from the star towards the center of the field of view.
For diffuse sources that extend across the field of view, such as dome screens or the night sky,
a continuum of superposed reflections accumulates near the center of the exposure. 

Light baffles for wide-field instruments can only be modestly effective.  
\citet{caa07} showed that the intensity of  a SuprimeCam flat field varies by 
$\pm$5\% in the corners of the focal plane due to light scattered from outside the field of view
and that the intensity of scattered light depends on the position of the telescope and whether exposures were taken in dark or twilight conditions. 
To reduce our susceptibility to these intensity variations in the periphery, we exclude regions
of the sensors that are more than 15$'$ from the center of the field.

Zeropoint variations across the focal plane that remain after dividing images by flat-field exposures
can be measured with two approaches.
A first technique is to image the same set of stars at different positions in the focal plane
by dithering the telescope or rotating the camera,
and find the spatially dependent corrections (the `star flat') that result in the smallest dispersion
in magnitudes for each star measured in different positions 
(\citealt{manfroid01}; \citealt{magnier04}; \citealt{caa07}; \citealt{padmanabhan08}; \citealt{rcg09}; and \citealt{wittman12}). 
A second approach is to compare the measured magnitudes of stars to
those in photometrically consistent catalogs, such as the SDSS catalogs (\citealt{koc03}). 
We apply both approaches 
and describe the process in detail below. 

The CFHT ELIXIR pipeline fits and corrects for a star flat for MegaPrime observations (\citealt{magnier04}), 
so we measure only the SuprimeCam star flat across nine years of observations. 
In addition to spatially varying zeropoint corrections, \citet{rcg09} find evidence for position-dependent color terms across the CFHT MegaPrime field of view, which the authors attribute to
an angular dependence of the transmission function for interference filters.

\subsection{Chip Configurations: SuprimeCam Sensor and Electronics Upgrades}
\label{sec:upgrades}

We have grouped our 2000-2009 SuprimeCam images into successive periods that correspond to 
upgrades of the CCD array (see Table 3 of Paper I). 
The CCD sensors in the early 8- and 9-chip configurations exhibited nonlinear response. We were
able to correct the nonlinearity except for two MIT/Lincoln CCD sensors in configuration 8
and three MIT/Lincoln CCD sensors in configuration 9, which we discarded.
The `10\_1' and `10\_2' configurations, installed March 27, 2001, 
feature ten MIT/Lincoln CCD sensors with
fewer cosmetic defects and linear response below saturation. 
The upper left chip had lower quantum efficiency than the other CCDs, but this
can be corrected by the flat field. 
The last set, the `10\_3' configuration installed in July 2008, 
consists of ten Hamamatsu Photonics CCD sensors.

\subsection{Flat-Field Correction Applied to Subaru Imaging}
\label{sec:correction}

Before fitting for the SuprimeCam star flat, we divide each image by a stack of dome-flat or twilight-flat exposures taken during the same
observing run (or adjacent runs if few flats are available), and then by a heavily smoothed stack of night-sky flats (or `superflat').
Flat images are normalized by their median pixel value before being stacked. 
The night-sky flat is constructed from object-subtracted, smoothed exposures, already divided by the stacked dome or twilight flats, 
with no bright stars or strong internal reflections (see \citealt{esd05}) and typically varies by (0.5-1.5)\% across the field of view.
While the stack of dome-flat or twilight-flat exposures corrects pixel-to-pixel sensitivity variations, 
the smoothed night-sky superflat makes adjustments for larger-scale features. 

\subsection{Dither Patterns and Camera Rotations}

Telescope dithers or camera rotations that move stars substantial distances
across the focal plane are helpful to constrain
spatial zeropoint variation. 
The 34$'$x27$'$ SuprimeCam exposures that we analyze have dither patterns that generally vary by a relatively modest  angle of 1$'$ to 2$'$.
For Subaru images that we took to measure the masses of the MACS (\citealt{ebeling01}; \citealt{ebeling07}; \citealt{ebeling10}) galaxy clusters,
we rotated the camera by $90^\circ$ to facilitate star-flat fits.
Subaru images that were taken by other groups and included in our analysis, which we 
accessed through the 
Subaru-Mitaka-Okayama-Kiso Archive (SMOKA)\footnote{http://smoka.nao.ac.jp/} \citep{byi02}, 
can have different dither sequences and rotations (e.g., rotations of 45$^\circ$ or no rotation).

\subsection{Measuring the Star Flat}

We perform a separate star-flat fit to each of \totalsets~sets of exposures corresponding to a given cluster field, filter, and observing run (e.g.,~Abell~68,~$B_J$,~18~July~2007).  The median number of exposures we use each star-flat fit is six, 
with a typical relative rotation of 90$^\circ$ after the first three exposures.

\subsubsection{Selecting the Star Sample}

To select stars, we choose objects that the SExtractor \citep{be96} neural network classifier, supplied with the image seeing (SEEING\_FWHM), suggests are stellar-like
(i.e., where \mbox{CLASS\_STAR $>$ 0.65}).
We also admit only those star candidates for which the flux within 2.5 Kron (\citeyear{kron80}) radii (MAG\_AUTO) has less than 0.1 mag uncertainty, as well as where the stellar image is unblended, has no bad or saturated pixels, has no bright neighbors, and is not truncated by a CCD sensor boundary (i.e., we include only objects with FLAG=0). 
To exclude objects that are saturated or affected by detector nonlinearity in our images,
we include only measurements of objects with a maximum pixel value less than 25,000 ADU above the $\sim$10,000 ADU bias level, 
well below the full-well capacity of $\sim$35,000 ADU above the bias level.
Both for fitting the star flats and for measuring galaxy photometry and shapes for the analysis of weak lensing, 
we exclude objects in each catalog that are more than 15$'$ from the center of the field. 
To identify and remove exceptionally discrepant magnitudes early, 
we use objects that appear in different exposures to estimate the relative zeropoint of each exposure, 
 and subtract these zeropoints from the magnitude measured for each object.
 For each star candidate, we remove any magnitude measurements $m_{\rm star}^{\rm exp}$ for which $|m_{\rm star}^{\rm exp} - \textrm{median}(m)| > 1$ mag, 
where $\textrm{median}(m)$ is the median of magnitudes $m_{\rm star}^{\rm exp}$ across all exposures.

\subsubsection{Star-flat Model}
\label{sec:starflatmodel}

We model the SuprimeCam position-dependent zeropoint with a spatially varying function $C(x,y,\textrm{chip},\textrm{rotation})$,
which is the sum of a separate function $f(x,y)_{\rm rot}$ for each rotation of the camera 
and a single set of chip-dependent offsets ${\cal O}_{\rm chip}$, 
shared across the rotations and dithered exposures:
\begin{equation}
C(x,y,\textrm{chip},\textrm{rotation}) = f(x,y)_{\rm rot} + {\cal O}_{\rm chip},
\label{eqn:surface}
\end{equation}
where $f(x,y)_{\rm rot}$ is the product of third-order Chebyshev polynomials in $x$ and $y$ coordinates on the focal plane.

For the stars in each exposure that meet the sample criteria, we express the measured magnitude of each star, 
$m_{\rm star}^{\rm exp}$, 
in terms of the spatially varying correction $C(x,y,\textrm{chip}, \textrm{rotation})$, and a magnitude $m_{\rm star}^{\rm model}$, which is a free parameter in the fit and corresponds to the stellar 
magnitude that would be measured if the exposures had no position-dependent zeropoint variation:
\begin{equation}
m_{\rm star}^{\rm exp} =  C(x,y,\textrm{chip},\textrm{rotation}) + ZP_{\rm exp} + m_{\rm star}^{\rm model},
\label{eqn:repeat}
\end{equation}
where $ZP_{\rm exp}$ is the zeropoint for each exposure.

We also introduce constraints from available magnitudes from photometrically consistent catalogs.
These are SDSS magnitudes or, alternatively, SuprimeCam or MegaPrime
magnitudes corrected by a previous successful star-flat fit.
We express each available 
catalog magnitude $m_{\rm star}^{\rm cat}$  in terms of 
the modeled magnitude $m_{\rm star}^{\rm model}$,  
a zeropoint offset ${\cal O}_{\rm cat}$, and a color term ${\cal S}_{\rm cat} \times c_{\rm star}^{\rm cat}$:
\begin{equation}
m_{\rm star}^{\rm cat} =  m_{\rm star}^{\rm model} + {\cal O}_{\rm cat} - {\cal S}_{\rm cat} \times c_{\rm star}^{\rm cat}.
\label{eqn:catalog}
\end{equation}
 
The star color $c_{\rm star}^{\rm cat}$ is calculated
from catalog magnitudes (e.g., $g'-r'$). We find the coefficient ${\cal S}_{\rm cat}$ of the color term before fitting for the other star-flat parameters. 

The free parameters in the model are the coefficients in the Chebyshev polynomials in $f(x,y)_{\rm rot}$ for each camera rotation, ${\cal O}_{\rm chip}$, 
 ${\cal O}_{\rm cat}$, $ZP_{\rm exp}$, and, for each star, $m_{\rm star}^{\rm model}$.
 
\begin{table*}
\caption{Diagnostic fit statistics averaged across acceptable star flats, 
sorted by chip configuration and whether SDSS stellar photometry was
available for at least 400 matched stellar objects.
The statistic $\langle\sigma_{\mathrm{jack}}\rangle$ shows 
that the star-flat correction is well constrained.
}
\label{tab:flatstat}
\begin{tabular}{cccc}
\hline
Chip Configuration&SDSS Matches&Cluster/Filter/Run Combinations&
 $\langle\sigma_{\mathrm{jack}}\rangle$\\ 
 \hline
10\_3 &None&3&0.003 mag \\
10\_3 &Yes&4&0.003 mag \\
10\_1 \& 10\_2&None&69 &0.004 mag\\
10\_1 \& 10\_2&Yes&96 & 0.003 mag\\
8~\&~9&None&6&0.009 mag \\
8~\&~9&Yes&5&0.021 mag\\
\hline
\end{tabular}
\end{table*}  
 
\begin{figure*}
\centering

\includegraphics[angle=0,width=3.25in]{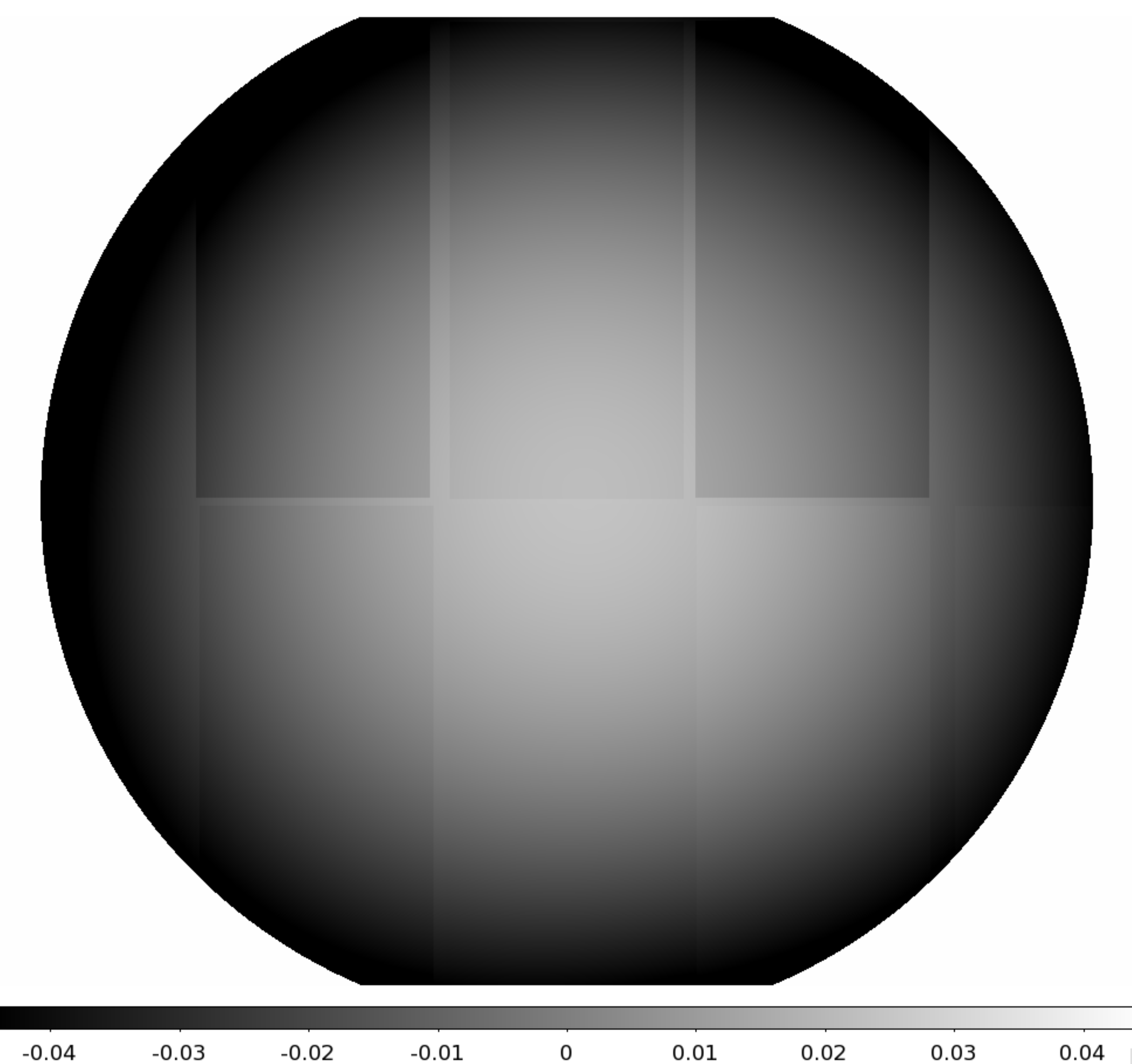}
\includegraphics[angle=0,width=3.25in]{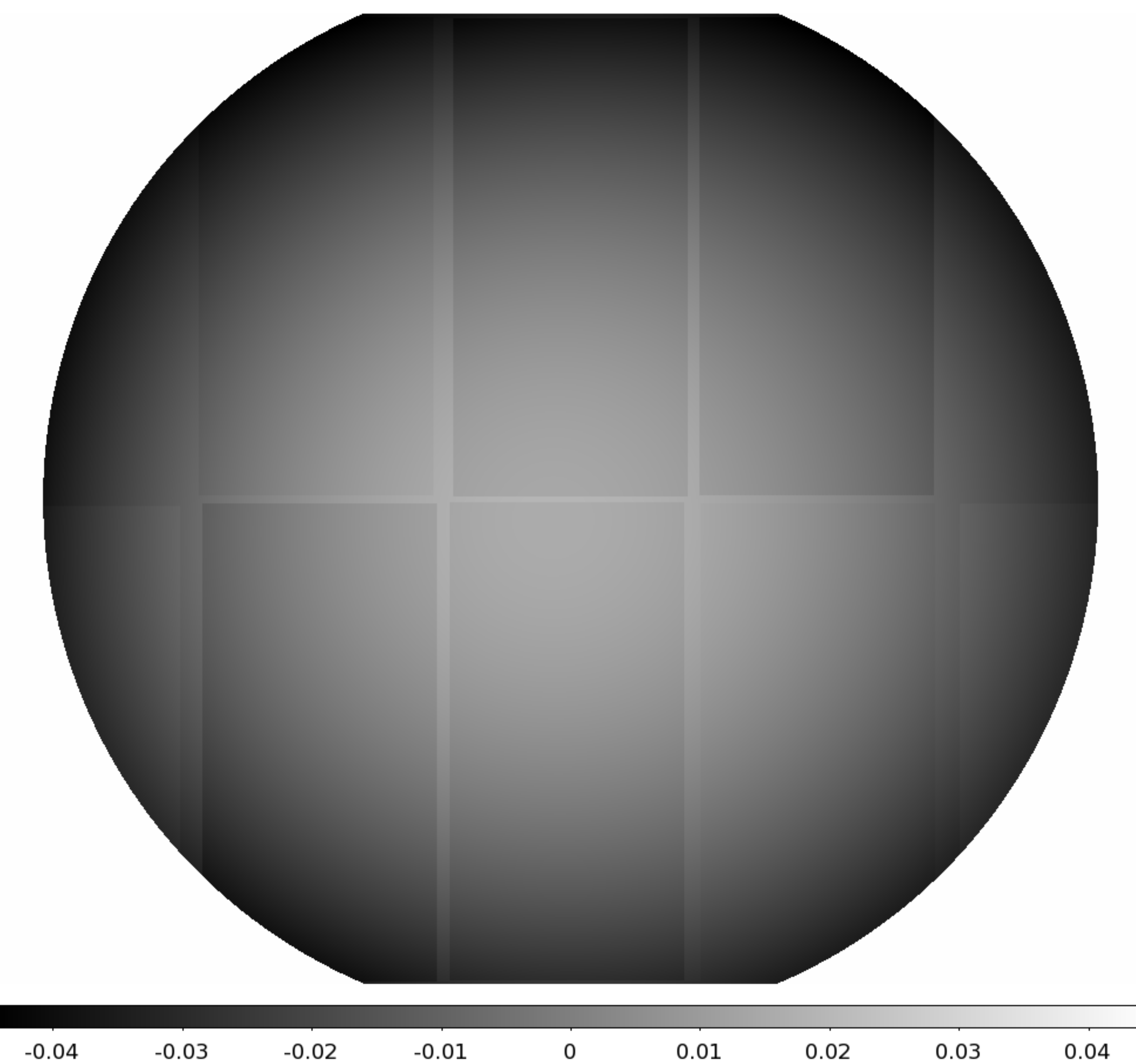}

\caption{
The average star-flat model for the Subaru SuprimeCam  $B_{J}$ (left panel) and $R_{C}$ (right panel) filters.
The star-flat models show our estimate of the  average, position-dependent zeropoint correction to stellar magnitudes 
measured from images corrected with a night-sky flat (the `superflat').
The best-fit star-flat model maximizes the agreement among flux measurements of the same stars
in multiple positions, as well as agreement with available magnitudes from photometrically consistent catalogs  
(i.e., from the SDSS or previously corrected SuprimeCam or MegaPrime photometry).
Dividing pixelated images (already corrected by the night-sky flat) by the star flat yields
corrected images with uniform photometry. 
The shaded horizontal bar below each figure shows the correction to the night-sky flat field in magnitudes. 
We restrict photometry and star-flat fits to the area within 15$'$ of the center of the field because of 
increased levels of light scattered from outside the field of view into the periphery, strong vignetting, and changes in the
point spread function at large radii. 
The model for the star flat consists of the
product of third-order Chebyshev polynomials in $x$ and $y$
plus a chip-dependent offset,
as described in Sec.~\ref{sec:starflatmodel}.
}
\label{fig:flatcorrection}
\end{figure*}

\begin{figure*}
\centering
$
\begin{array}{c}
\includegraphics[angle=0,width=6.5in]{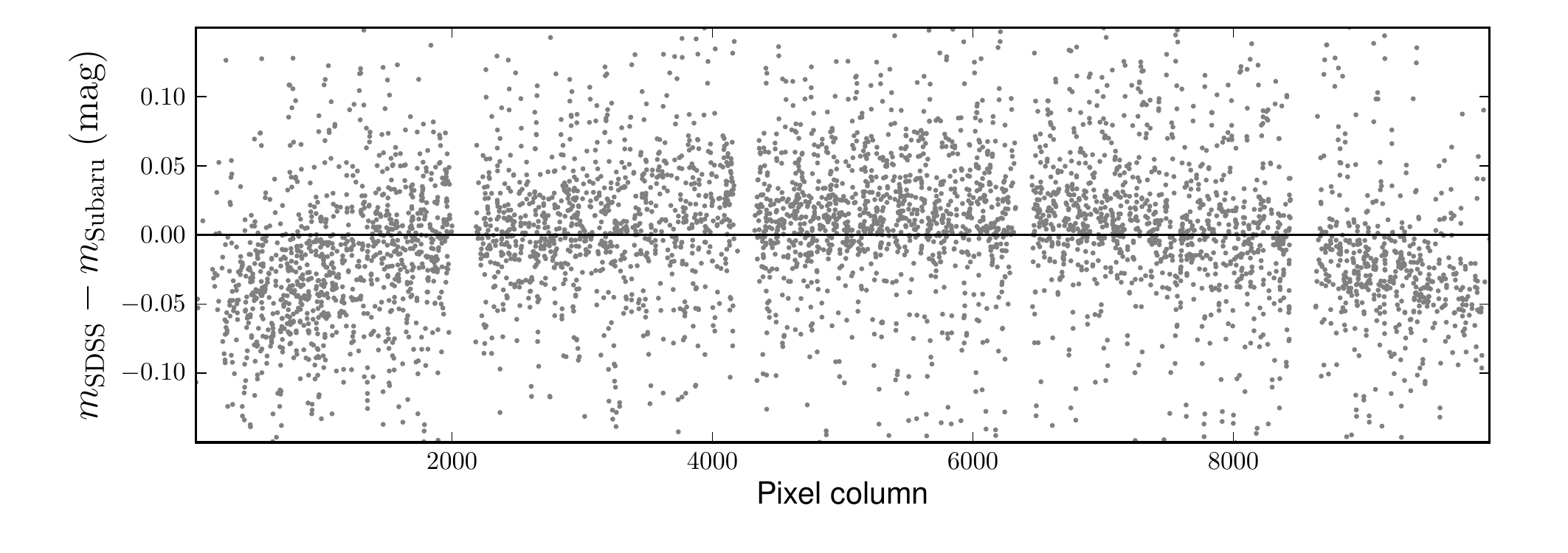}\\
\includegraphics[angle=0,width=6.55in]{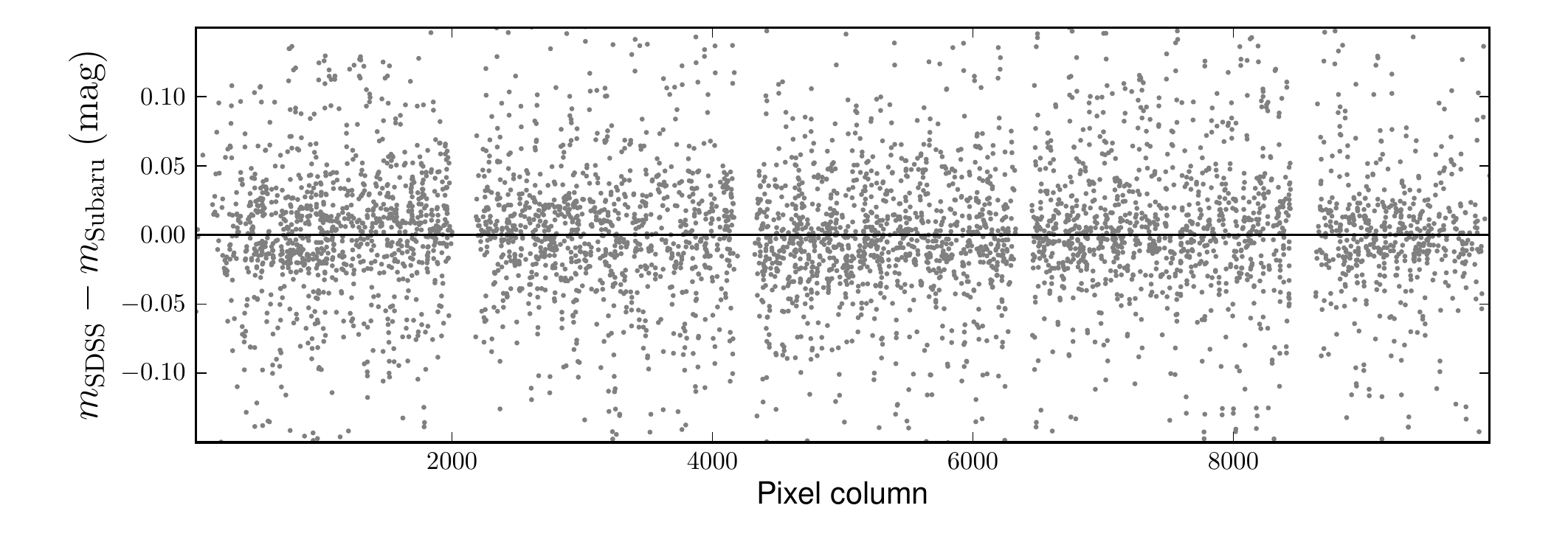}
\end{array}
$
\caption{
{\it Star}-flat correction removes strong spatial zeropoint variation that remains after {\it night sky}-flat correction.
The top panel shows the comparison between SuprimeCam $V_J$ instrumental magnitudes (extracted from {\it sky}-flat corrected images) 
and uniform SDSS stellar photometry in the RXJ1720.1+2638 galaxy cluster field, while the bottom shows the improved agreement after star-flat correction. 
A first-order color term is used to transform SDSS $g'$ magnitudes to expected SuprimeCam $V_J$ magnitudes. 
Flux measurements of the same stars at multiple positions on the focal plane, in particular before and after 90$^{\circ}$ camera rotations,
place the strongest constraints on the star-flat model, except when fields have high stellar density and comparison magnitudes from photometrically consistent catalogs are available. 
Comparison SDSS magnitudes are important for constraining linear terms in the Chebyshev polynomials. (Without SDSS photometry of the field, linear terms are degenerate with variations in the zeropoint offsets between dithered exposures; in those cases, we exclude first-order terms from the Chebyshev polynomials.)  
}
\label{fig:flatresiduals}
\end{figure*}

\subsubsection{Fitting Procedure}

We construct a set of linear equations from two sets of equations: 
Equation \ref{eqn:repeat} for each observation of each stellar candidate, 
and Equation \ref{eqn:catalog} for each star with a catalog magnitude.
The matrix representing the system of equations where each equation corresponds to a row is sparse because only a handful of
equations include $m_{\rm star}^{\rm model}$ for each star.
Each row in the matrix is weighted by the inverse-square of the uncertainty in $m_{\rm star}^{\rm exp}$ or $m_{\rm star}^{\rm cat}$. 
When the statistical uncertainty is less than 0.04 mag, 
we set the uncertainty to be 0.04 mag for purposes of the fit, 
so that a small number of objects will not dominate the solution.
We use CXSparse \citep{dav06}, a C library for sparse matrix algebra, 
to apply a QR decomposition computed with a Householder transformation.

After computing an inital solution, measured magnitudes $m_{\rm star}^{\rm exp}$ that are more than
5$\sigma$ from the corresponding model magnitudes $m_{\rm star}^{\rm model}$  are removed.
We then refit the remaining  measured and catalog magnitudes, 
$m_{\rm star}^{\rm exp}$ and
$m_{\rm star}^{\rm cat}$. 

When fewer than 400 stars in the exposure have a 
catalog magnitude 
$m_{\rm star}^{\rm cat}$, 
we remove linear terms in the Chebyshev polynomials in $f(x,y)_{\rm rot}$ because the set of exposure zeropoints ${\cal O}_{\rm chip}$ exhibit a degeneracy with the linear Chebyshev terms. 
Consider, for example, exposures taken after three successive telescope dithers to the north by 1$'$.
If all stars show fluxes that diminish by 10\% with each consecutive exposure,
these changes in flux  could reflect either worsening atmospheric transparency (i.e., $ZP_{\rm exp}$) 
or a linear spatial gradient in the camera's sensitivity.  
The addition of a sufficient number of $m_{\rm star}^{\rm cat}$ magnitudes 
from an external photometrically consistent catalog for stars spanning the field of view, however, breaks this degeneracy.

\subsubsection{Evaluating the Star-Flat Fit to Each Set of Exposures}

To identify star-flat solutions that have small statistical 
uncertainty and are robust to outliers, we calculate a 
statistic, which we call $\sigma_{\rm jack}$, using star-flat fits to
ten separate jackknife samples,
each containing a randomly selected set of half the stars.
The statistic, $\sigma_{\rm jack}$,  is a measure of the 
variation in the star-flat solution across the samples. 

We create a pixelated image of the best-fit star-flat model (in magnitudes) across the focal plane for each jackknife sample, where each pixel cell corresponds to an area of ~\mbox{$20'' \times 20''$}, 
or $100 \times 100$ CCD pixels.
Each jackknife correction map can then be represented by $A_{ij}^n$,
where $i$ and $j$ are the pixel coordinates, $n$ denotes the $n$th  jackknife sample,
and $A_{ij}^n$ is the mean of the star-flat model within pixel $(i,j)$.
We use the $A_{ij}^n$ maps to assess the uncertainty of the star-flat fits. 
Since we are interested only in the relative spatially varying corrections, 
we subtract the median correction across the image from the correction in each pixel:
$\delta A_{ij}^n = A_{ij}^n - \mathrm{median}(A^n)$.
We then calculate the standard deviation 
$\sigma(\delta A_{ij})$ for each pixel across the jackknife fits.

To assess the uncertainty of the correction calculated from each star flat, 
we use the median of $\sigma(\delta A_{ij})$ across all pixels in the image,
which we call $\sigma_{\mathrm{jack}}$.

\subsubsection{Required Value of $\sigma_{\rm jack}$}

Selecting a maximum acceptable value for $\sigma_{\rm jack}$ presents a tradeoff between 
two objectives. 
On the one hand, we want to measure the star flat for as many 
observing runs and pointings as possible to correct for any variation with position in the sky and with time.
On the other hand, we want to apply a correction only when the statistical uncertainty on 
the correction is small.
We adopted the criteria outlined in the following paragraph but,
as shown in Table~\ref{tab:flatstat}, the average values of the
$\sigma_{\mathrm{\rm jack}}$ statistic for each chip configuration are
substantially better than these thresholds.

For chip configurations 8 and 9 (see Section \ref{sec:upgrades}), the \textit{minimal} requirement for 
using the star-flat correction is that
\mbox{$\sigma_{\rm jack} < 0.03$ mag} (i.e., star flat constrainted to $\approx$ 0.03 mag).
For chip configurations 10\_1, 10\_2, and 10\_3, where greater numbers of objects are generally available, we consider a
fit acceptable if \mbox{$\sigma_{\rm jack} < 0.01$ mag}.
We attempted star-flat fits to \totalsets~sets, and \successfulsets~of the solutions satisfied these requirements. 

A star-flat solution may be too poorly constrainted to meet the minimal $\sigma_{\rm jack}$ requirement for 
several reasons. These include low stellar density in the galaxy cluster field, 
exposures taken without camera rotation,
no overlap with the SDSS footprint,
and minimal telescope dithers between exposures.
When the star-flat fit does not meet the $\sigma_{\mathrm{jack}}$ criterion, 
we look for a satisfactory star flat for a different filter of the same field and
use the corrected $m_{\rm star}^{\rm model}$ magnitudes as reference $m^{\rm cat}_{\rm star}$ magnitudes.   
When even these additional  constraints from $m^{\rm cat}_{\rm star}$ 
do not yield an acceptable solution, we 
correct the data by the satisfactory star flat for data taken closest in time (with the same filter and chip configuration).

\subsection{The Measured Star-Flat Correction}

Figure~\ref{fig:flatcorrection} shows the star-flat shape for the $B_{J}$ and $R_{C}$ bands, averaged across the fits that meet the $\sigma_{\mathrm{jack}}$ 
criteria for each band.
The star-flat corrections averaged over satisfactory fits show a maximal variation 
across the 15$'$-radius field of $\sim$0.06 mag in the $B_J$ and $z^+$ bands and $\sim$0.05 mag for $V_J$, $R_C$, and $I_C$. 
Within a $13'$-radius field, the relative corrections are $\sim$0.05 mag for $B_J$ and $z^+$
and $\sim$0.03-0.04 mag for $V_J$, $R_C$, and $I_C$.
Figure~\ref{fig:flatresiduals} shows how the star-flat correction improves the agreement between the SuprimeCam $V_J$ and SDSS magnitudes for the RXJ1720.1+2638 field. 

\subsection{Comparison with \citet{caa07} SuprimeCam Star Flat Analysis}

To be able to extract accurate photometry for the COSMOS project, \citet{caa07} measured a star flat for Subaru SuprimeCam imaging taken from 2004 through 2005 (chip configuration $10\_2$). 
Before fitting for the star flat, \citet{caa07} first normalized images by a dome flat and then divided  the dome-flat--normalized photometry by the Jacobian of the astrometric solution to explicitly correct for geometric distortion ($\sim$3\% at a radius of 13$'$).
By contrast, we divided dome- or twilight-flat--normalized images by a night-sky flat, which 
shows variation of $\sim$0.005 mag between field center and 15$'$,
and made no explicit correction for pixel size changes. 

The COSMOS correction, including the $\sim$3\% explicit adjustment for pixel size, 
agrees with our corrections to  $\sim$(1-2)\% within 13$'$ of the center of the field.
The only difference with the COSMOS total correction 
that we find is that the $B_J$ and $z^+$ corrections may be $\sim$1\% 
greater than one would expect from only the geometric effect within 13$'$.

Since the images we analyze were taken between 2000 and 2009, 
and span three major SuprimeCam upgrades,
we must characterize the star-flat corrections across all configurations. 
Chip-dependent zeropoints are substantially larger for data 
taken during the early  8- and 9-chip SuprimeCam configurations 
because the array included chips from different manufacturers. 

\subsection{Implementing Correct SWarp Image Resampling} 

Resampling the exposures of each cluster field taken using different camera rotations, in separate dithered positions, and multiple filters
to a common pixel grid and coordinate system
makes it possible to measure the magnitudes and colors of objects easily. 
As described earlier, geometric distortions through the optical paths of most wide-field telescopes
yield a pixel scale (arcsecond pixel$^{-1}$) that decreases with increasing distance from the field center.

We find that the commonly used SWarp program (\citealt{swarp02}, version 2.19.1) requires 
a patch to its source code to preserve relative fluxes of point sources imaged by separate CCD sensors. 
The SWarp `VARIABLE' FSCALASTRO\_TYPE setting (in contrast to `NONE' and `FIXED') 
preserves the flux ratios of sources on the same CCD sensor during resampling, but we find that the ratios among the 
fluxes of 
objects on \textit{different} CCD sensors become altered. 

This problem arises from the fact that, in the SWarp source code, the pixel flux 
is multiplied by a factor that depends on the average pixel scale of 
each resampled image in the `VARIABLE'  FSCALASTRO\_TYPE. 
This chip-dependent factor yields chip-dependent zeropoint offsets with the current SWarp version.  
A simple modification to the SWarp source code can repair this problem by 
removing the unneeded factor for each chip.

We discovered the problem with SWarp after all images were coadded, so
we applied the appropriate position-dependent flux correction to the
photometry of objects measured from resampled exposures.

\begin{figure*}
\centering
\includegraphics[angle=0,width=3.25in]{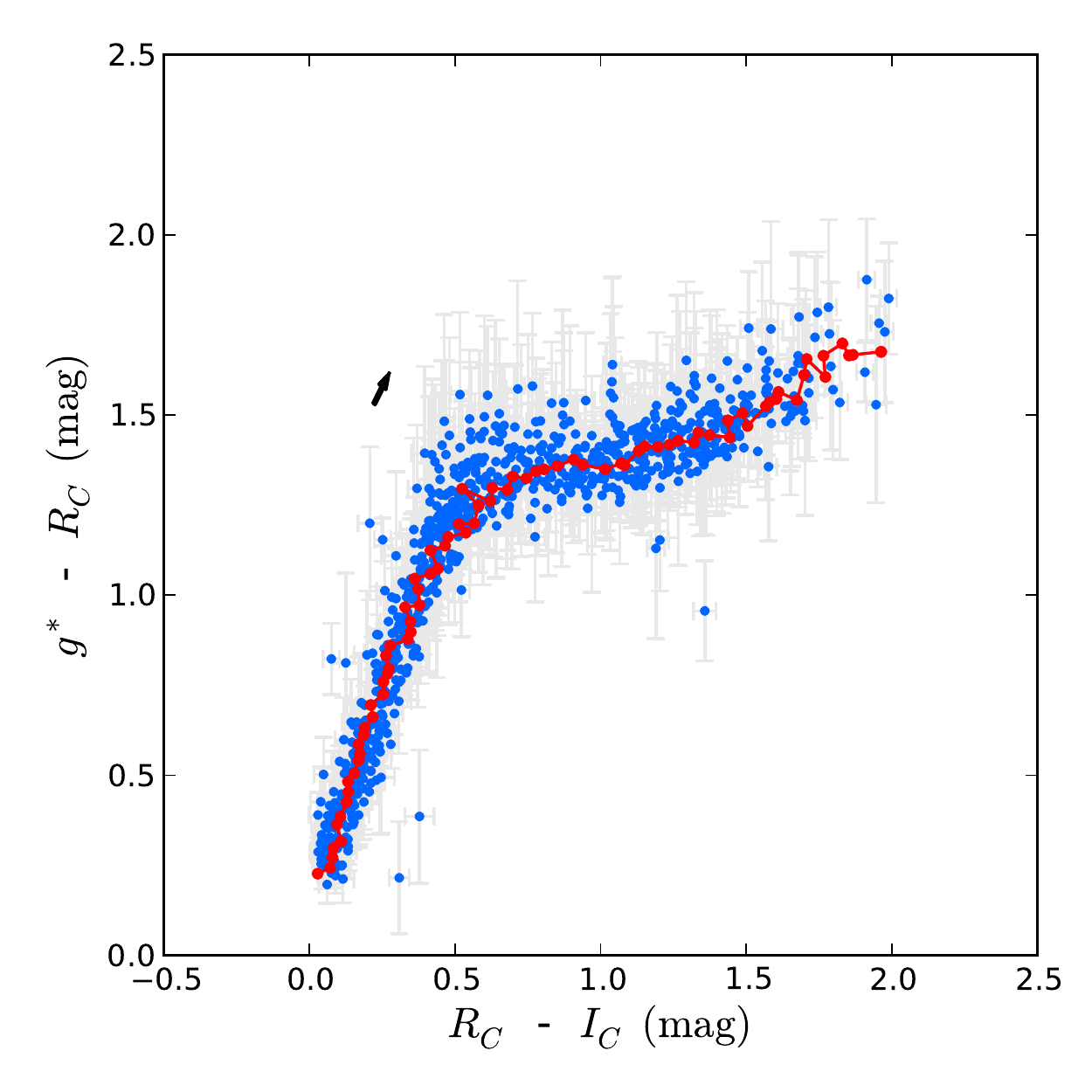}
\includegraphics[angle=0,width=3.25in]{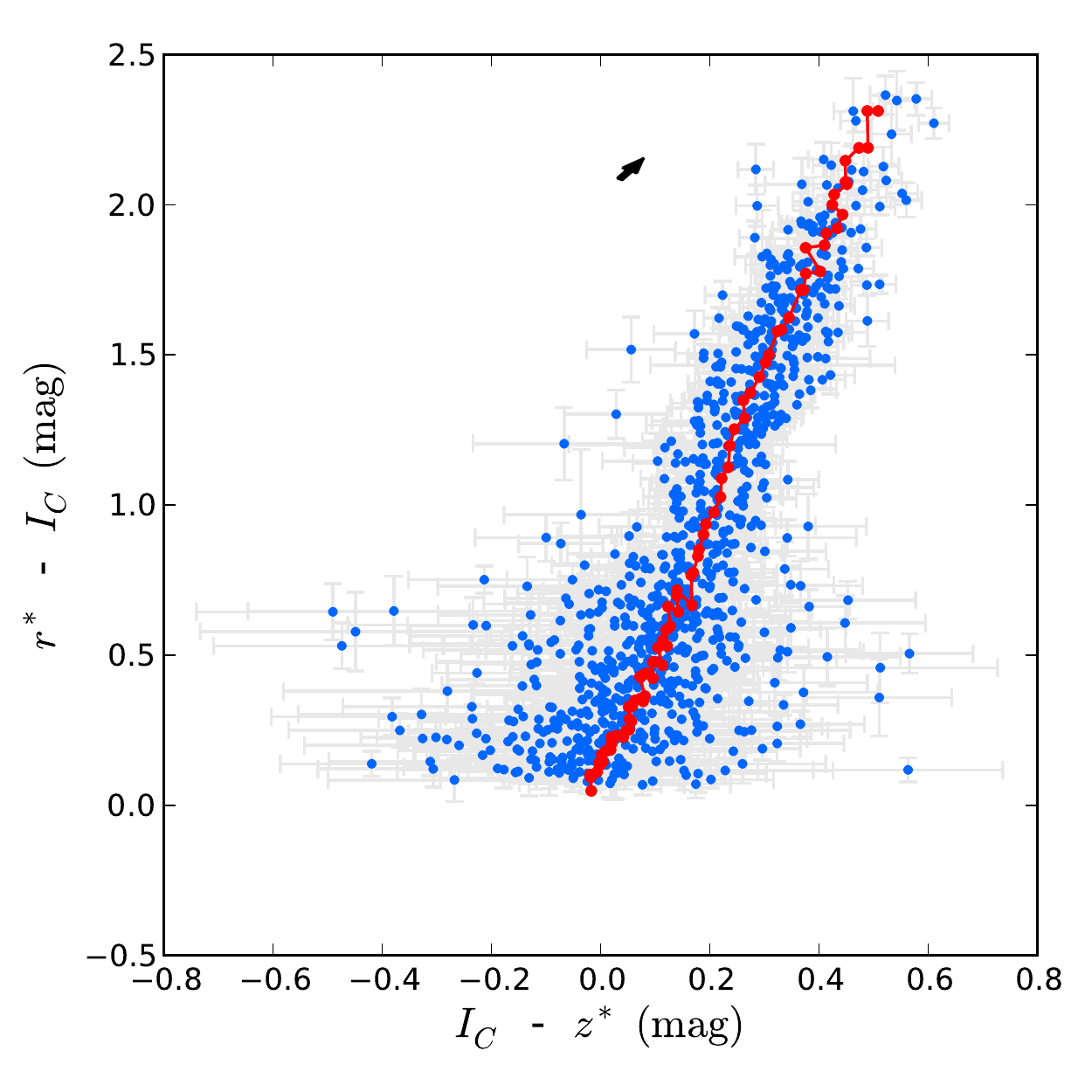}

\caption{
Color calibration of photometry through nine Subaru SuprimeCam ($B_JV_JR_CI_Cz^+$) and CFHT MegaPrime ($g^*r^*i^*z^*$)
filter bands of the RXJ1347-11 galaxy cluster field.
We \textit{simultaneously} vary eight zeropoints and hold one constant (here $R_C$) during the fit to maximize
agreement between the instrumental stellar locus and the model (dereddened) stellar locus.
The red points and lines show the model for the dereddened stellar locus in color-color space, and
the blue points show the colors for the individual stars in the field after applying the best-fit photometric zeropoints. 
For each point along the SDSS $u'g'r'i'z'$ stellar locus (indexed by $g'-i'$ color), we construct a best-fit model spectrum 
that can reproduce the point's color (see Figure~\ref{fig:stellarspec}). 
To calibrate photometry, we compute synthetic magnitudes from each model spectrum and the total response functions of the filters and instrument used.
These synthetic magnitudes form the model stellar locus that we use in each fit for unknown zeropoints shown. 
The arrow shows, for reference, the direction and magnitude of reddening due to dust predicted by the \citet{sfd98} dust extinction map.  
}
\label{fig:slr}
\end{figure*}

\section{Using Stellar Locus Matching as a Calibration Tool}
\label{sec:slr}

The traditional approaches to calibrating photometry through standard-star observations or from overlap with well-calibrated surveys (e.g., SDSS)
are not practical or sufficiently reliable for our analysis of SuprimeCam and MegaPrime imaging.
A substantial fraction of nights have no
standard-star observations and, even when they are available, 
robust calibration is possible only in the most favorable atmospheric conditions. 
Bright SDSS stars ($R_C \lesssim19$ mag) saturate in our exposures, and
Sloan coverage is therefore not useful for high precision calibration.
To fit for accurate zeropoints, we apply an improved stellar locus matching method. 

Investigations at several wavelengths show that the Galactic dust sheet extends only $\sim$50-100 pc from the midplane (\citealt{drimmel01}; \citealt{marshall06}; \citealt{kalberla09}; \citealt{jones11}), and the vast majority of stars visible in SDSS, Subaru, and CFHT 
imaging are at distances beyond the edge of the Galactic dust sheet (e.g., \citealt{high09}). 
Star colors  lie predominantly in a narrow band in color-color space, called the stellar locus. 
The locus has only modest sensitivity to metallicity variations 
among Milky Way stars at wavelengths redder than $\sim$5000 \AA~(\citealt{ivezic07}; \citealt{high09}).
Therefore, by shifting zeropoints for each band until the measured stellar locus matches the dereddened SDSS stellar locus,
we can establish a photometric calibration that includes a correction for the Galactic dust.

As an example of a typical stellar locus and our model, we show in Figure~\ref{fig:slr}  the results of a fit for the instrumental zeropoints of nine SuprimeCam and MegaPrime bands for the RXJ1347-11 galaxy-cluster field.
Below, we discuss significant improvements to the zeropoint accuracy of stellar locus matching that can be achieved by first dereddening the measured SDSS stellar locus,
then constructing a spectroscopic model for the dereddened locus,
and finally solving for consistent 
zeropoints in multiple bands simultaneously (e.g., $B_JV_JR_CI_Cz^+$). In the following section, we first review the astrophysical basis for using the stellar locus as a calibration tool and then describe our improved technique.

\subsection{Stellar Populations Along the Locus}

Several teams have used the stellar locus as a tool to 
calibrate optical photometry (\citealt{macdonald04}; \citealt{ivezic04}; \citealt{high09}; \citealt{gilbank11}; \citealt{coupon12}).
Recently, \citet{high09} applied the calibration technique to $g'r'i'z'$ Magellan imaging of 11 galaxy clusters.  
For these $z < 0.25$ galaxy clusters, they fit the 
average color of the cluster red-sequence galaxies and 
estimated the cluster redshifts with 0.6\% accuracy in 
\mbox{$\sigma((z_p - z_s)/(1+z_s))$} 
(see also \citealt{high10}).
To understand the effectiveness of the stellar locus technique as well as its limitations, 
they also scrutinized the astrophysical understanding of the Milky Way stellar locus. 
Here we discuss the constituents of the locus, their distances from the Earth, and
the effect of metallicity variation in the Milky Way.

Blue stars are generally visible to much greater distances than red stars, because blue stars 
are predominately more luminous. 
As shown in Figure~\ref{fig:slr}, the path of the locus of stars shows a sharp change in 
direction in the plot of $g^* - R_C$ versus $R_C - I_C$ at $R_C - I_C \approx 0.6$ (or SDSS $r'-i' \approx 0.7$). 
The part of the locus to the blue side of the kink includes A through K stars 
(members of the halo and disk main sequence),
as well as an evolved halo population (\citealt{high09}).
The origin of the `kink' is the fact that the disk-metallicity M-dwarf stars 
redward of the kink have substantially nonthermal spectra due to absorption by oxides and metal hydrides, including TiO, VO, CaH, and FeH (e.g., \citealt{high09}; \citealt{west11}).
By contrast, low-metallicity M dwarfs have thermal spectra, and their colors place them along an 
extrapolation of the locus to the blue side of $R_C - I_C \approx 0.6$. 
The absence of such a second branch in the stellar color-color plots for the Subaru and CFHT images analyzed here
suggests that we detect few halo M dwarfs in our exposures. This absence is consistent with
the faint intrinsic luminosities of M dwarfs ($10 < r < 15$ mag).
SDSS  identified only a few thousand M subdwarfs with high confidence across the entire SDSS 
footprint to their 22-mag limit in the $r'$ band (\citealt{lsst09}).

\subsection{Sensitivity of Stellar Colors to Metallicity}
\label{sec:metallicity}

The stellar populations with useful photometry in an astronomical image will depend, at the bright end, on 
the exposure's saturation limit and, at the faint end, on the detection limit.
Our SuprimeCam imaging generally saturates at $R_C \approx 19$ mag and yields detections to 
$R_C \approx 26$ mag. 
For SDSS exposures, the range of useful magnitudes in $r'$ is between $\sim$14 and $\sim$22.5 mag.  

Catalogs made from deeper exposures will include a higher fraction of halo members in the stellar locus 
to the blue side of the kink in the color-color plot in Figure~\ref{fig:slr}(a). 
Although stars with $0.2 < g-r < 0.4$ are approximately evenly split between the disk and halo populations
at SDSS depths, our Subaru and CFHT catalogs contain a higher fraction of 
halo stars, given the deeper limiting magnitudes. 
Since we use the SDSS stellar locus as a model for the locus in our deeper Subaru and CFHT exposures,
we need to consider whether the properties of the stellar locus may change with a larger population of
halo-metallicity main sequence stars. 

Using SDSS photometry and spectroscopy, \citet{ivezic08} find that the halo population has a 
Gaussian metallicity distribution
with mean  [Fe/H]$_{\rm halo} \approx -1.5$, 
while the metallicity of disk stars decreases with distance $Z$ from the midplane
according to
\mbox{$\mathrm{[Fe/H]_{\rm disk}} = (-0.78 + 0.35,\mathrm{\exp}(-|Z|,{\rm kpc}^{-1})$})~dex. 

\citet{high09} use model stellar atmospheres to study how the metallicity difference between the 
disk and halo main sequence populations affects the color of their stellar population. 
They assign a metallicity of [Fe/H]$=-0.5$ for the disk population and 
a  metallicity of [Fe/H]$=-1.5$ for the halo population.  
The difference in $g'-r'$ color for main sequence stars in the two populations, in the region of the stellar 
locus to the blue end of the kink,
is only $\sim$0.01 mag. 
However, the $u'-g'$ colors for the disk main sequence stars are $\sim$0.1 mag redder than those for their halo counterparts.

For this reason, we elect not to use the stellar locus to calibrate near-UV bands for our `Weighing the Giants' study of Subaru and CFHT imaging.
Near-UV emission of stars has comparatively strong sensitivity to the difference between halo and disk metallicities (\citealt{high09}), and the SDSS \textit{u'}-band filter leak near 7100~\AA\ additionally affects measurements of the stellar locus in colors that include the $u'$ band.
There are, however, strategies that may enable use of the stellar locus measured from SDSS magnitudes
to calibrate near-UV photometry in future efforts. 

\subsection{SDSS Stellar Locus Corrected for Extinction}

We measure the stellar locus from SDSS Data Release 8 (\citealt{aihara11}) photometry corrected for Galactic dust extinction.
Stars are drawn from fields with comparatively low Galactic extinction ($A_V < 0.2$ mag) and their
magnitudes corrected by the  \citet{sfd98} extinction map.
We place stars into bins according to their $g'-i'$ color (a proxy for effective temperature) and, for a series of colors (e.g., $g'-r'$, $r'-i'$), measure the median stellar color 
within each bin following the general approach used by \citet{covey07}.

\citet{high09} instead used the \citet{covey07} stellar locus, which is not corrected for extinction by Milky Way dust. We find that $g'-z'$ locus color and photometric 
calibration change by $\sim$0.05 mag after correcting the stellar locus for Milky Way extinction.

\begin{figure}
\centering
\subfigure[]{\includegraphics[angle=0,width=3.25in]{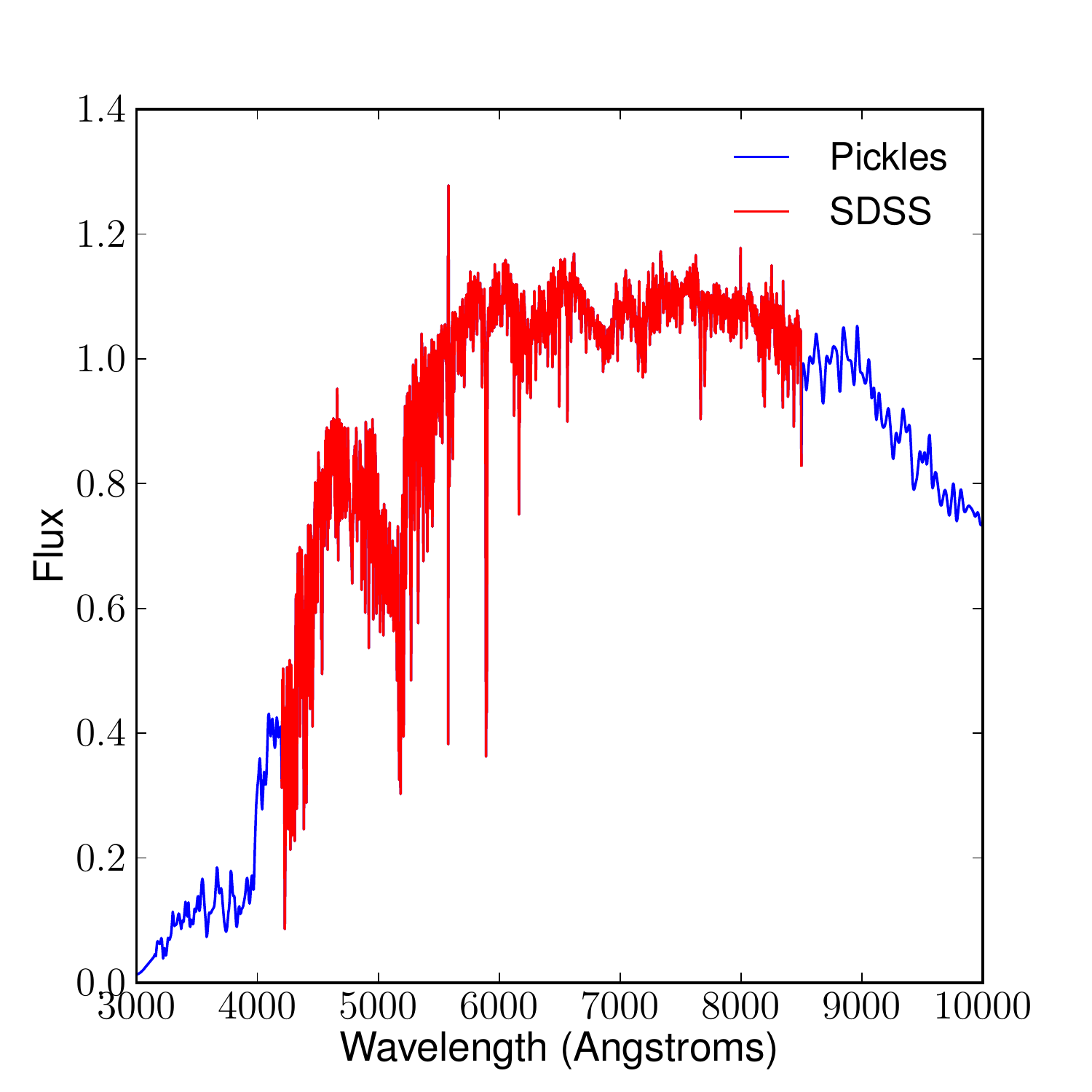}}
\caption{
Best-fit stellar spectrum model for a sample point along the SDSS stellar locus:
$g' - i' \approx 1.6$ mag. 
The red curve in the middle region shows the SDSS spectroscopic data 
for a K5 main sequence dwarf whose color best matches the color of 
this point on the stellar locus. 
The range of wavelengths covered by SDSS spectra \mbox{(from 3900 to 9100 $\AA$)}
must be extended in the blue and red regions to cover the 
Subaru/SuprimeCam and CFHT/Megaprime bandpasses that we wish to calibrate. 
The curves shown in blue correspond to the spectrum from the \citet{pic98} 
library with a shape most similar to the SDSS spectrum in the overlap region.
These are multiplied by functions that are linear in wavelength to reproduce the full 
$u'g'r'i'z'$ color of the point on the stellar locus, as described in the text.
}
\label{fig:stellarspec}
\end{figure}  

\subsection{Defining Model Spectra Along the Locus}

The broadband filters used for the Subaru and CFHT observations, especially the SuprimeCam $B_JV_JR_CI_C$ bandpasses,
have substantially different transmission functions than SDSS $u'g'r'i'z'$ filters. 
The color transformations are nonlinear, and we found that 
a simple, linear color term yielded poor zeropoint solutions. 
To perform improved color transformations between magnitudes in different filter sets,
we construct a comprehensive spectroscopic model for the SDSS stellar locus.
An advantage of this spectroscopic model is that the locus for any filter system can be 
easily computed, requiring only the total response function for each filter.
The total response function is the product of the transmission function for the atmosphere, reflectivity of the telescope mirrors, transmission function for the optics and filter, and the CCD-sensor wavelength-dependent response.

For each of 75 points along the SDSS locus, indexed by $g'-i'$ color, we identify the SDSS stellar spectrum 
(from spectroscopic data)
whose synthetic $g'r'i'$ colors (i.e., $g'-r'$, $r'-i'$, and $g'-i'$) 
best match the $g'r'i'$ colors for this point on the stellar locus.
For this step, we use the CasJobs\footnote{http://casjobs.sdss.org/CasJobs/} tool to query the SDSS database. 
The wavelength range covered by SDSS spectra (3900 - 9100 $\AA$) does not extend through the $u'$ and $z'$ bands, so
we identify the \citet{pic98} main sequence dwarf stellar spectrum that best cross-correlates with the SDSS spectrum
in the range of overlapping wavelengths.
With the \citet{pic98} spectrum, we extend the SDSS spectrum at both the blue and red end.
Figure \ref{fig:stellarspec} shows an example composite spectrum, constructed from SDSS and \citet{pic98} K5 main sequence dwarf spectra, for a stellar locus point with 
$g' - r' \approx 1.1$ mag and $r' - i'  \approx $ 0.4 mag. 

We assemble a continuous spectrum by scaling the \citet{pic98}  blue and red  segments
separately to match the SDSS spectrum in the overlapping wavelength ranges of 4100 to 4600~\AA~ 
and 8000 to 9000~\AA, respectively. 
The spliced spectrum is then comprised of the \citet{pic98} blue segment up to 
4200~\AA, the SDSS segment between 4200~\AA~and 8500~\AA, and the \citet{pic98} red segment above 8500~\AA, as can be seen in Figure~\ref{fig:stellarspec}. 
We next introduce functions that are linear in wavelength that allow us to adjust the \citet{pic98} 
segments to reproduce the locus point's $u'g'r'i'z'$ colors. 
We multiply the blue \citet{pic98} segment by a linear function, $F_{\rm blue}(\lambda) = A\lambda + B$,
where $F_{\rm blue}(4200) = 1$, and then the red \citet{pic98} segment by a second linear function, $F_{\rm red}(\lambda)$, with $F_{\rm red}(8500) = 1$. 
We fit for the slope of each line to match the $u'-g'$ and $i'-z'$ colors, respectively, of the point on the stellar locus.

\subsection{Fitting the Stellar Locus}
The objective of our fitting algorithm is to find the set of
zeropoints that yields the best match between the observed and model stellar locus.
We perform a search for these zeropoints using $\chi^2$ minimization and the downhill simplex method \citep{nme65}.

Calculating the $\chi^2$ for a given model locus and set of filter zeropoints $ZP_f$ is a two-step process.
We follow a similar strategy to that of \citet{high09} but employ an improved
methodology that enables us to fit simultaneously and self-consistently for the complete set of unknown zeropoints.

For a given set of zeropoints $ZP_f$,  we search, for each observed star, through all 75 points along the model locus 
to find the point with the best $\chi^2$  (i.e., the best match to  the observed stellar spectral energy distribution (SED)). 
Keeping the zeropoints fixed, we repeat this process for all stars, summing an overall $\chi^2$ to calculate a goodness-of-fit (GOF) statistic.
We repeat this overall process for each new set of zeropoints until the best set of zeropoints is found.

\citet{high09} instead solved for the zeropoints of any two filters (e.g., $ZP_g$ and $ZP_i$ from
$g-r$ and $r-i$ colors) by minimizing the weighted, perpendicular color-distance residual:
\begin{equation}
d^w_{\alpha \beta} = \frac{|{\boldsymbol d_{\alpha \beta}}|}{|{\boldsymbol \sigma_{\alpha}} \cdot  \hat{d}_{\alpha \beta}|},
\end{equation} 
where $\boldsymbol \sigma_{\alpha}$ is the vector of measurement errors,
$\boldsymbol d_{\alpha \beta}$ is the vector distance in color space between 
star $\alpha$ and locus point $\beta$,
and $\hat{d}_{\alpha \beta}$ 
is a unit vector with the same orientation as $\boldsymbol d_{\alpha \beta}$. 
Calculating distances in color space, however, becomes increasingly cumbersome
for growing numbers of filters such as we have. 
In particular, $n$ magnitude measurements produce $(n^2 - n)/2$ colors (e.g., 10 colors for 5 bands, 15 colors for 6 bands) with correlated errors. 

To improve the accuracy of the calibration, we instead employ a simple $\chi^2$ method that
enables simultaneous fitting for large numbers of consistent zeropoints, and avoids
correlated input errors that arise when calculating distances in color space. 
We follow several steps to measure the $\chi^2_{\rm tot}$ GOF for each set of $ZP_f$ fit parameters.
For each star $\alpha$ and locus point $\beta$, a common, relative zeropoint $O_{\alpha\beta}$ between $m^{\alpha(obs)}_f$ and $m^{\beta(model)}_f + ZP_f$ is shared across all filters $f$. This relative zeropoint $O_{\alpha\beta}$ accounts for the difference between the normalization of the star's instrumental magnitudes, 
which depends on the star's apparent magnitude, and the arbitrary normalization of the model SED. 
The $\chi^2_{\alpha\beta}$ agreement between $m^{\alpha(obs)}_f$ and $m^{\beta(model)}_f - ZP_f$ is a function of $O_{\alpha\beta}$:   
\begin{equation}
\chi^2_{\alpha\beta}(O_{\alpha\beta}) =    \sum_{f=1}^{n}  \Bigg( \frac{m_{f}^{\alpha(obs)}   - (m_{f}^{\beta(model)} - ZP_f + O_{\alpha\beta}) }{e_{f}^{\alpha(obs)}} \Bigg)^2.
\end{equation}
We identify the relative zeropoint $O_{\alpha\beta}$ that minimizes
$\chi^2_{\alpha\beta}(O_{\alpha\beta})$ (i.e., maximizes the likelihood).
This is simply the weighted mean of the differences between $m_{f}^{\alpha(obs)}$ and $m_{f}^{\beta(model)} - ZP_f$:
\begin{equation}
O_{\alpha\beta}^{*} = \frac{\sum_{f=1}^{n} (m_{f}^{\alpha(obs)}  - (m_{f}^{\beta(model)} - ZP_f))/(e_{f}^{\alpha(obs)})^2}{\sum_{f=1}^{n} {1}/{(e_{f}^{\alpha (obs)})^2}}.
\label{eqn:offset}
\end{equation}

For each star $\alpha$, we loop over the locus points to 
find the locus point $\beta$ with the smallest $\chi^2_{\alpha\beta}(O_{\alpha\beta}^{*})$ statistic, 
$\chi^{2,\mathrm{min}*}_{\alpha\beta}$. We then sum $\chi^{2, \rm min *}_{\alpha\beta}$ over all stars $\alpha$ to calculate the overall GOF 
statistic for the given set of input zeropoints $ZP_f$: 
$\chi^2_{\rm tot}=\sum_{\alpha}^{} \chi^{2, \rm min *}_{\alpha\beta}$.

To find the best-fitting zeropoints, we use the amoeba downhill simplex method \citep{nme65}.
We solve for zeropoints in all filters (except one) simultaneously.  
This yields more robust and accurate zeropoints than previous implementations and improves the 
accuracy of our photometric redshifts. 

The open-source Python code is available at \codeurl. Photometric 
calibration requires only a catalog of measured stellar magnitudes and the total transmission function
for each filter.

\begin{figure*}
\centering
\includegraphics[angle=0,width=3.25in]{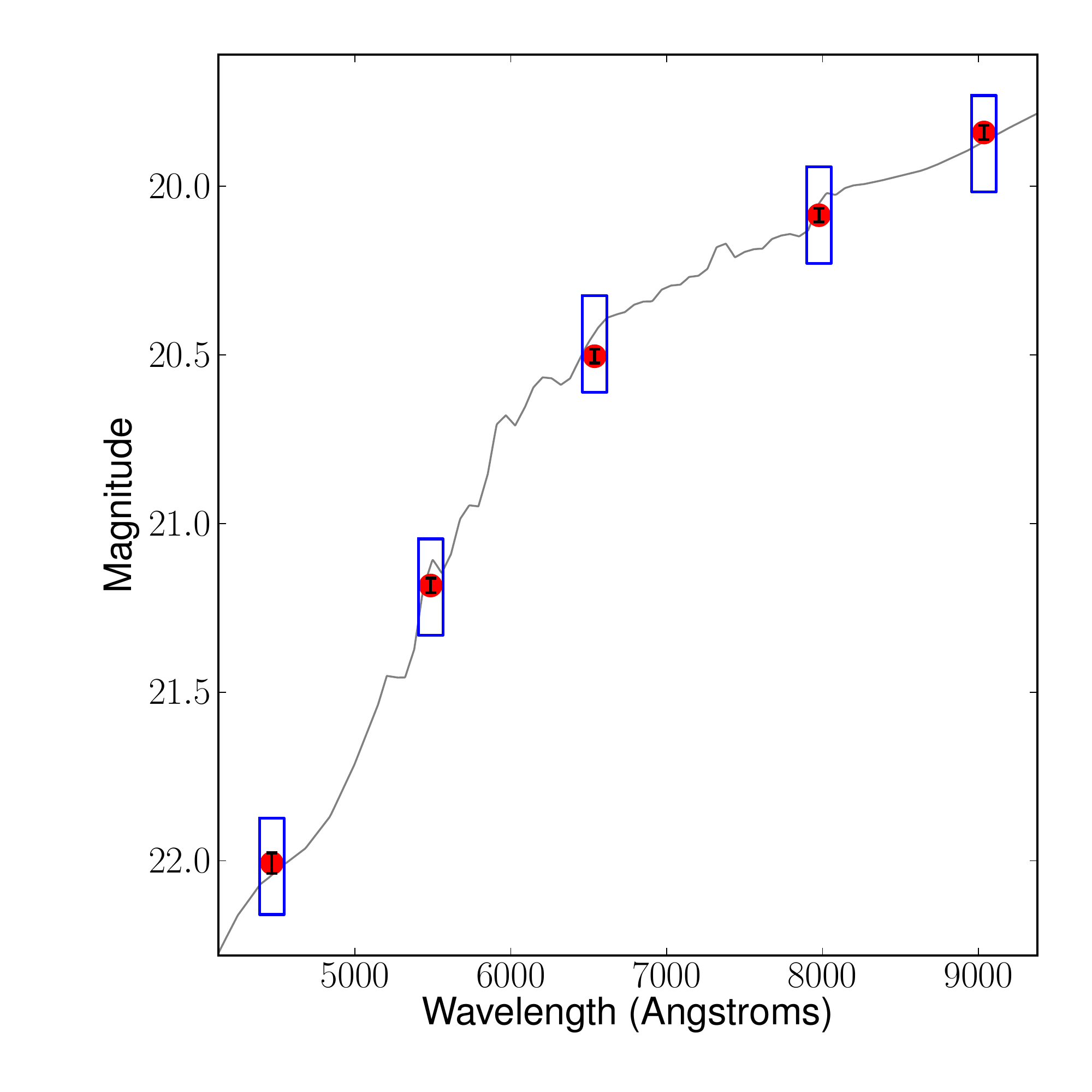}
\includegraphics[angle=0,width=3.25in]{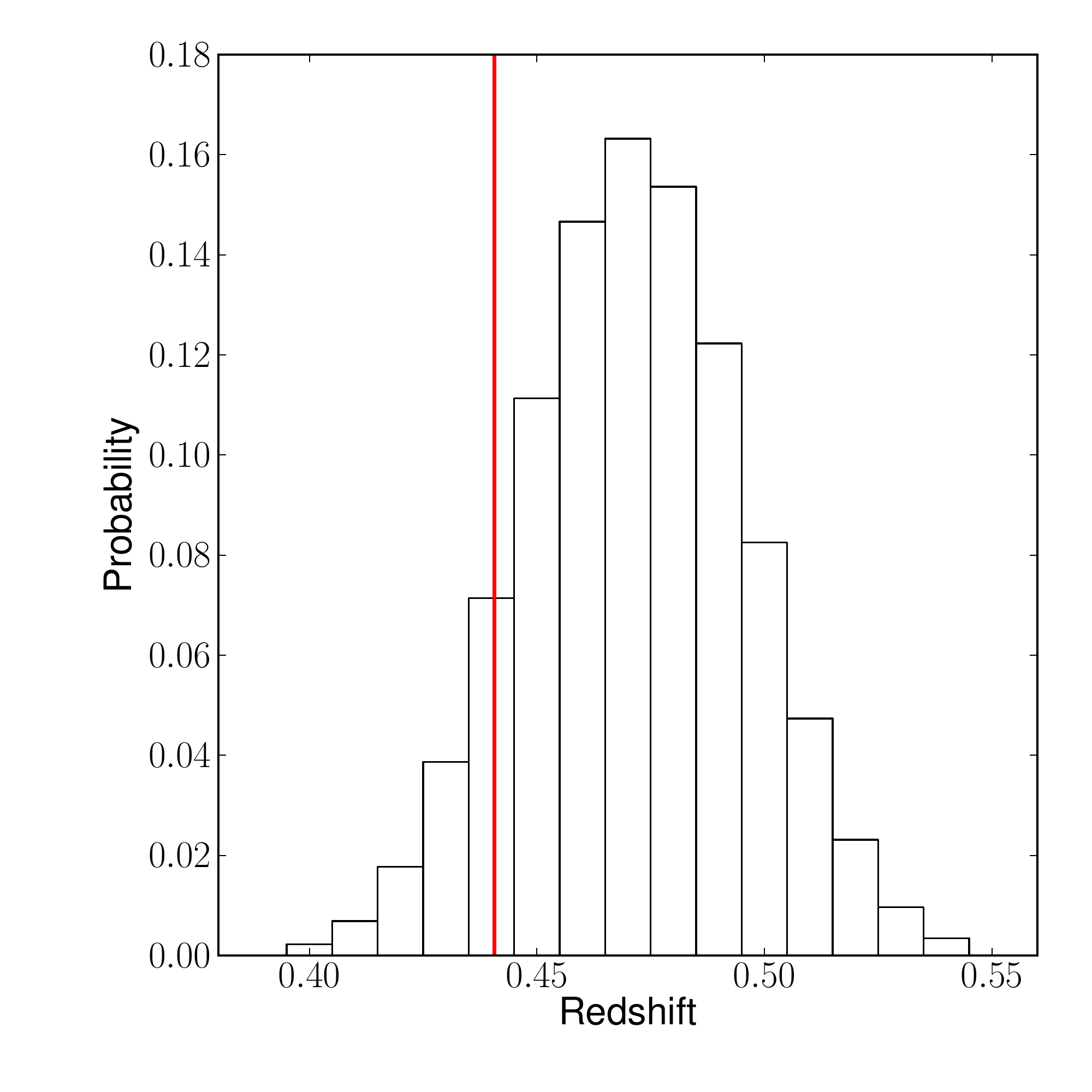}
\caption{Best-fit SED model for a sample galaxy's broadband magnitudes from the BPZ photometric redshift code (left panel) and
posterior redshift probability distribution (right panel). In this example, the template with greatest posterior probability is 
the elliptical galaxy spectrum from \citet{cww80}, modified by \citet{capak04}.
The red points in the left panel show, in order of increasing central wavelength, the calibrated galaxy magnitude measured for 
the SuprimeCam $B_J,V_J,R_C,I_C,$ and $z^{+}$ bands. The blue rectangles show the expected flux for the 
elliptical SED in these photometric bands. 
In the right panel, the horizontal coordinate of the vertical red line is the spectroscopic redshift of this
elliptical galaxy. 
}
\label{fig:sedfit}
\end{figure*}

\begin{figure*}
\begin{center} $
\begin{array}{cc}
\includegraphics[angle=0,width=3.25in]{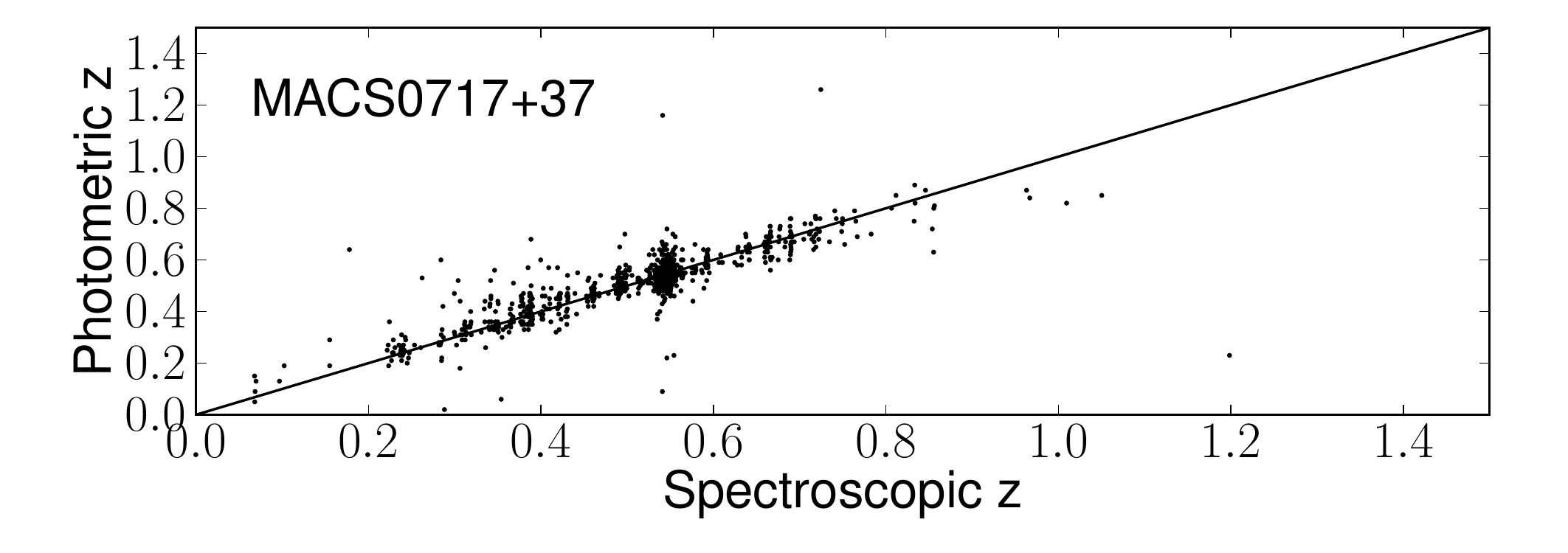} &
\includegraphics[angle=0,width=3.25in]{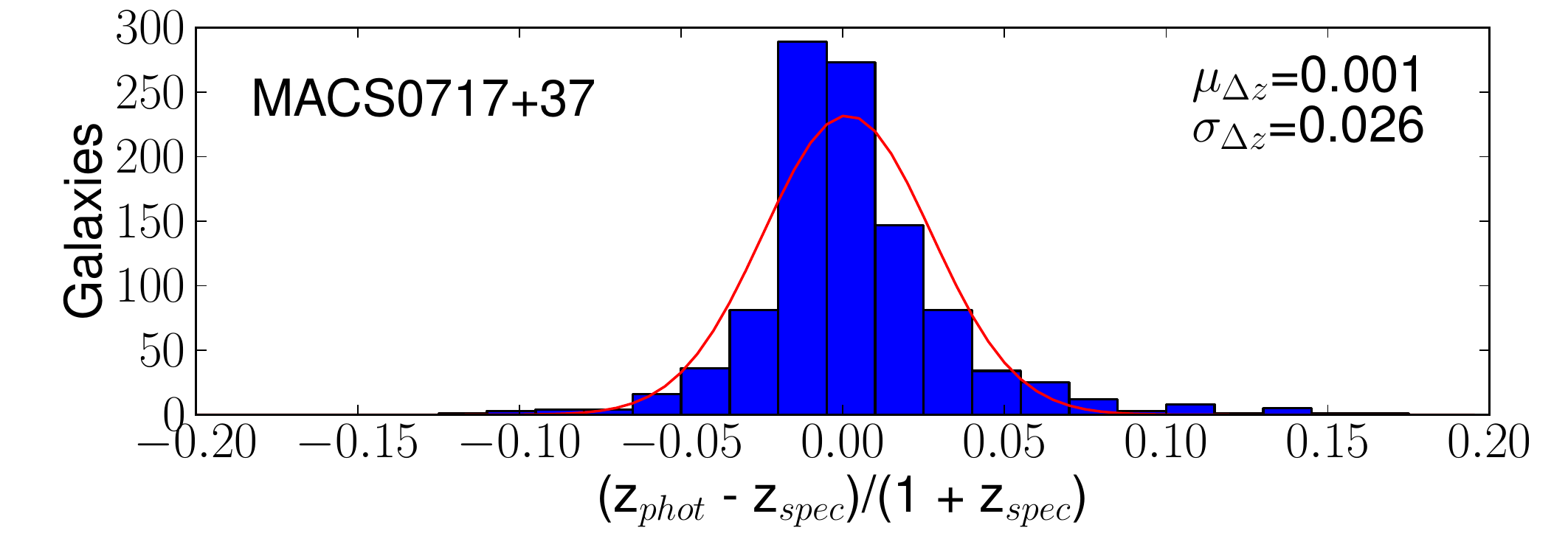} \\
\includegraphics[angle=0,width=3.25in]{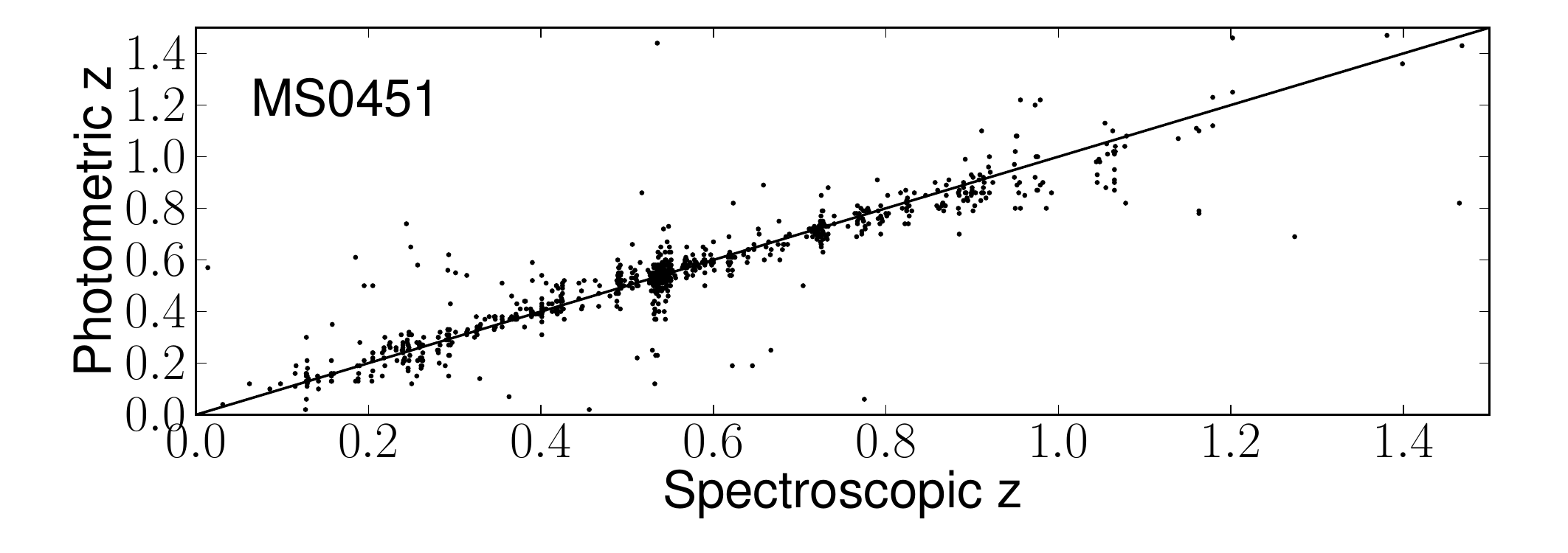} &
\includegraphics[angle=0,width=3.25in]{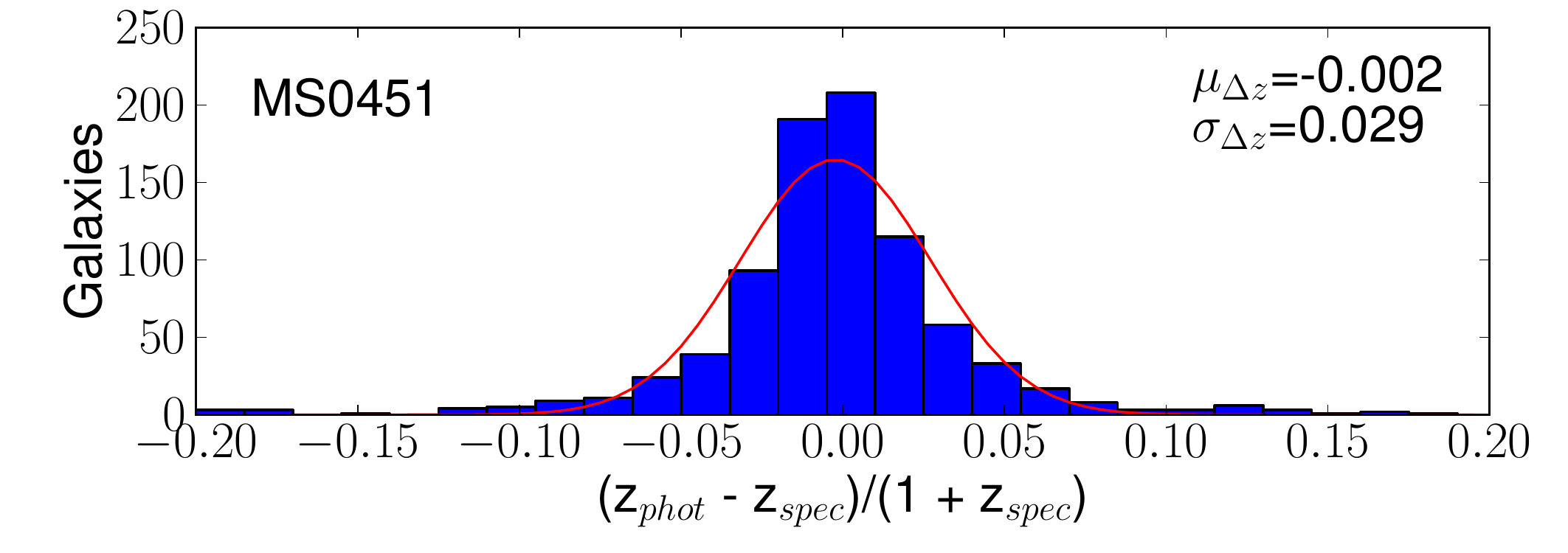} \\
\includegraphics[angle=0,width=3.25in]{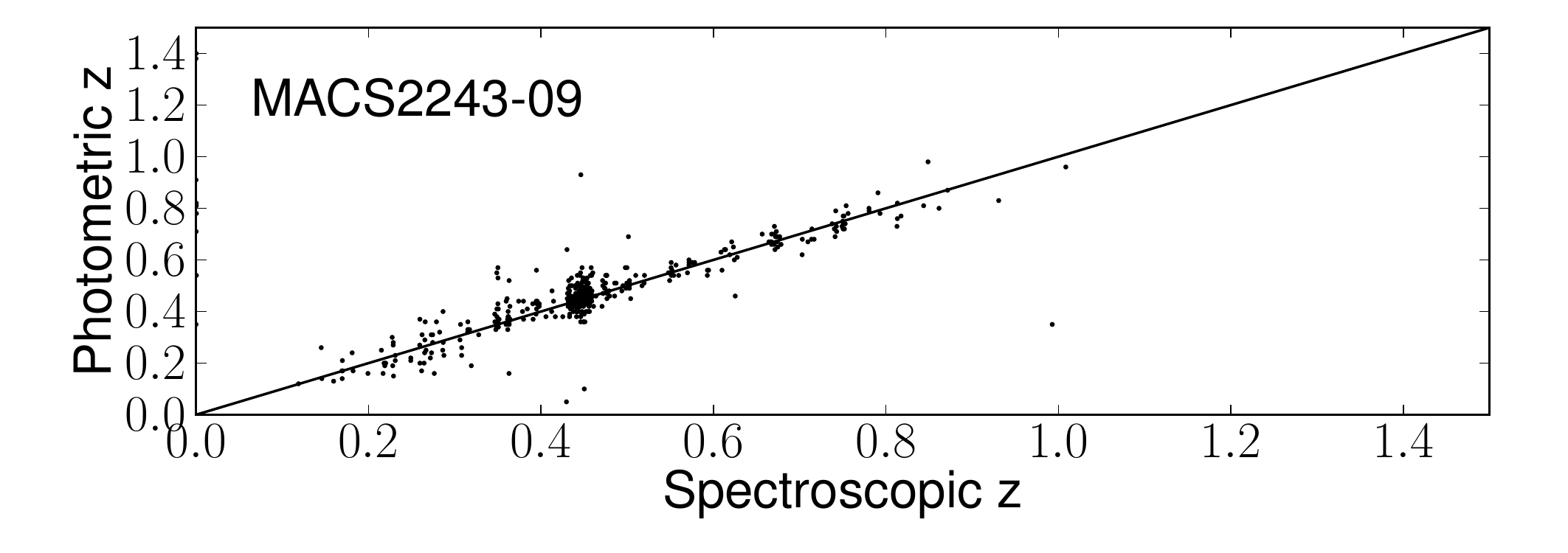} &
\includegraphics[angle=0,width=3.25in]{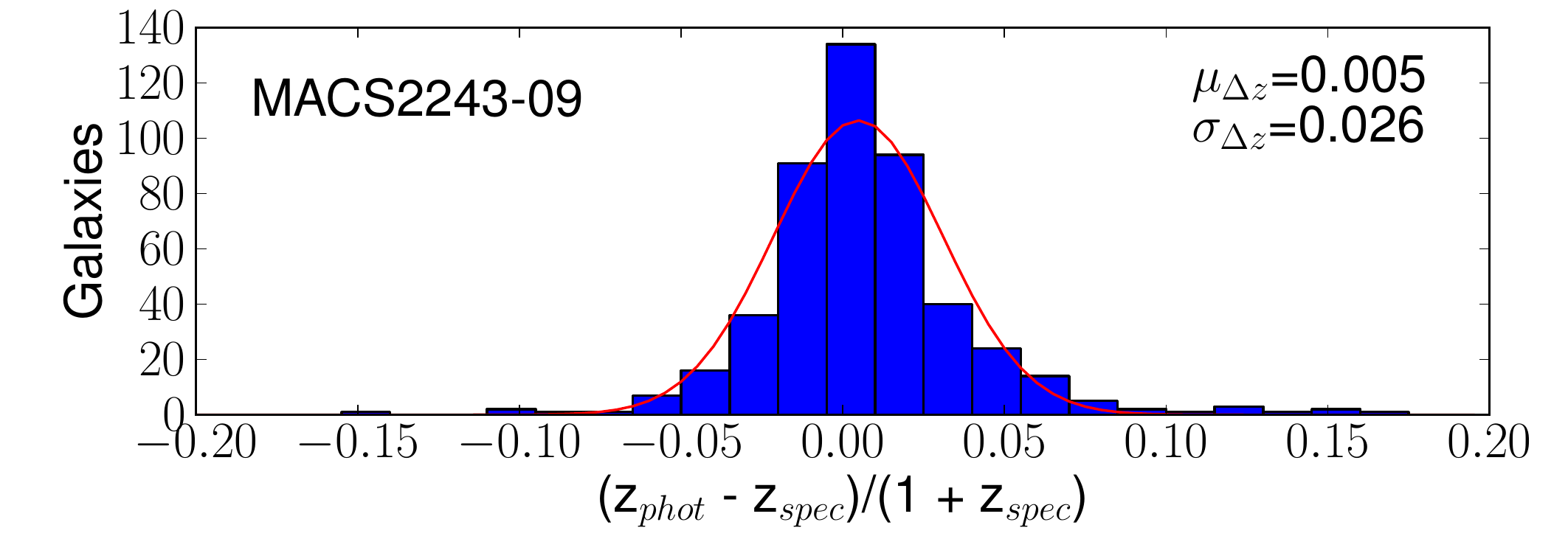} \\
\includegraphics[angle=0,width=3.25in]{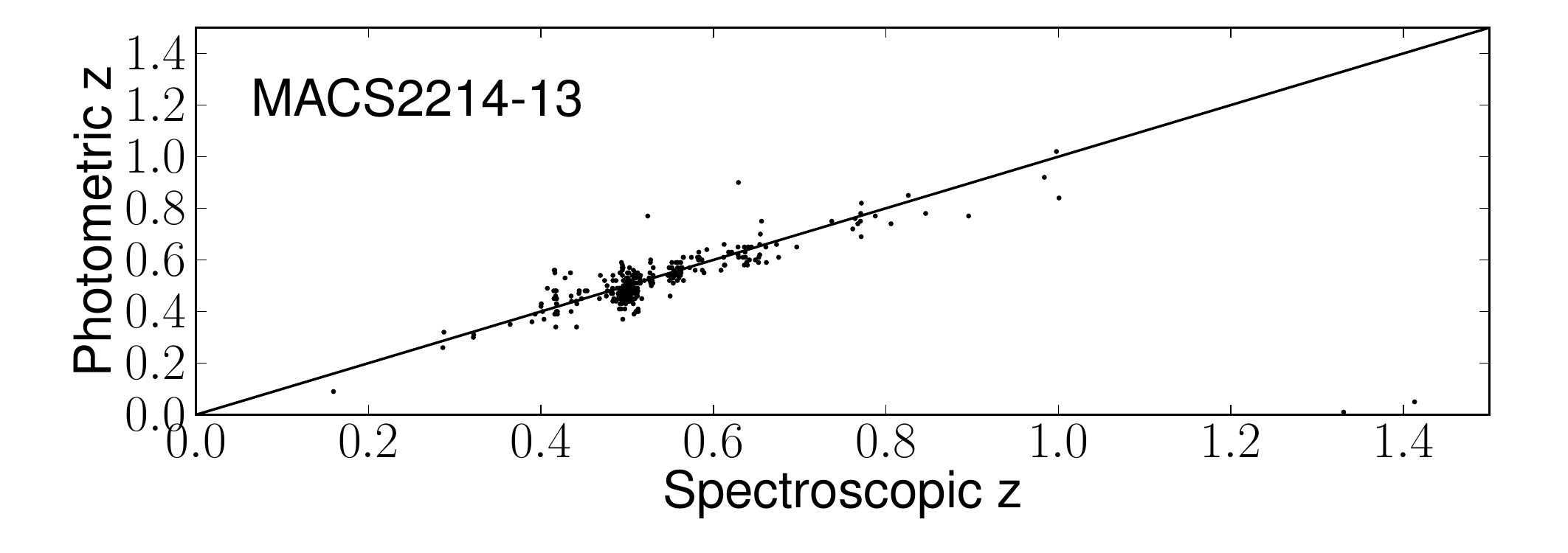} &
\includegraphics[angle=0,width=3.25in]{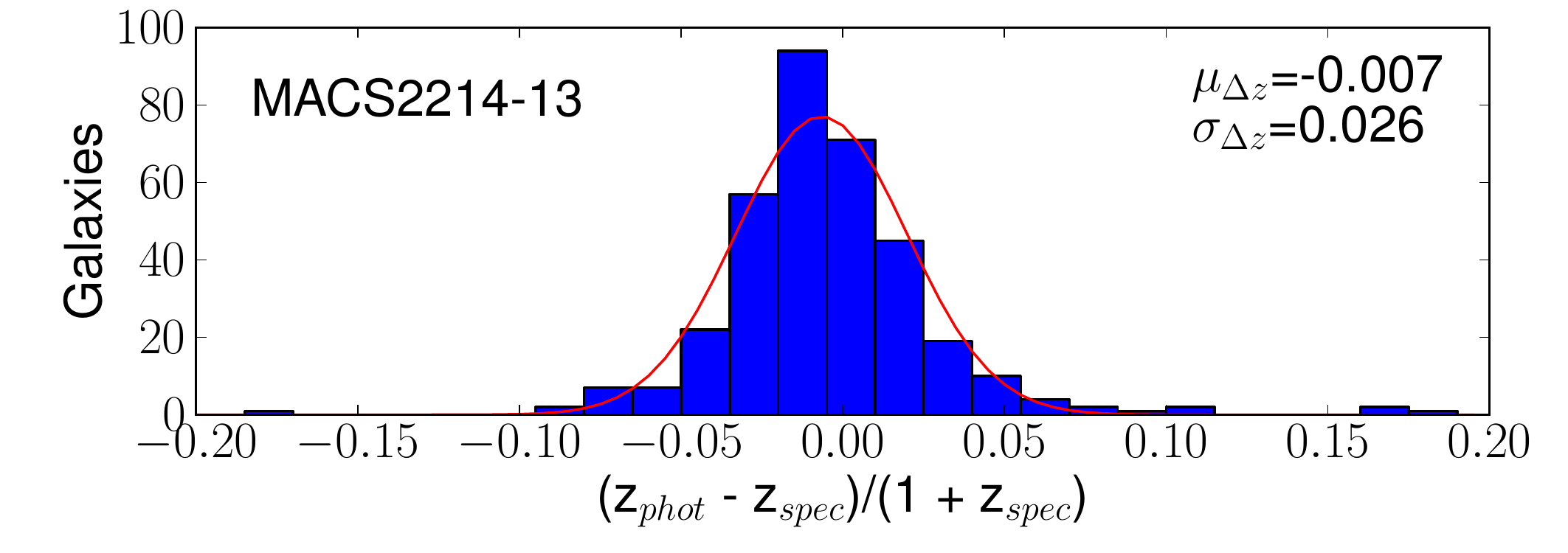} \\
\includegraphics[angle=0,width=3.25in]{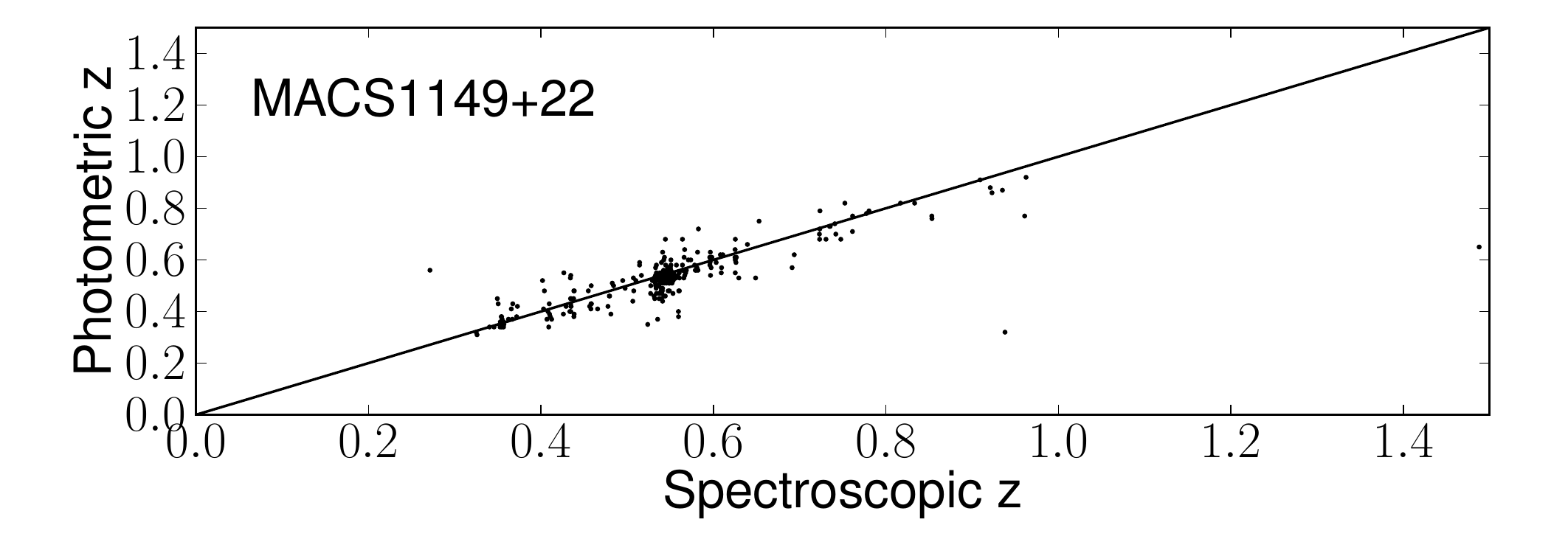} &
\includegraphics[angle=0,width=3.25in]{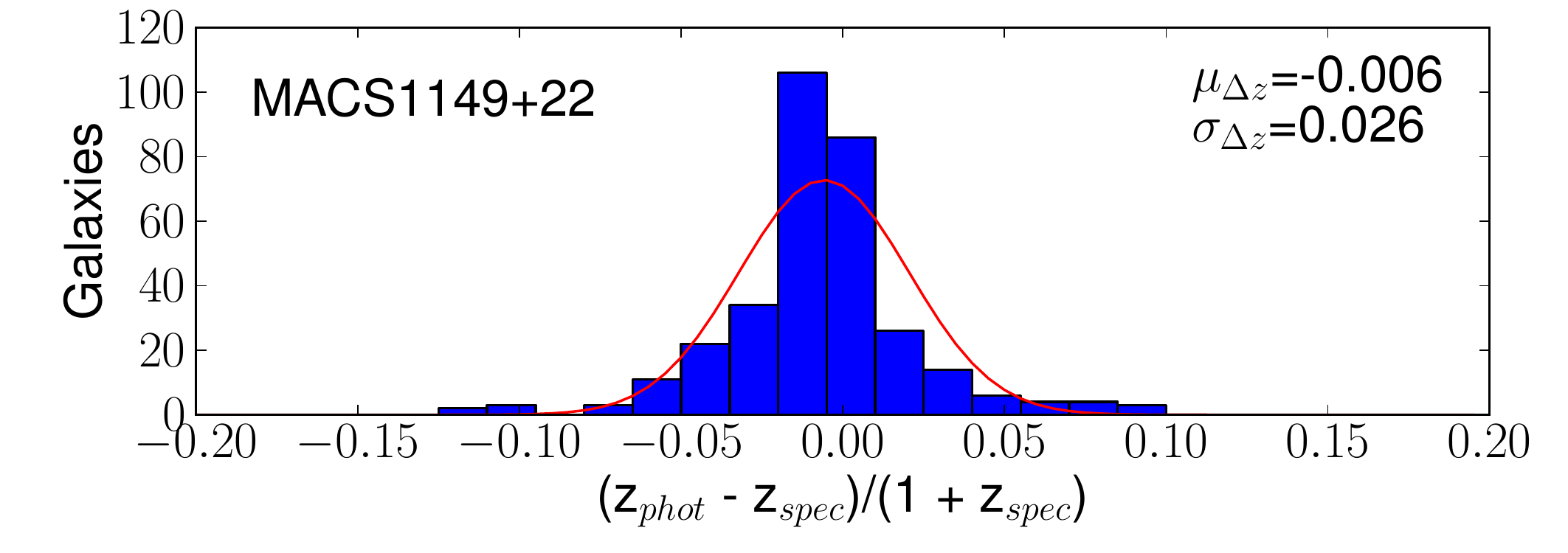} \\
\includegraphics[angle=0,width=3.25in]{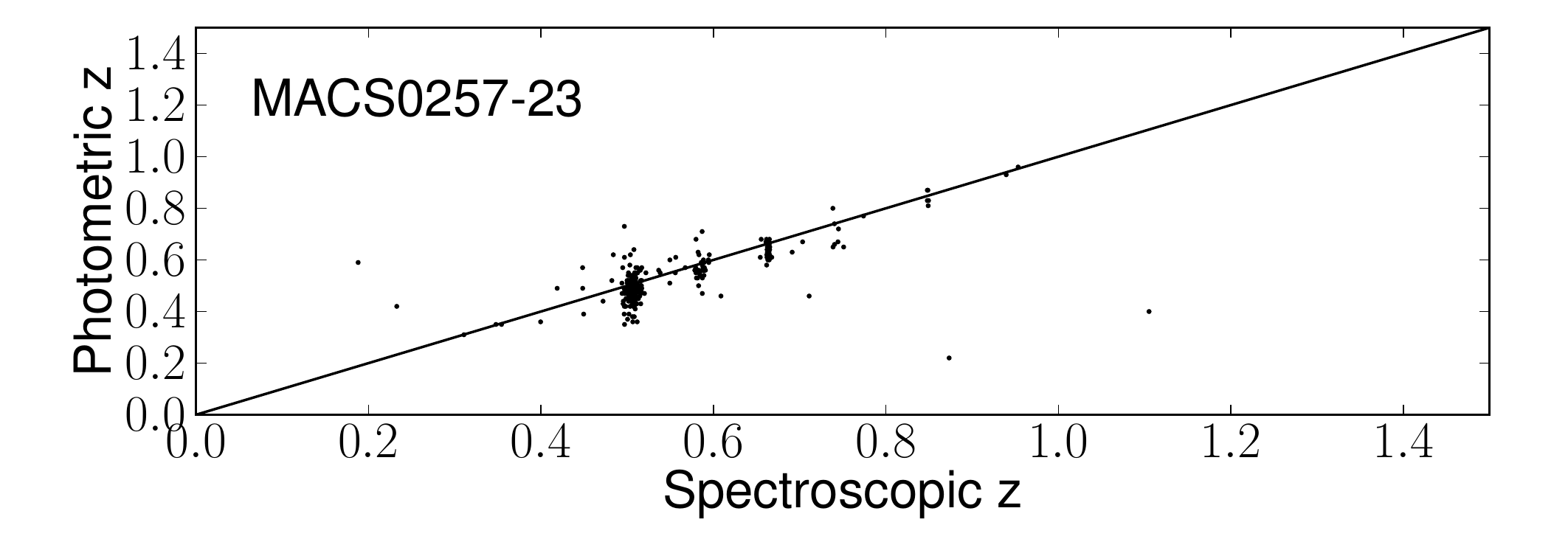} &
\includegraphics[angle=0,width=3.25in]{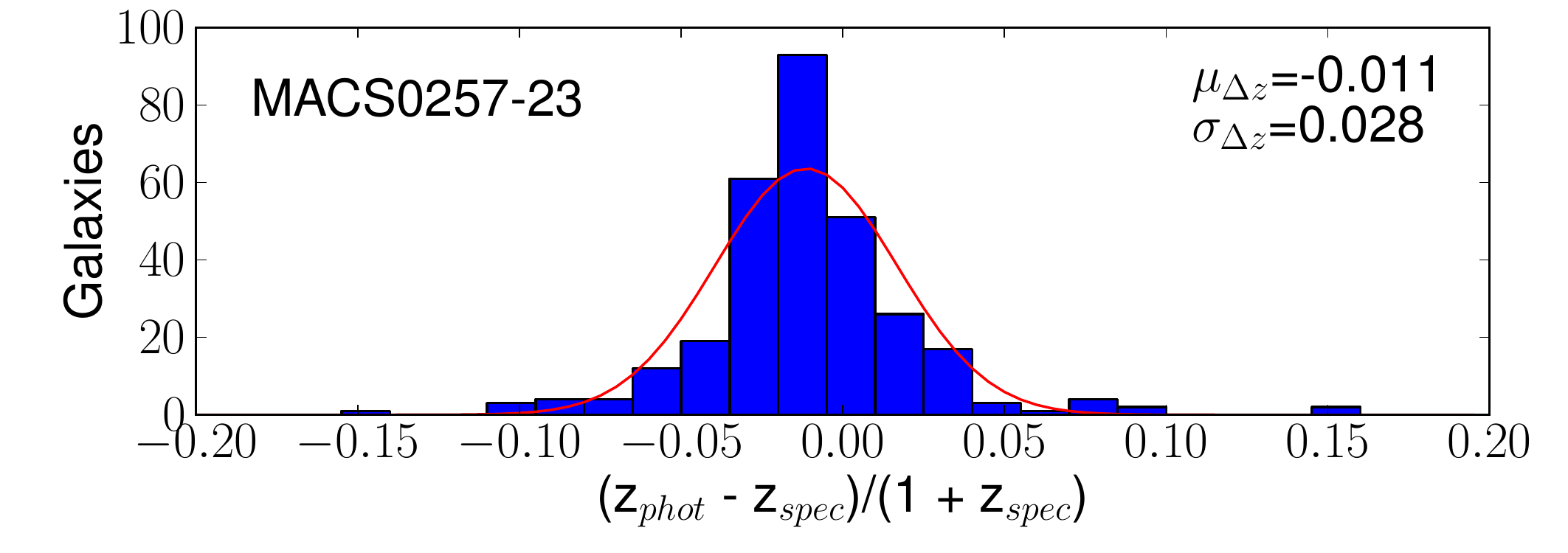} \\
\includegraphics[angle=0,width=3.25in]{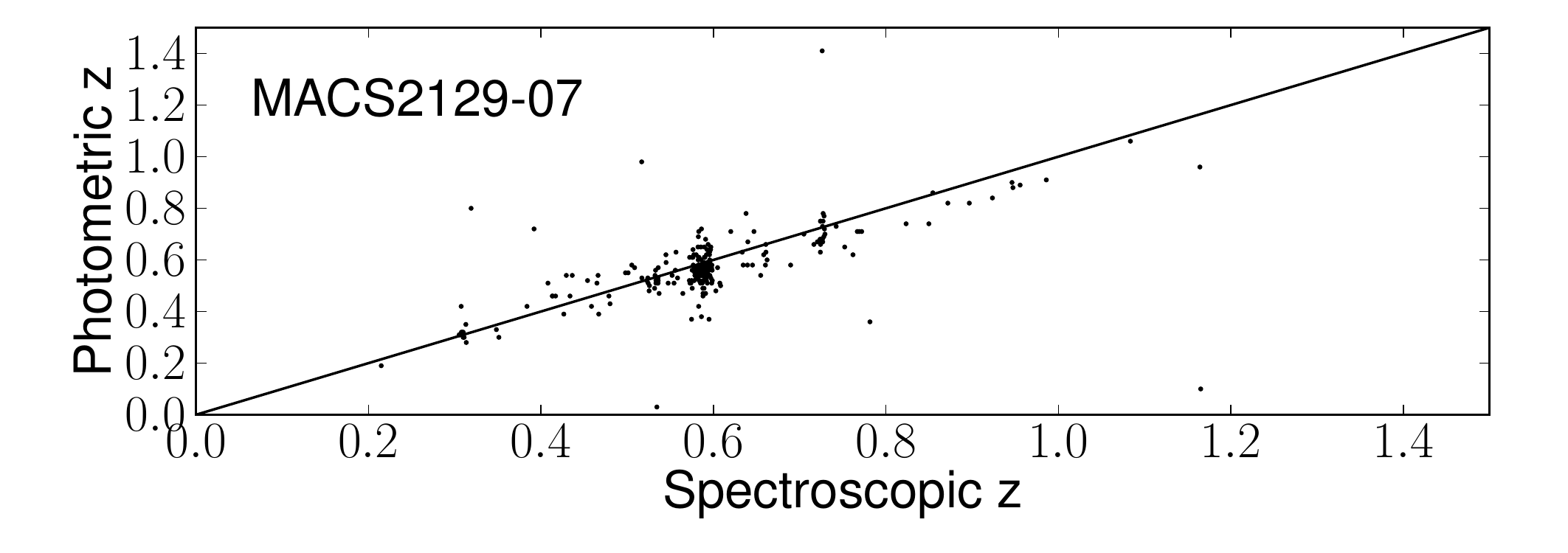} &
\includegraphics[angle=0,width=3.25in]{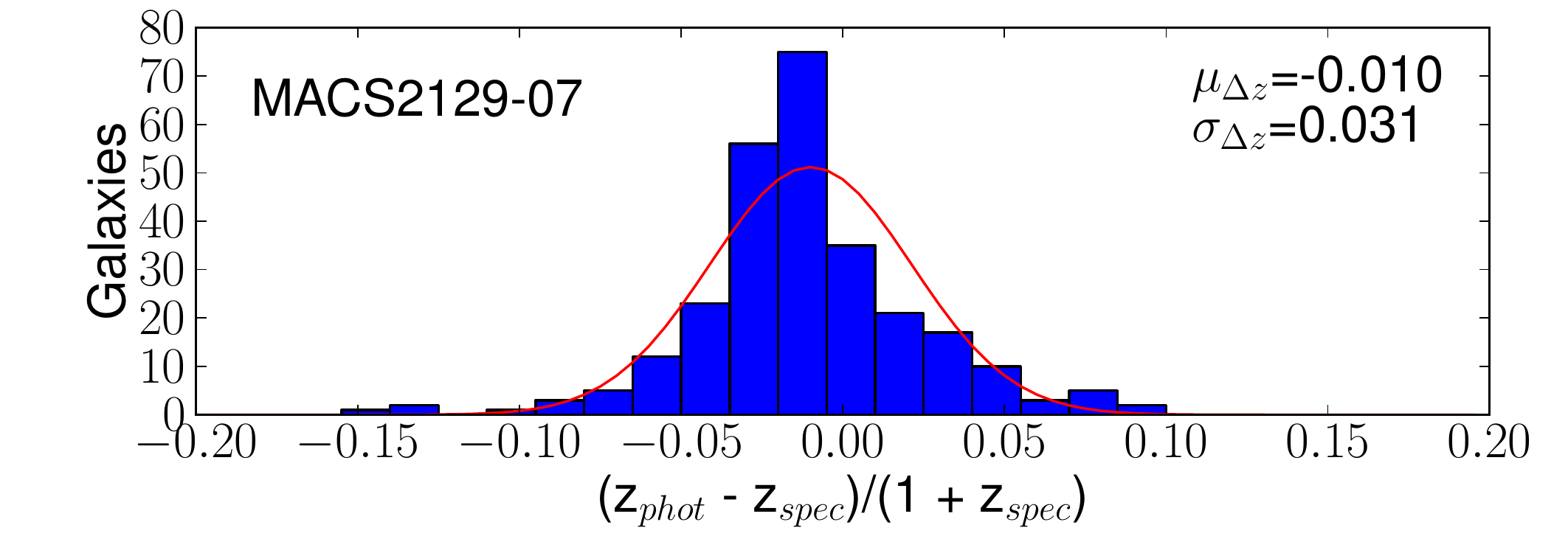} 

\end{array} $
\end{center} 
\caption{
(Plots for seven clusters are shown on next page.)
Left panels show the most probable photometric redshift $z_p$ plotted against
spectroscopic redshift $z_s$ for galaxies in fourteen cluster fields with at least five bands of photometry.
Spectroscopic redshifts are principally from Keck multi-fiber spectroscopy of cluster fields, with additional measurements collected from the NASA Extragalactic Database (NED). 
We show only photometric redshift estimates for galaxies for which the $p(z)$ probability distribution shows substantial weight close to the most probable redshift $z_p$
(BPZ \mbox{ODDS $>$ 0.9}). 
Right panels show histograms of $\Delta z$ = ($z_p - z_s$)/($1 + z_s$);
 the overlaid curves (in red) and the statistics for the mean $\mu_{\Delta z}$ and  
 standard deviation $\sigma_{\Delta z}$ correspond to 
 Gaussian fits that exclude outliers with $|\Delta z| > 0.1$.
}
\label{fig:clusterz1}
\end{figure*}

\begin{figure*}

\begin{center} $
\begin{array}{cc}

\includegraphics[angle=0,width=3.25in]{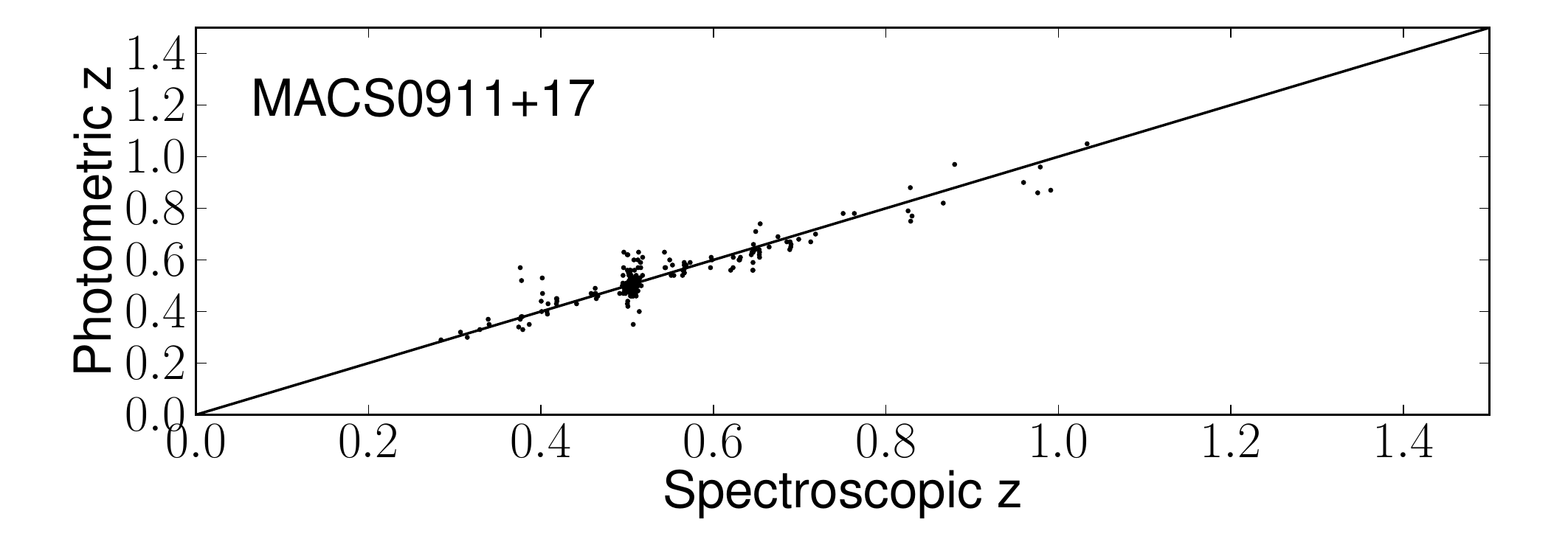} &
\includegraphics[angle=0,width=3.25in]{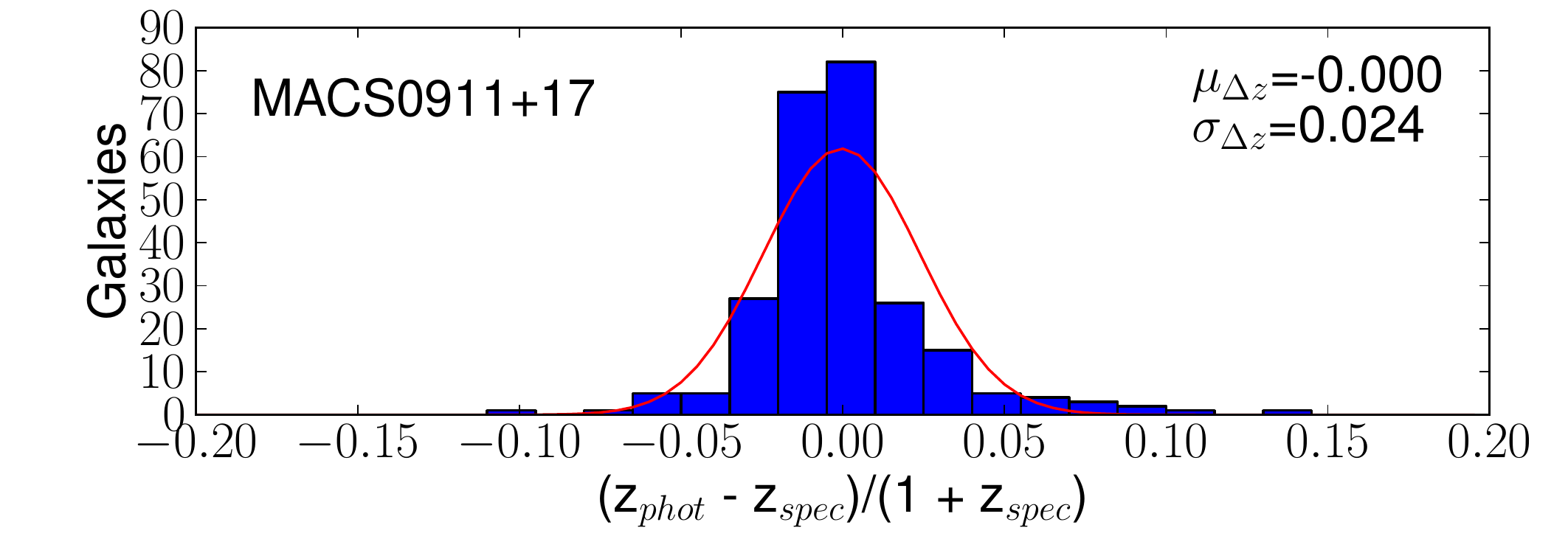} \\
\includegraphics[angle=0,width=3.25in]{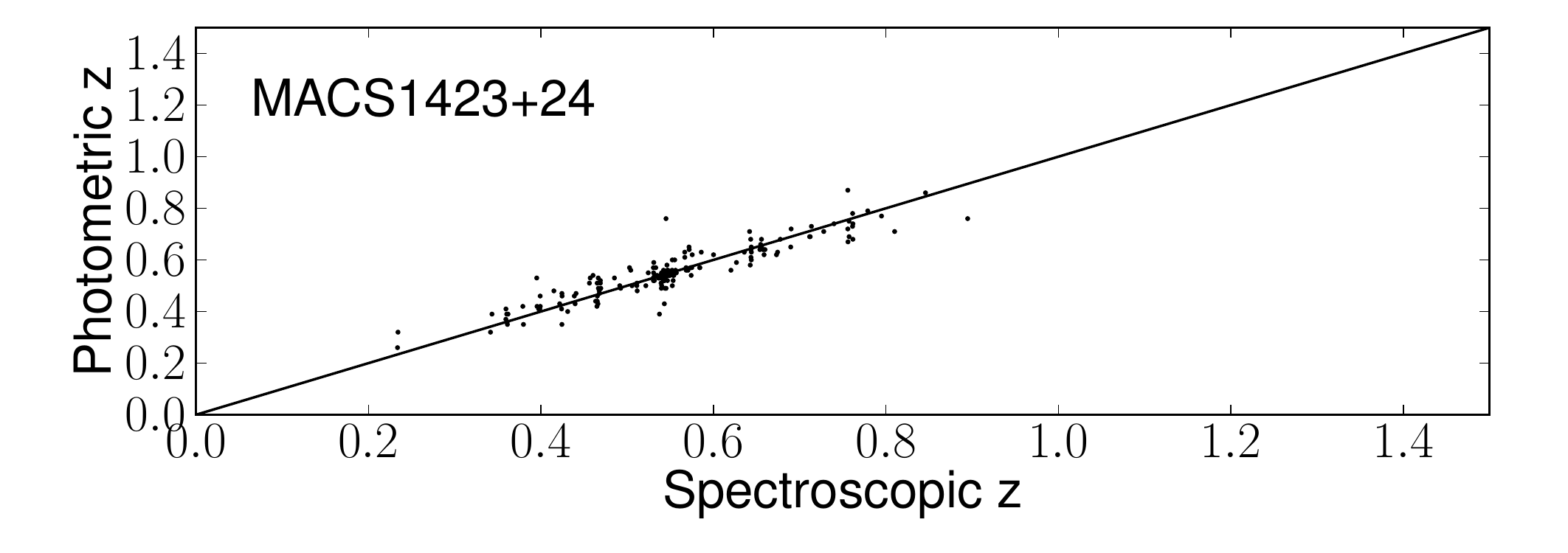} &
\includegraphics[angle=0,width=3.25in]{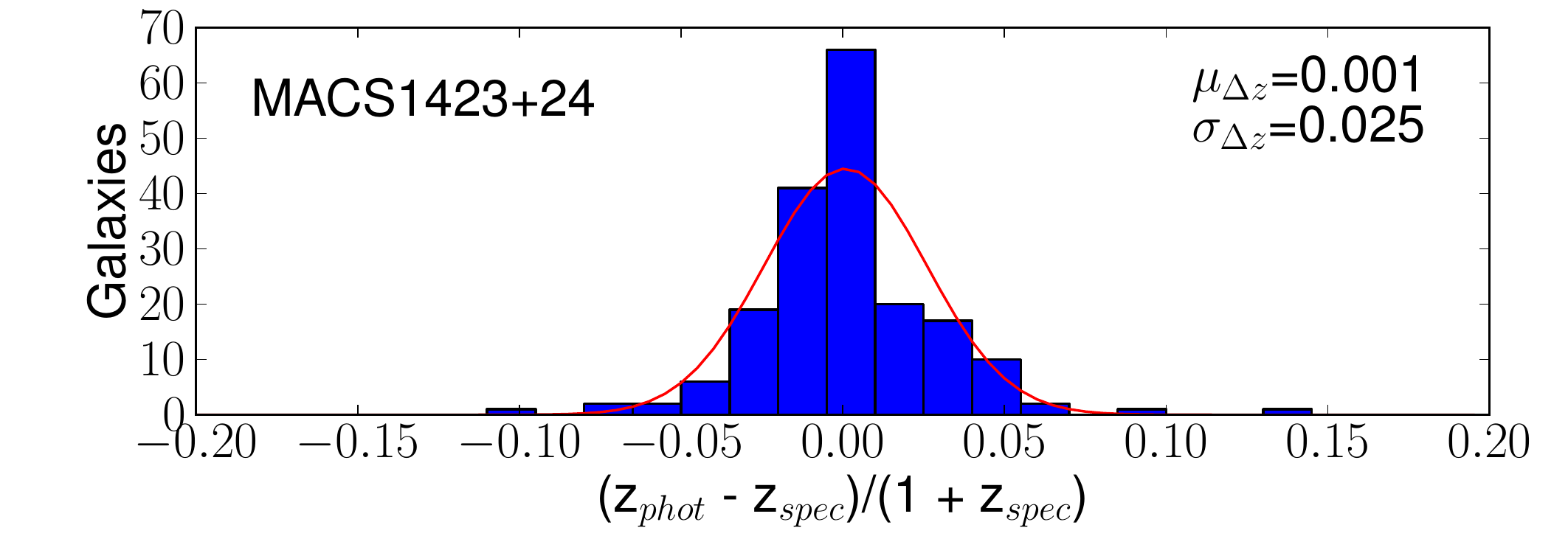} \\
\includegraphics[angle=0,width=3.25in]{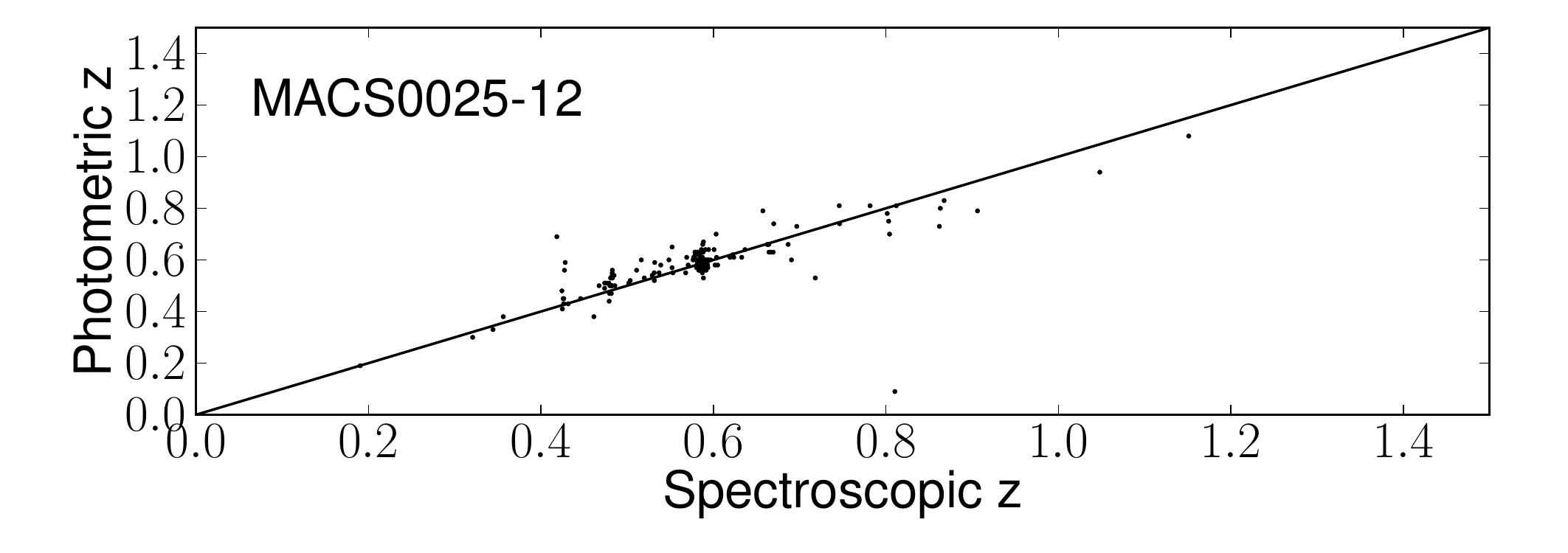} &
\includegraphics[angle=0,width=3.25in]{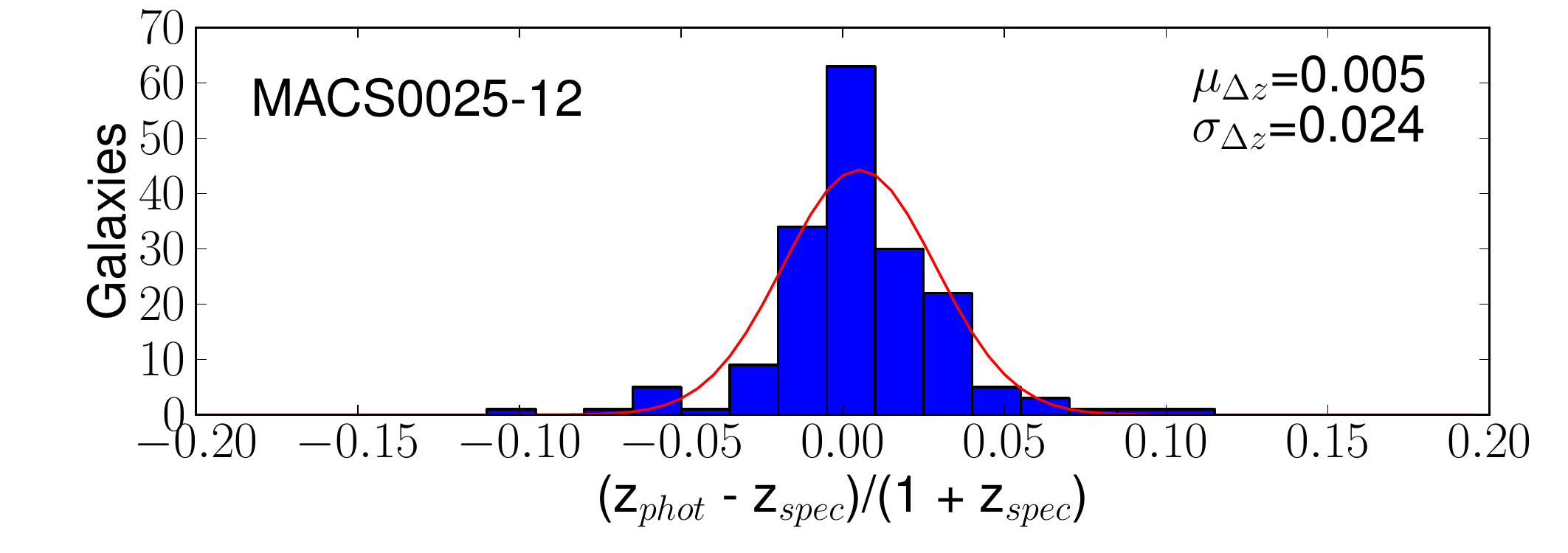} \\
\includegraphics[angle=0,width=3.25in]{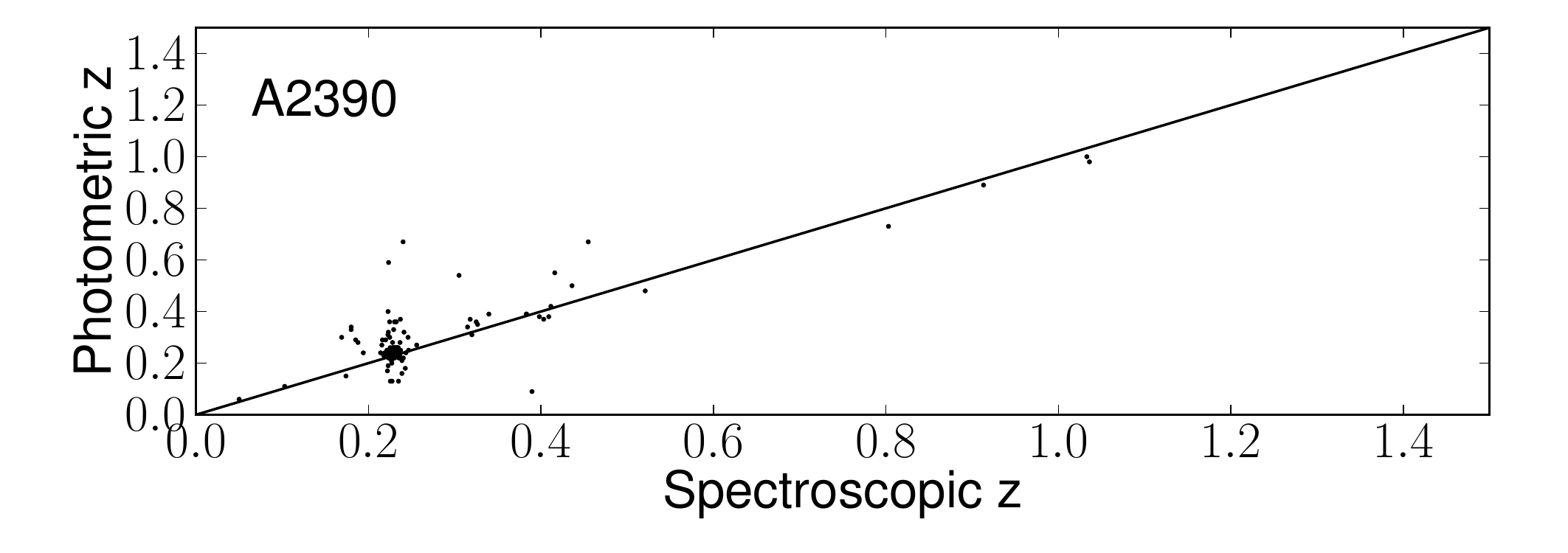} &
\includegraphics[angle=0,width=3.25in]{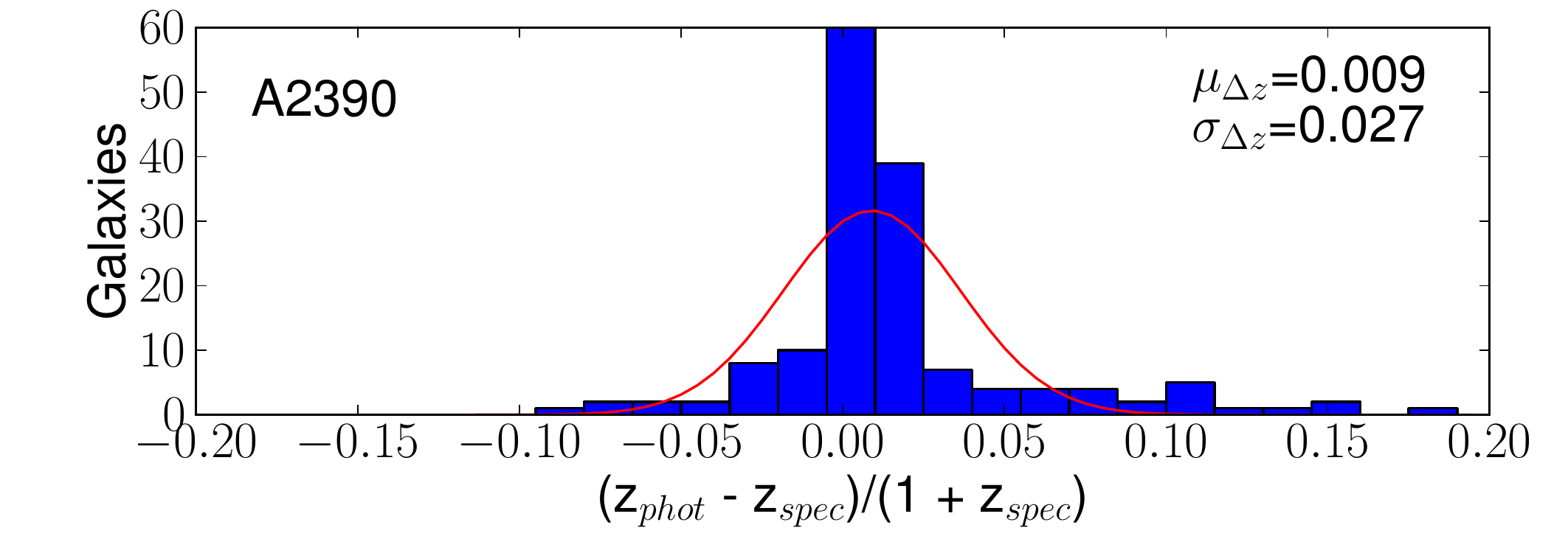} \\
\includegraphics[angle=0,width=3.25in]{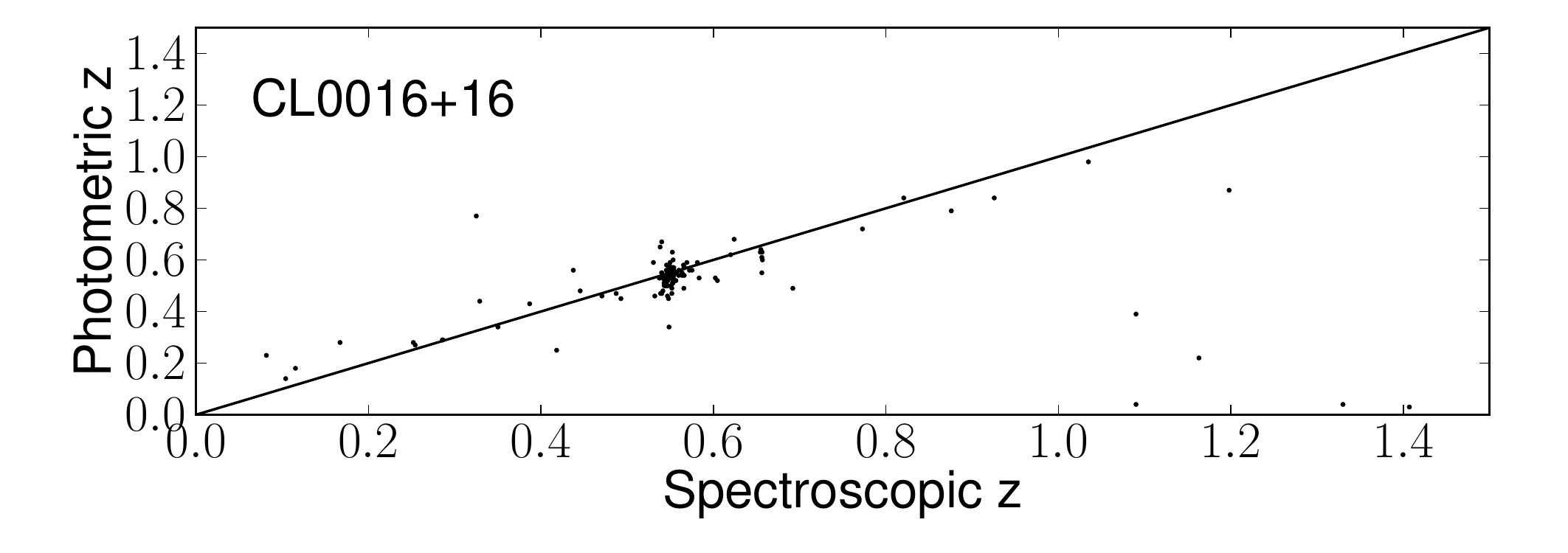} &
\includegraphics[angle=0,width=3.25in]{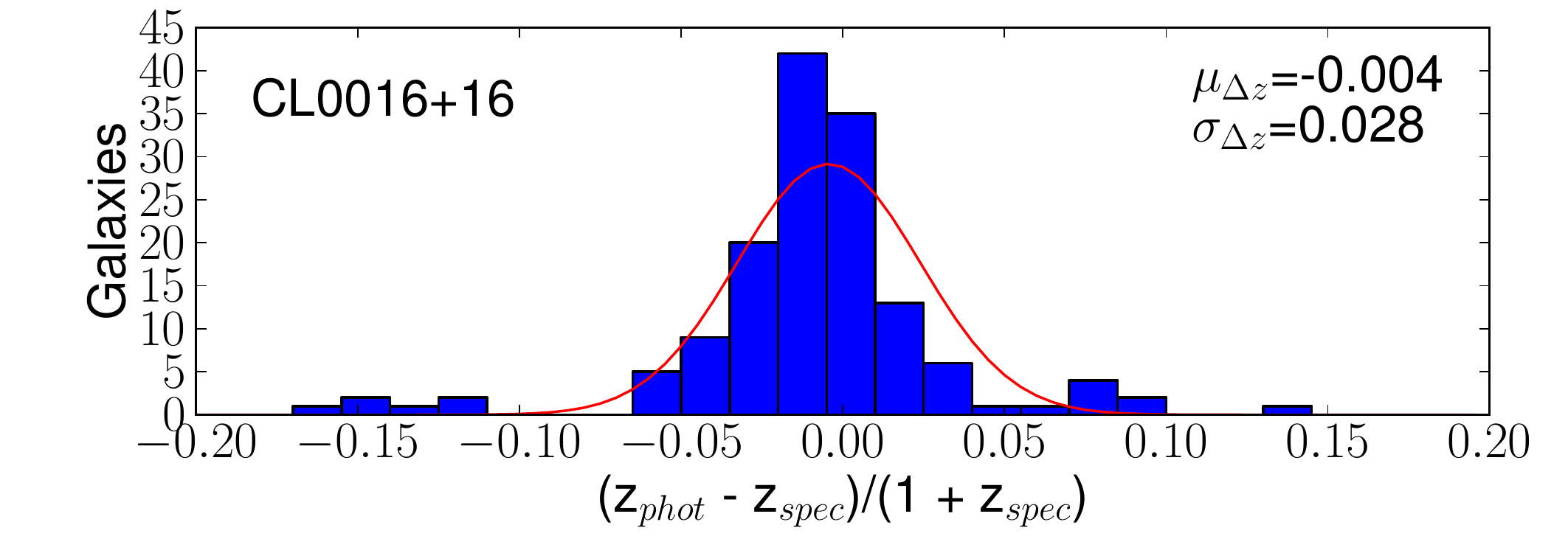} \\
\includegraphics[angle=0,width=3.25in]{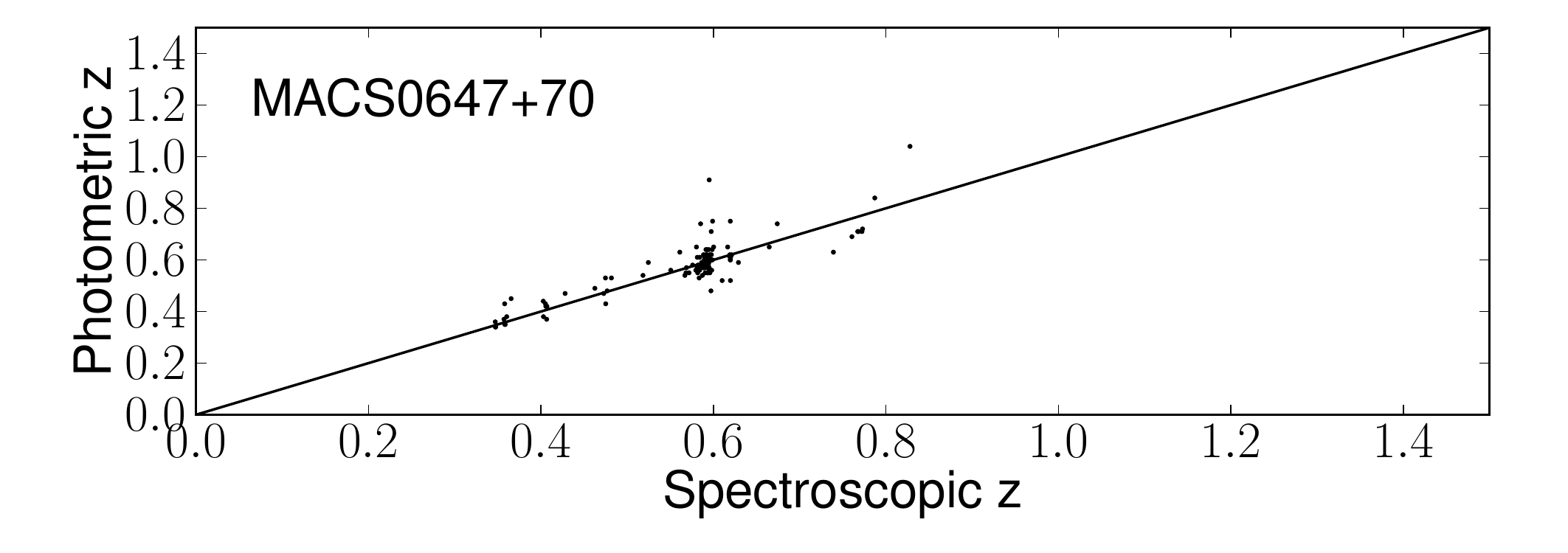} &
\includegraphics[angle=0,width=3.25in]{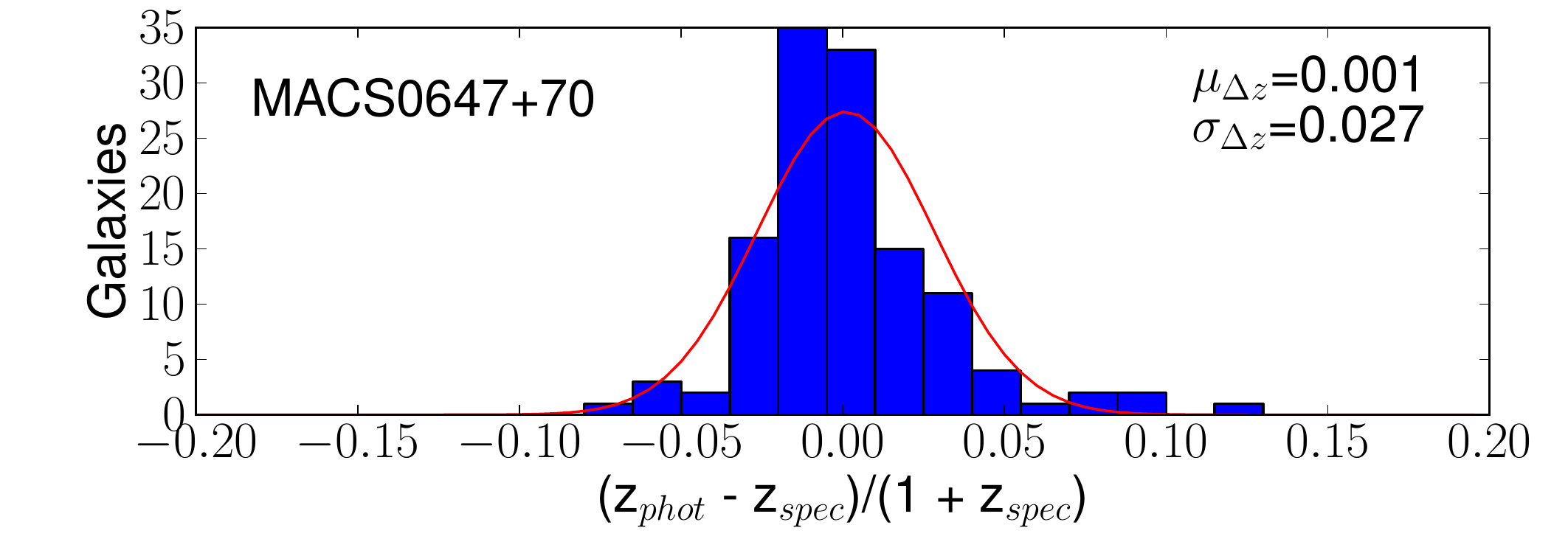} \\
\includegraphics[angle=0,width=3.25in]{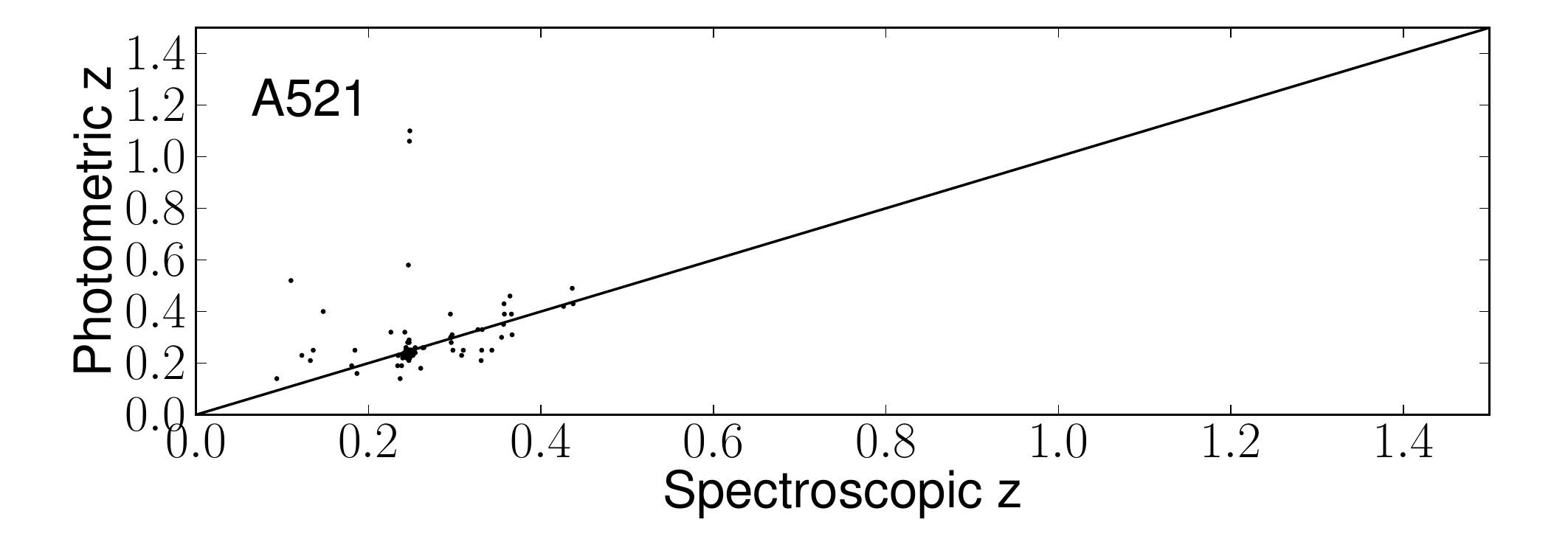} &
\includegraphics[angle=0,width=3.25in]{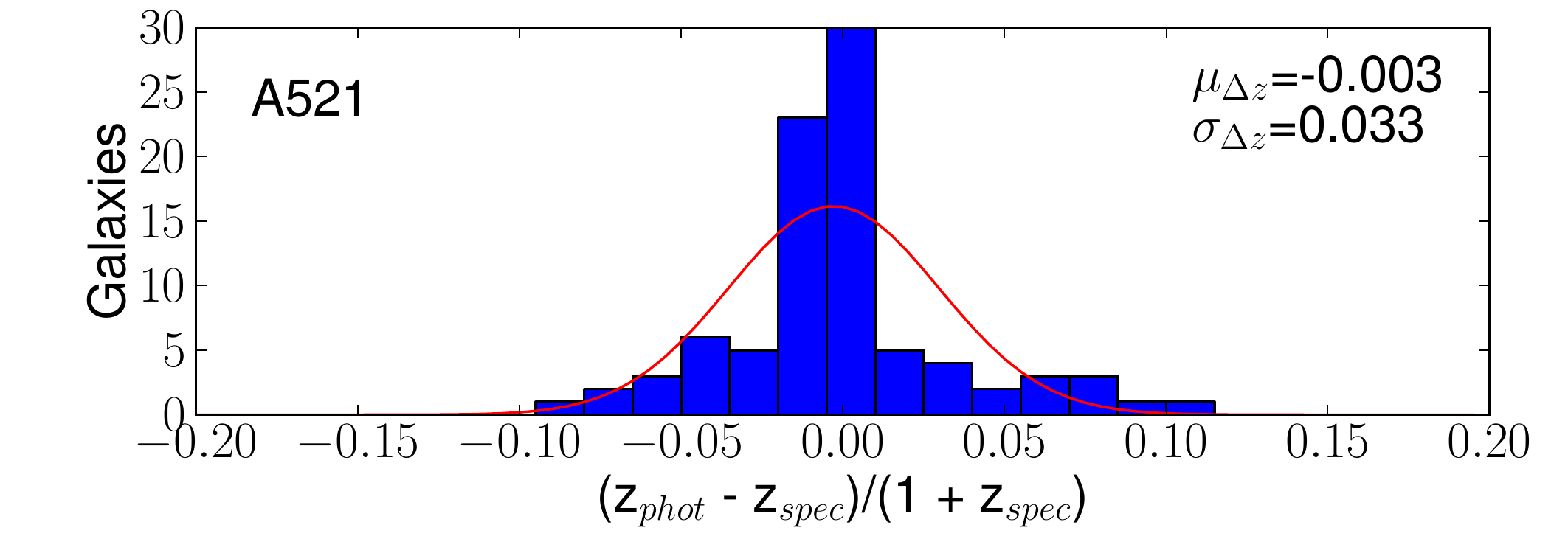} 
\end{array} $
\end{center} 
\contcaption{}

\end{figure*} 

\subsection{Testing Stellar Locus Matching Against SDSS Calibration and SFD Dust Extinction}

Of the \clusterfields~galaxy-cluster fields in the sample,
\sdssfields~have available SDSS photometry. The galaxy-cluster fields
in our sample span a wide range in RA and DEC (see Paper I), as well
as Galactic latitude and longitude, and therefore the ranges of stellar metallicity and 
dust extinction are wide. We can use the available SDSS
photometry to perform a consistency check on the spectroscopic locus
model.  We first correct the available SDSS stellar $g'r'i'z'$
photometry in each galaxy-cluster field by the Schlegel, Finkbeiner \& Davis \citeyear{sfd98} (SFD)
extinction, and then perform stellar locus matching (SLM) on the corrected
SDSS photometry.  The residual color $(g'_{\rm SLM} - r'_{\rm SLM}) -
(g'_{\rm SDSS} - r'_{\rm SDSS})$ has mean 0.006 mag and standard
deviation 0.014 mag, while the residual color $(r'_{\rm SLM} - z'_{\rm
  SLM}) - (r'_{\rm SDSS} - z'_{\rm SDSS})$ has mean $-0.005$ mag
and standard deviation 0.019 mag. 
This excellent agreement demonstrates the effectiveness of our stellar locus matching 
technique. 

\section{Photometric Redshift Algorithms and Templates}
\label{sec:photoz}

We estimate photometric redshifts with the Bayesian BPZ code (\citealt{benitez00}; \citealt{coe06}), which uses  
spectral templates as a model for the rest-frame SEDs of galaxies.
Constraining the possible redshifts of galaxies based on their broadband colors \citep{bau62} is feasible because of regular features (e.g., the \mbox{4000 \AA~break}) in
rest-frame galaxy spectra. 
Given galaxy photometry in $n$ broadband filters, the BPZ code defines $n-1$ colors $C = \{c_f \}$
relative to a `base' filter (e.g., the band in which the galaxies were selected).
BPZ computes the likelihood $p(C|z,T)$ of the observed galaxy colors $C$ 
at a discrete series of possible redshifts $z$ (e.g., 0.01, 0.02, 0.03, ...) for each galaxy template $T$.

\citet{benitez00} introduced the use of prior probability density functions $p(z,m,T)$ into galaxy photometric redshift estimation, which depend on galaxy redshift $z$, apparent magnitude $m$, and template type $T$. 
He calibrated the parameters of these prior functions using
737 galaxies, including $\sim$130 with spectroscopic redshifts, with $20 < I < 27$ from the HDFN (\citealt{williams96}), and 591 additional galaxies with spectroscopic redshifts and $20 < I < 22.5$ from the Canada-France Redshift Survey catalog (\citealt{lilly95}) with spectral classification from 
$V-I$ colors. 
When the colors $C$ are consistent with the SED templates at 
multiple redshifts, the prior function can alleviate or break redshift degeneracies. 
Here we use the \citet{benitez00} prior functions with a modest adjustment that
increases the probability at low redshift  ($z < 0.2$; see \citealt{erben09}).
The posterior redshift probability distribution is
\begin{equation}
p(z) \equiv
p(z|C,m) \propto \sum_{T}p(C|z,T)p(z,m,T).
\end{equation}

The posterior probability $p(z)$ distribution is then smoothed, using the BPZ code, with a Gaussian that we specify to have \mbox{$\sigma(z/(1+z))$=0.03}.

\subsection{Most Probable Redshift $z_p$}
\label{sec:odds}

The peak of the posterior redshift probability distribution $p(z)$ is, by definition, the most probable redshift of the galaxy, $z_p$ (the BPZ parameter BPZ\_Z\_B). 
We use BPZ to compute, from the estimated $p(z)$ distribution, the probability 
that the galaxy redshift is within $\pm$0.1(1+$z_p$) of $z_p$,

\begin{equation}
{\rm ODDS} = \int_{z_p - 0.1(1+z_p)}^{z_p + 0.1(1+z_p)} p(z) dz.
\end{equation}
For example, the ODDS statistic for a Gaussian $p(z)$ centered at $z_p$ with standard deviation $\sigma$ = 0.05(1 + $z_p$) would be 0.95. 

\subsection{Template Set}

The performance of the BPZ code depends on the template library.
For this study, we use the template set
assembled by \citet{capak04}, 
which comprises elliptical, Sbc, Scd, and Im spectra from \citet{cww80} and two starburst templates from \citet{kinney96}.
\citet{capak04} adjusted these templates both to increase agreement between 
spectroscopically determined redshifts and the BPZ most probable redshift $z_p$ and to match
the colors of galaxies 
fainter than the limiting magnitudes of spectroscopic samples. 

The \citet{capak04} templates are ordered by type (see list above), and, 
before fitting, we generate additional templates 
by interpolating eight times between adjacent templates in this ordered list (i.e., INTER=8).

Figure \ref{fig:sedfit} shows an example photometric redshift fit to a galaxy with $B_JV_JR_CI_Cz^+$ 
magnitudes.
We show the best-fit template and measured galaxy magnitudes, as well as the
$p(z)$ distribution and spectroscopic redshift value.

\section{Galaxy SED Extrapolation to Find \MakeLowercase{$u^*$} and \textit{$B_J$} Zeropoints}
\label{sec:extrapolate}
The photometric calibration of very blue filters, particularly $u^*$, from stellar locus matching is more challenging than for redder bands. 
Absorption by metals at near-UV and blue-optical wavelengths causes the $u^*$ and $B_J$ magnitudes of main sequence stars to depend 
more strongly on the chemical abundances in the stellar atmospheres than $V_J,R_C,I_C,$ and $z^{+}$ magnitudes (see Section~\ref{sec:metallicity}). 
Consequently, the stellar locus shows more variation across the sky in the $u^*$ and $B_J$ bands than in redder bandpasses. 

We have therefore developed a technique to adjust Megaprime $u^*$-band and SuprimeCam $B_J$-band zeropoints after 
stellar locus matching calibration. 
We use the BPZ photometric redshift code to calculate for each galaxy the expected $u^*$ and $B_J$ magnitudes for each galaxy 
given the calibrated photometry at redder observer-frame wavelengths (e.g., $V_J,R_C,I_C,$ and $z^{+}$),
based on the best-fit SED template model (e.g., elliptical, Sbc) and redshift. 
Our method employs a training set of 10,000 galaxies, drawn randomly from 
the galaxies whose $u^*$ and $B_J$ magnitudes have uncertainties
less than 0.05 mag.
We modify the photometry catalog so that $B_J$ magnitudes are assigned an 
uncertainty of 0.2 and every $u^+$ magnitude is assigned an uncertainty of 90 mag (effectively removing any constraint from $u^+$ magnitudes).

The zeropoint corrections for the $u^*$ and $B_J$ bands are estimated as
the median difference $\mu_{1/2}$ between the predicted and measured magnitudes for galaxies,
\begin{equation}
ZP_{\rm new} - ZP_{\rm old} = \mu_{1/2}(m_{\rm gal}^{\rm BPZ} - m_{\rm gal}^{\rm exp}),
\end{equation}
although the peak as opposed to median difference could possibly yield improved accuracy.
After calculating photometric redshifts for the training set of galaxies, we 
then reject any galaxy with BPZ ODDS parameter less than $0.95$ and most probable photometric
redshift smaller than $z_p=0.4$ so that the 4000 \AA~break does not fall near the $u^*$ or $B_J$ bandpasses.
These criteria yield a sample of $\sim$1000 training galaxies.
We apply this technique to fields for which we have $u^*$ or $B_J$ observations in addition to observations in at least four other bandpasses.

The zeropoints we estimate with this technique will of course depend on the accuracy of the
BPZ template set used to compute the expected $u^*$ and $B_J$ magnitudes. 
This procedure generally yields small $\sim$0.01 mag adjustments to the $B_J$-band
zeropoint calibrated through stellar locus matching. 
We do not use available $u^*$-band  photometry for measuring cluster masses (Paper III).

\section{Photometric calibration tests}
\label{sec:confirm}

We demonstrate the reliability of the stellar-locus-matching
zeropoint calibration, as well as the performance of the redshift estimation, by comparing
our estimates for the most probable photometric redshift $z_p$ and 
the posterior probability distribution $p(z)$ to redshifts derived from spectra 
in both our galaxy cluster fields and the Cosmic Evolution Survey (COSMOS; \citealt{scoville07}) field, as well as accurate 
photometric redshifts from thirty-band photometry of COSMOS galaxies (COSMOS-30; \citealt{caa07}; \citealt{ics09}). 

\begin{figure}
\centering
\includegraphics[angle=0,width=3.25in]{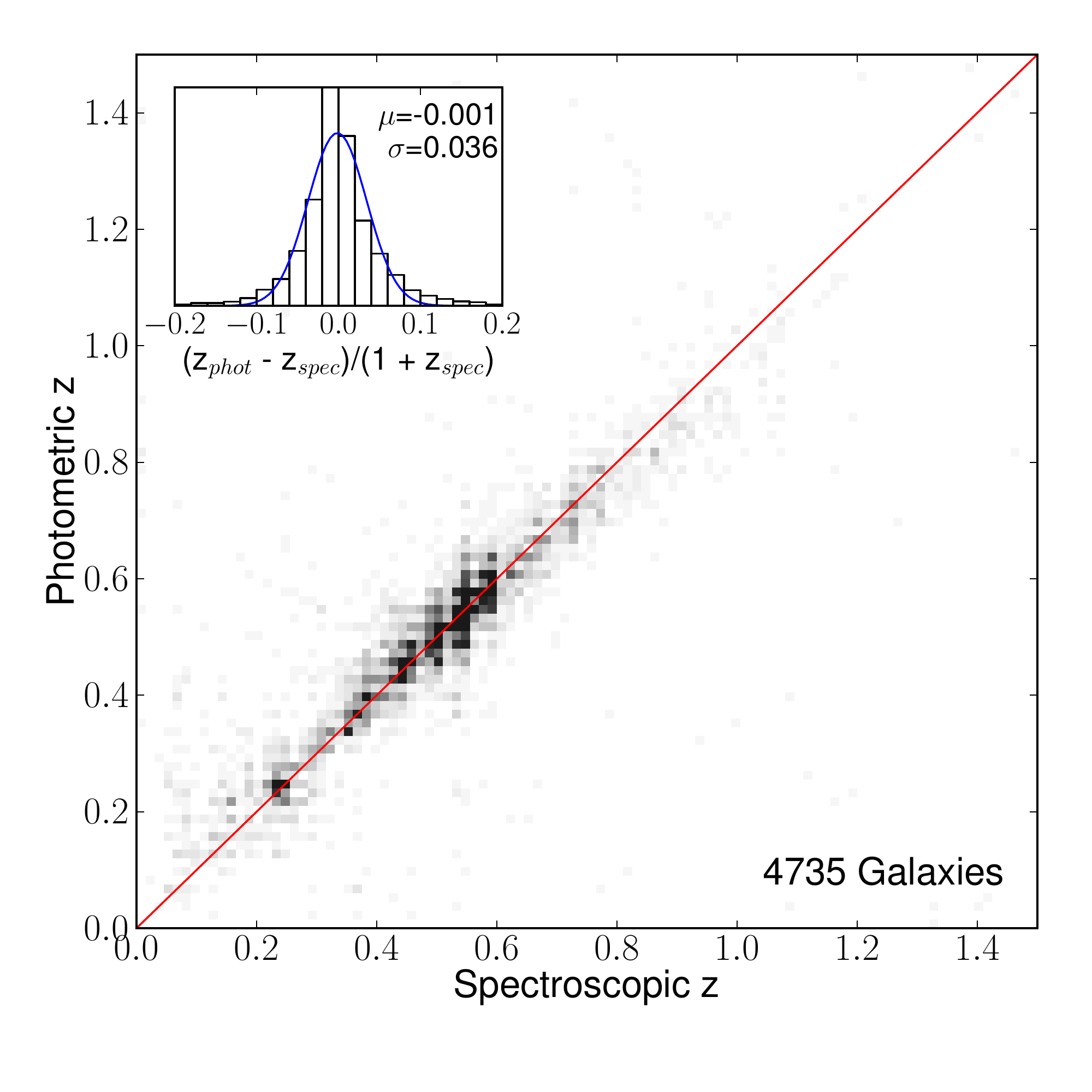} 
\caption{
Most probable photometric redshift $z_p$ plotted against
spectroscopic redshift $z_s$ for 4735 galaxies with BPZ ODDS $>$ 0.9, in cluster fields with at least five bands of photometry.
The grayscale intensity is proportional to the number of galaxies in each bin. 
Spectroscopic redshifts are from Keck, Gemini, and VLT spectroscopy of cluster fields, with additional measurements collected from the NASA Extragalactic Database. 
The histogram in the top left corner shows the distribution of
$\Delta z$ = ($z_p - z_s$)/($1 + z_s$);
the overlaid curve (in blue) and the statistics for the mean $\mu$ and  
 standard deviation $\sigma$ correspond to 
 Gaussian fits that exclude outliers with $|\Delta z| > 0.1$.
}
\label{fig:clusterstack}
\end{figure}

\subsection{Spectroscopic Redshift Comparisons in the Cluster Fields}

\subsubsection{Galaxy Samples in Cluster Fields}

We have measured, or compiled from the NASA Extragalactic Database\footnote{http://ned.ipac.caltech.edu/}, spectroscopic redshifts for 5007~galaxies in \speczfields~cluster fields
with five bands of photometry. 
We acquired spectra for galaxies in our cluster fields  (e.g., \citealt{barrett06}; \citealt{ma071708}) 
with the Deep Imaging Multi-Object Spectrograph (DEIMOS) on the Keck-II telescope, 
the Low Resolution Image Spectrometer (LRIS) on Keck-I, the Gemini Multi-Object Spectrographs (GMOS) on the Gemini
telescope, and the Visible Multi-Object Spectrograph (VIMOS) on the VLT.
Of the 5007 galaxies, 4735 (95\%) have BPZ ODDS $>$ 0.9.
In Figure \ref{fig:clusterz1}, we plot the most probable photometric redshift $z_p$ versus the spectroscopic redshift $z_s$ for these galaxies with BPZ ODDS $>$ 0.9 in the 14 cluster fields with the largest numbers of available spectroscopic redshifts. 
The photometric redshifts $z_p$ for each of these 14 fields have typical accuracy of \mbox{$\sigma((z_p - z_s)/(1+z_s)) \lesssim$~\accuracy} and bias 
$(z_p - z_s) / (1 + z_s) \lesssim 0.01$, 
with \mbox{$|(z_p - z_s)/(1+z_s)| > 0.1$} outliers excluded.

\subsubsection{Combined Sample of Galaxies in Cluster Fields}

Figure \ref{fig:clusterstack} compares the most probable photometric redshift $z_p$ to the spectroscopic redshift $z_s$ for
the combined sample of galaxies with spectroscopic redshifts and BPZ \mbox{ODDS $>$ 0.9}, across all cluster fields.
The photometric redshifts have accuracy of \mbox{$\sigma((z_p - z_s)/(1+z_s)) = 0.036$} and bias of \mbox{$\langle(z_p - z_s) / (1 + z_s)\rangle = -0.001$}, 
with \mbox{$|z_p - z_s|/(1+z_s) > 0.1$} outliers  (8.2\% of galaxies) excluded.

The \mbox{$\sigma((z_p - z_s)/(1+z_s))=0.036$} distribution is tighter than the \mbox{$\sigma ((z_p - z_s)/(1+z_s))\approx0.05$} that might be expected for galaxies with BPZ \mbox{ODDS $\approx$ 0.95} (as defined in Section~\ref{sec:odds}). 
A discrepancy, here as well as in comparisons that follow, can reasonably be attributed to the fact that the BPZ $p(z)$ distribution is smoothed with a \mbox{$\sigma(z/(1+z))$=0.03} normal distribution before the ODDS parameter is calculated, 
and the fact that we reject \mbox{$|(z_p - z_s)/(1+z_s)| > 0.1$} outliers
when we compute \mbox{$\sigma((z_p - z_s)/(1+z_s))$}.

\subsection{Photometric Redshifts of Galaxies on the Red Sequence}
\label{sssec:redseq}
\begin{figure}
\centering
\includegraphics[angle=0,width=3.25in]{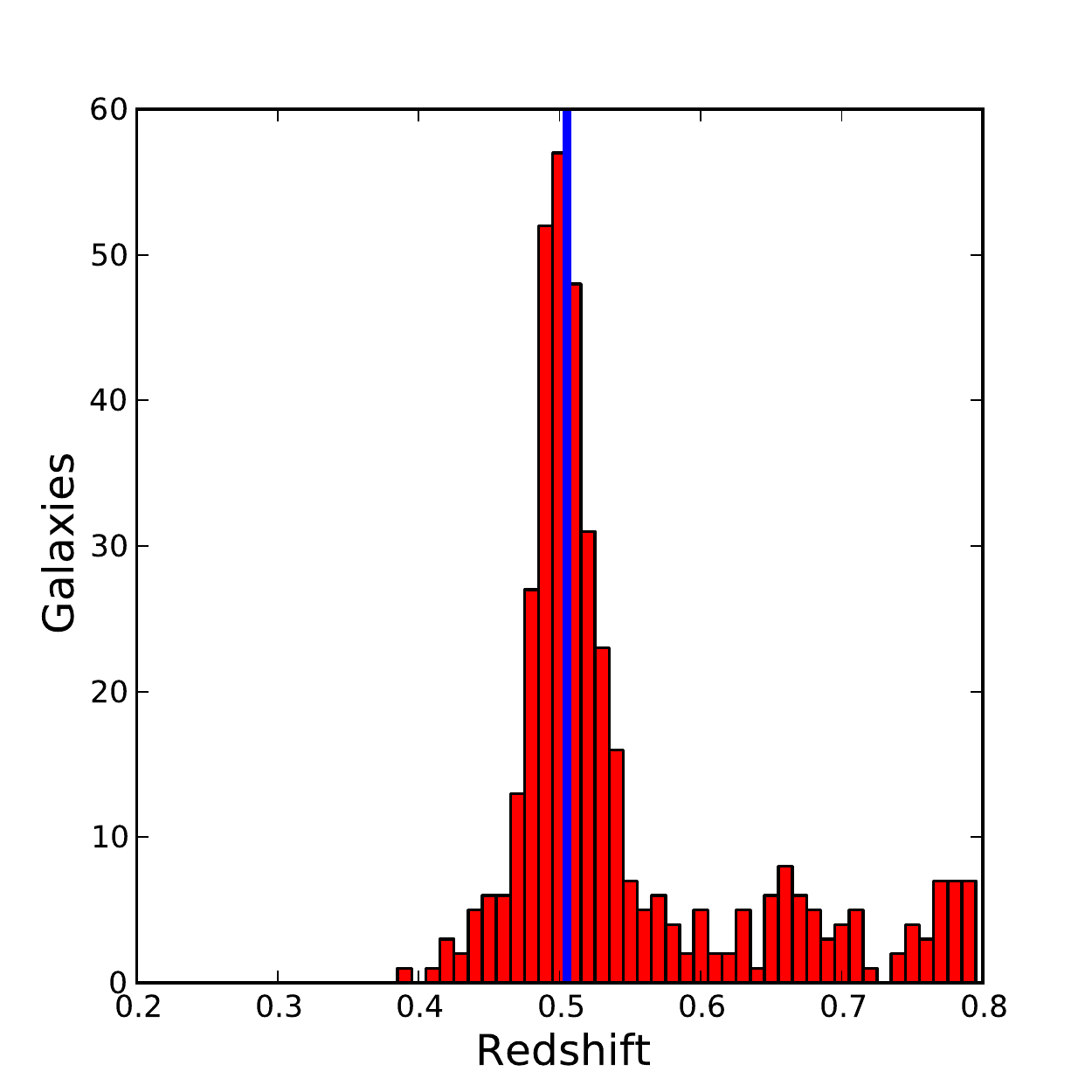} 
\caption{
Histogram of the most probable photometric redshift $z_p$ for members of the red sequence in MACS0911+17.
The spectroscopic redshift of the cluster ($z_s=0.51$) is shown as a blue vertical line.
See Sec.~\ref{sssec:redseq} for details on the selection of red-sequence galaxies.
These plots provide a diagnostic check of the photometric redshift quality available even
when few spectroscopic redshifts are available. 
}
\label{fig:red sequence}
\end{figure}

Using plots of galaxy color versus magnitude for each cluster field (e.g., $V_J - R_C$ versus $V_J$), we find the best-fit slope, intercept, and width of the distributions of red sequence galaxies for each cluster.
We use these parameters to identify the galaxies along the red sequence (see Paper III for details).
These samples may include foreground and background galaxy
populations that happen to have the same colors and magnitudes as the red sequence -- 
a contamination that becomes increasingly significant at fainter magnitudes. 
Figure~\ref{fig:red sequence} shows the histogram of most probable photometric redshifts $z_p$ 
with BPZ ODDS $>$ 0.90, for 
galaxies along the MACS0911+17 red sequence.

\begin{figure}
\centering
\includegraphics[angle=0,width=3.25in]{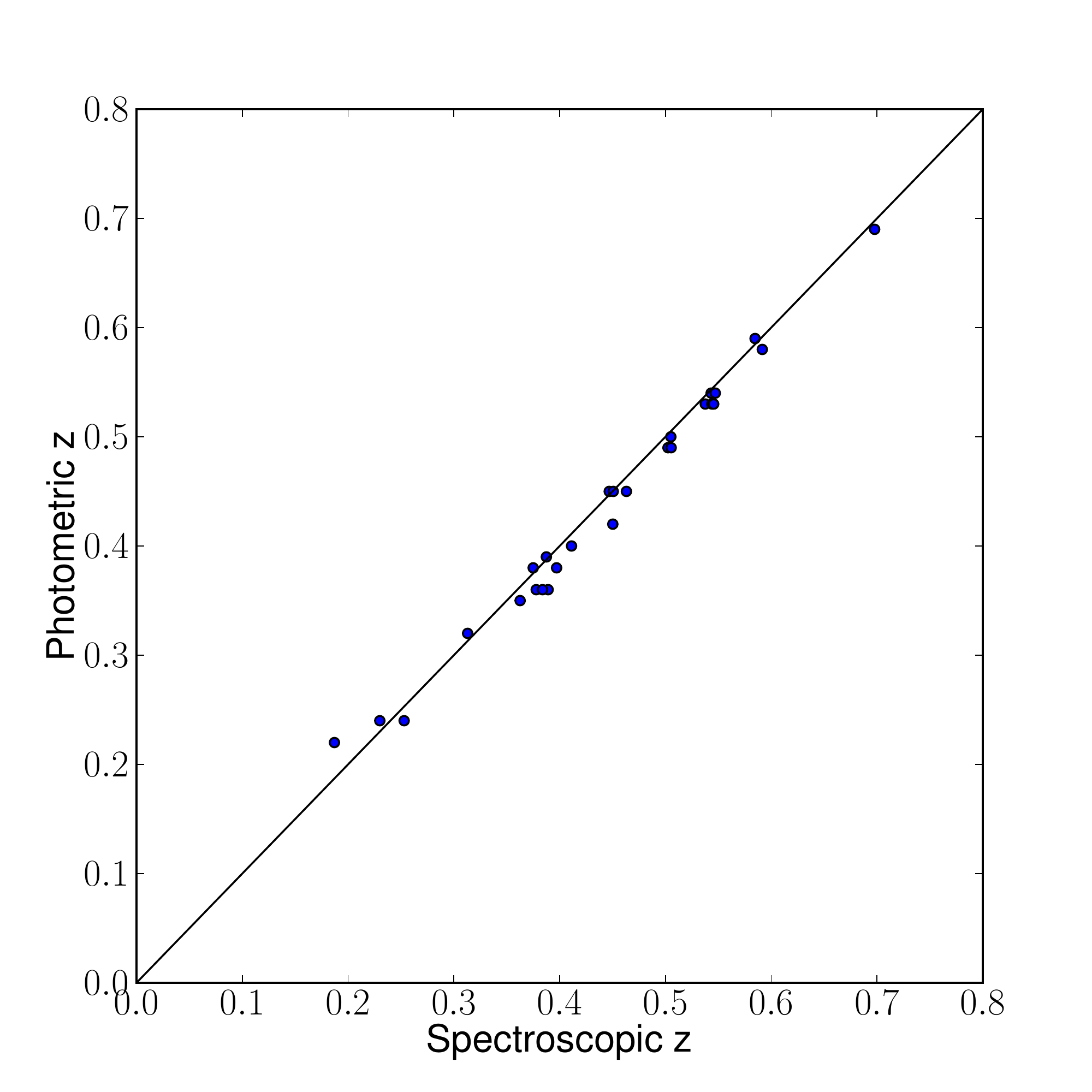} 
\caption{Peak photometric redshift $z_{\rm peak}$, for galaxies on the red sequence in each 
cluster field, plotted against the spectroscopic cluster redshift $z_{\rm cluster}$. 
The redshift estimate $z_{\rm peak}$ corresponds to the center  of the $z_p$ redshift bin (with bin width $\Delta z = 0.01$) with the greatest number
of red sequence members. 
Figure~\ref{fig:red sequence} shows an example $z_p$ histogram, for the MACS 0911+17 red sequence. 
}
\label{fig:redseqclusters}
\end{figure}  

For cluster fields imaged through five or more filters (e.g., $B_JV_JR_CI_Cz^+$), we plot  in Figure~\ref{fig:redseqclusters} the peak redshift in the histogram of most probable redshifts, $z_{\rm peak}$, against the galaxy cluster spectroscopic redshift $z_{\rm cluster}$.
While we have not otherwise made use of available near-UV photometry when computing photometric redshifts, 
we include CFHT MegaPrime $u^*$ photometry to estimate $z_{\rm peak}$ for the lowest redshift cluster in the sample, Abell 383, since near-UV photometry, combined with blue optical photometry, bracket the 4000 \AA~break of $z < 0.2$ galaxies. 
The dispersion of $z_{\rm peak}$ is \mbox{$\sigma((z_{\rm peak} - z_{\rm cluster}) / (1 + z_{\rm cluster})) = 0.011$} and the bias is  \mbox{$\langle(z_{\rm peak} - z_{\rm cluster}) / (1 + z_{\rm cluster})\rangle = -0.005$}. 
The histogram bins have $\Delta z = 0.01$ width, which places a lower limit on the accuracy of the cluster redshift 
estimates. 
A more accurate estimate of the peak cluster redshift could be
obtained by applying, for example, a weighted Gaussian-kernel density estimator
to interpolate between adjacent bins and by estimating photometric redshifts with greater precision than $\Delta z = 0.01$.

\begin{figure*}
\centering
\includegraphics[angle=0,width=3.25in]{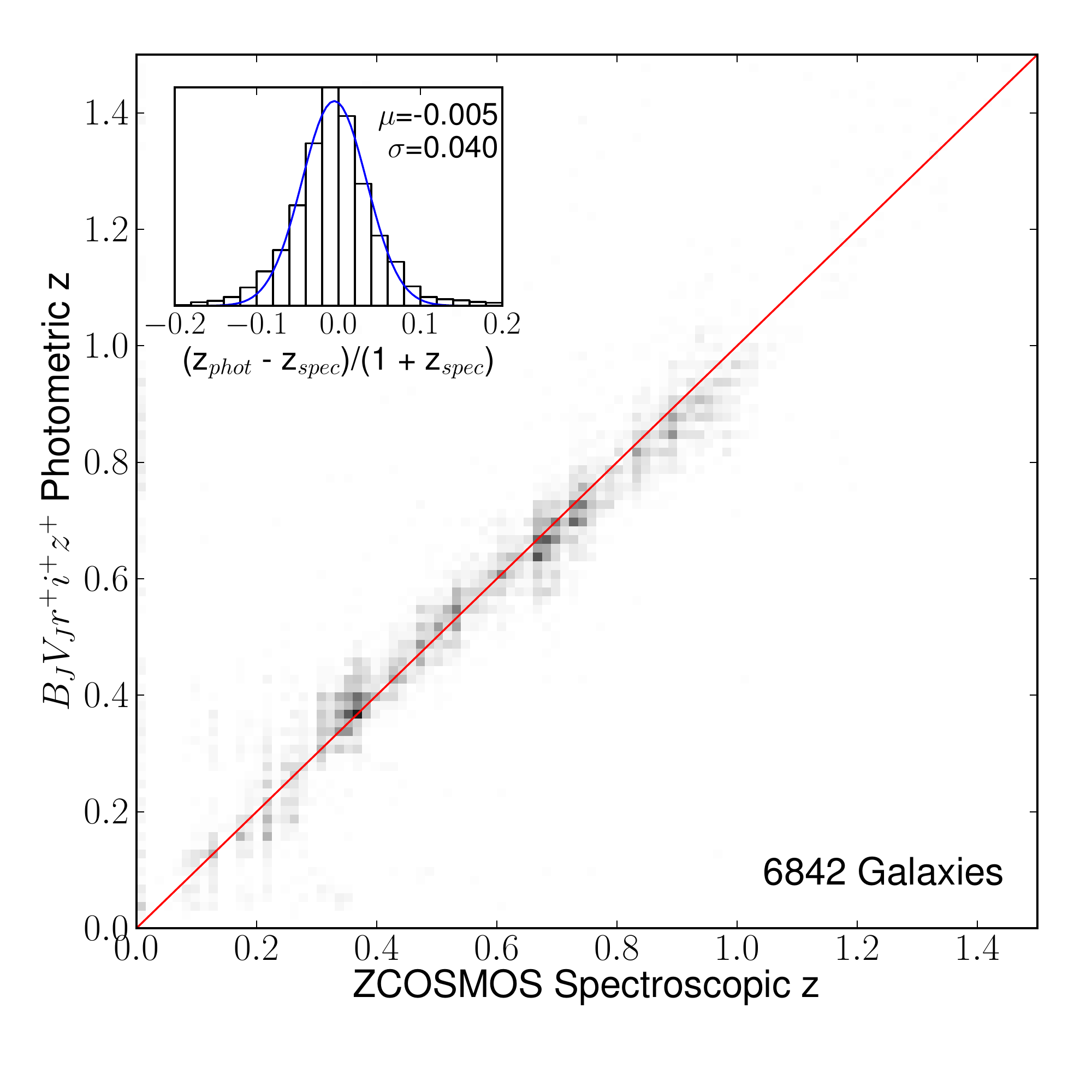}
\includegraphics[angle=0,width=3.25in]{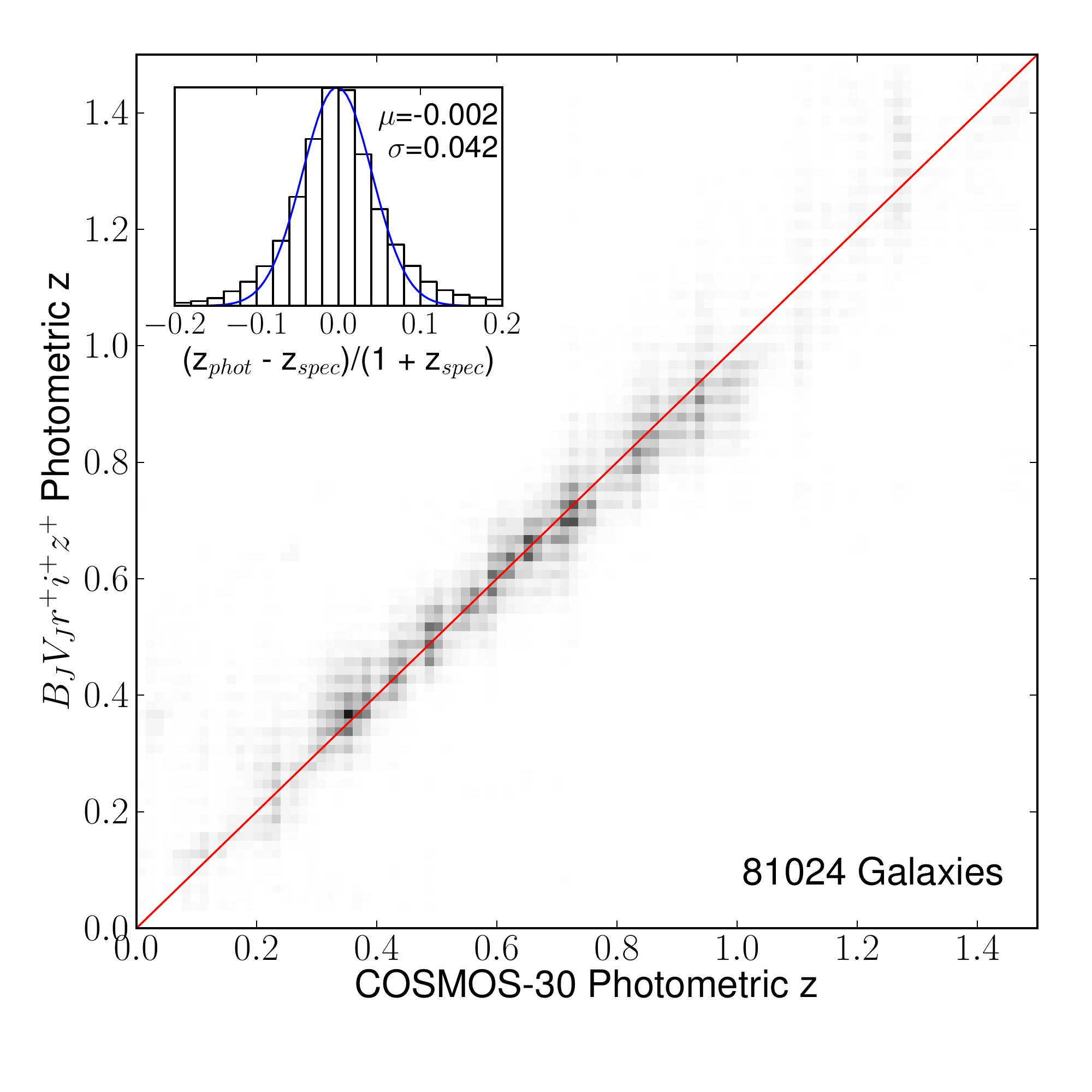}
\caption{
The left panel shows the most probable redshift $z_p$ calculated from five Subaru bands ($B_JV_Jr^+i^+z^+$) in the COSMOS
catalog plotted against the zCOSMOS spectroscopic redshifts for 6842 galaxies.
In the right panel, $z_p$ is plotted against the COSMOS-30 photometric redshifts for 81,024 galaxies.
The five-band $z_p$ values are computed from the COSMOS photometry recalibrated using 
stellar locus matching.
We include a galaxy in the plot only if BPZ ODDS $>$ 0.9. 
The histogram in the top left corner of each plot shows the distribution of
$\Delta z$ = ($z_p - z_{\rm COSMOS}$)/($1 + z_{\rm COSMOS}$);
the overlaid curve (in blue) and the statistics for the mean $\mu$ and  
 standard deviation $\sigma$ correspond to 
 Gaussian fits that exclude outliers with $|\Delta z| > 0.1$.
}
\label{fig:cosmosonepoint}
\end{figure*}

\begin{figure*}
\centering
\begin{center} $
\begin{array}{cc}
\includegraphics[angle=0,width=3.25in]{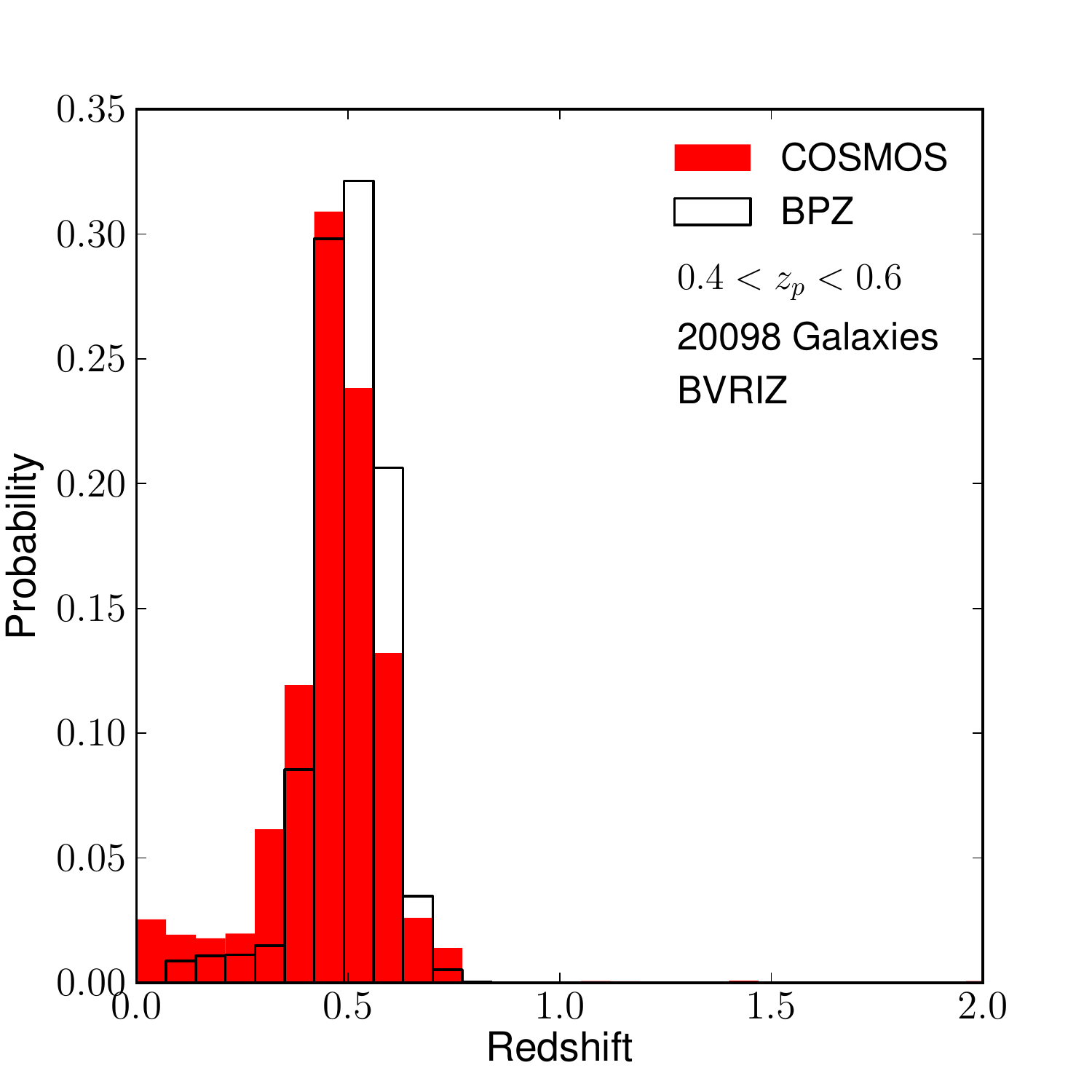} &
\includegraphics[angle=0,width=3.25in]{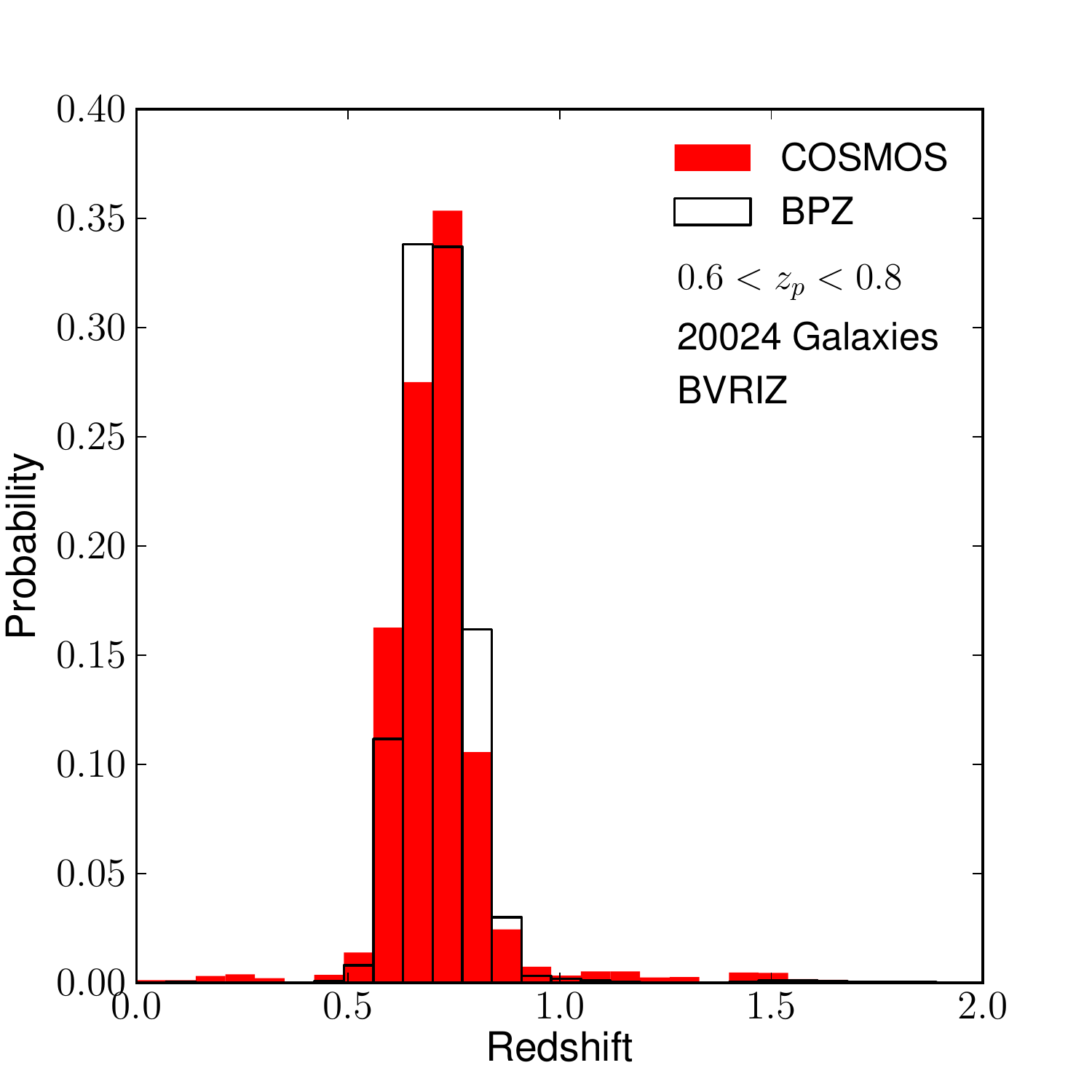} \\
\includegraphics[angle=0,width=3.25in]{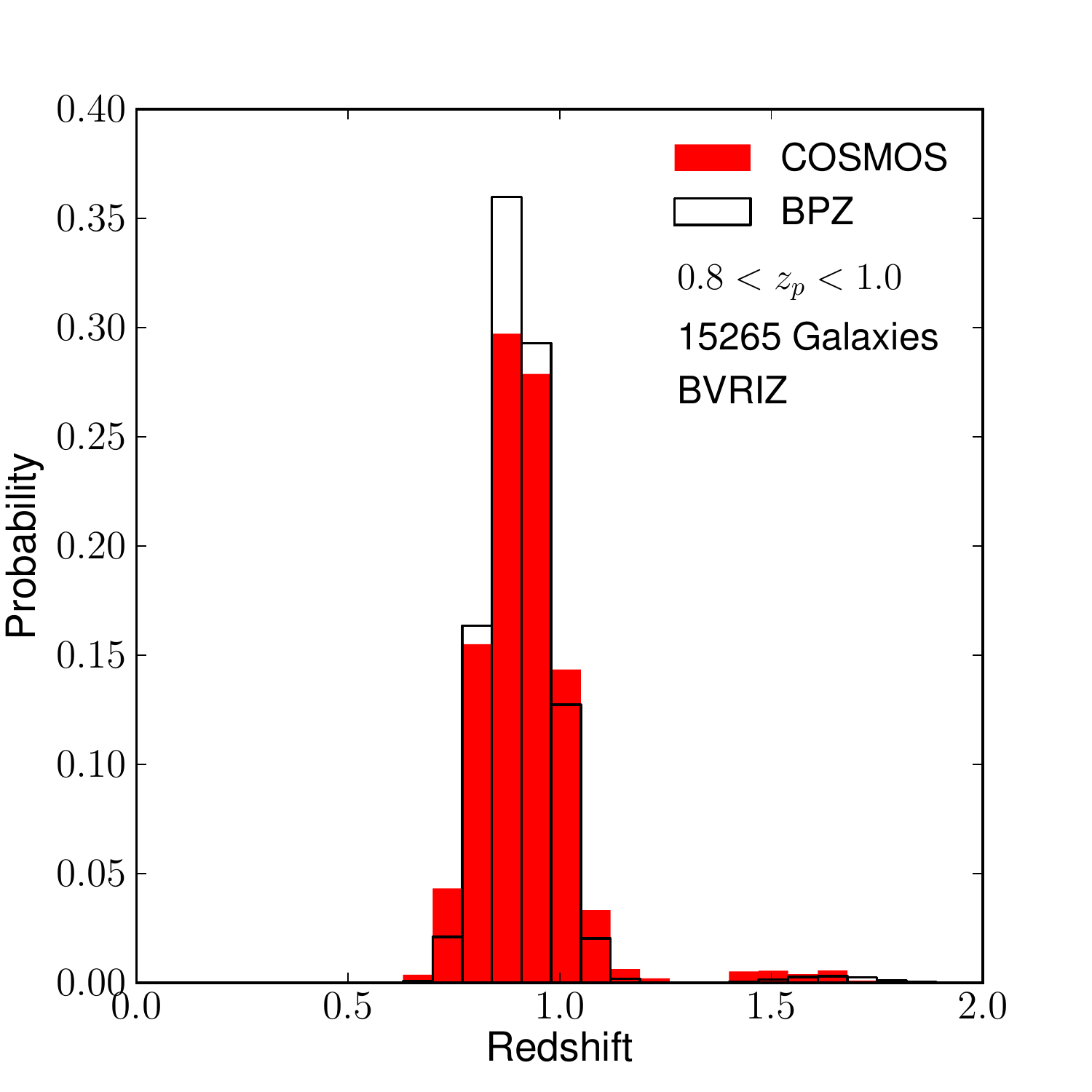} &
\includegraphics[angle=0,width=3.25in]{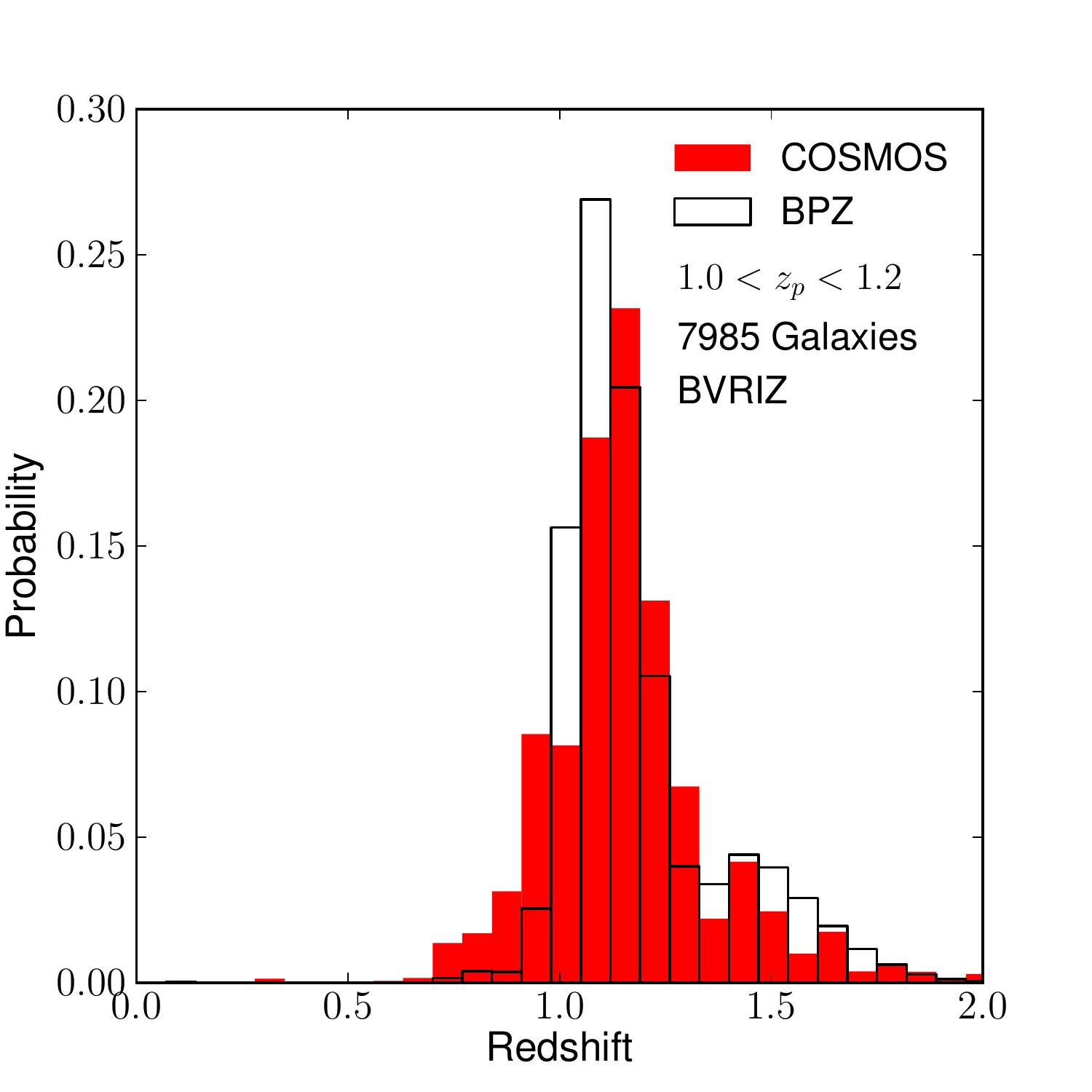} 
\end{array} $
\end{center}
\caption{ 
Stacked $p(z)$ distribution, $\sum_{gal} p(z)$, from $B_{J}V_{J}r^{+}i^{+}z^{+}$ photometry (open histogram) 
compared with the redshift distribution for thirty-band \citet{ics09} COSMOS-30 photometric 
redshifts (red histogram) for galaxies with $i^+ < 24.5$ mag and with $z_p$ in four redshift ranges:  
$0.4 < z_p < 0.6$, $0.6 < z_p < 0.8$, $0.8 < z_p < 1$, and  $1 < z_p < 1.2$.
We include all galaxies with BPZ ODDS $>$ 0.5.
The comparisons show good agreement between the $p(z)$ distribution from our analysis
and the \citet{ics09} estimates that use substantially more complete photometry.
We use the more informative $p(z)$ distribution, rather than the most probable redshift $z_p$ to measure galaxy-cluster masses.
As discussed in Paper III, this leads to significantly improved accuracy in our weak-lensing mass results. 
}
\label{fig:pdztest}
\end{figure*} 

\subsection{Galaxies in the COSMOS Survey}
\label{sec:prob}

Accurate redshifts based on thirty-band photometry (\citealt{caa07}; \citealt{ics09}) are available for galaxies to $i^+ \approx 25$ mag in a 2 deg$^{2}$ field imaged for the Cosmic Evolution Survey (COSMOS; \citealt{scoville07}). Spectroscopic redshifts are also available for a sample of COSMOS galaxies with $I < 22.5$ mag (zCOSMOS; \citealt{lilly09}).
The photometry of the COSMOS field includes SuprimeCam $B_JV_Jr^+i^+z^+$ magnitudes, which we use
to estimate photometric redshifts in an identical manner as for the photometry of the galaxy cluster fields, after recalibrating their zeropoints.

The magnitude limit of $i^+ \approx 25$ for the COSMOS catalog is comparable to that of our Subaru and CFHT galaxy cluster imaging, and beyond the completeness limit of today's spectroscopic samples. 
To test the five-filter photometric redshift estimates, we apply to the COSMOS-30 photometry catalog the same stellar locus matching and photometric redshift estimation procedure as we do for the cluster fields.
We compare our most probable $B_JV_Jr^+i^+z^+$ photometric redshifts $z_p$ to zCOSMOS 
spectroscopic redshifts and COSMOS-30 photometric redshifts estimated from 
thirty broad, intermediate, and narrow bands spanning the UV to the mid-IR (\citealt{ics09}).
The accurate COSMOS-30 photometric redshifts enable us to also
test the  posterior redshift probability distributions $p(z)$ based on five bands of photometry. 

\subsubsection{zCOSMOS Spectrosopic Redshifts with \MakeUppercase{$I$}~$< 22.5$}
\label{sec:zcosmos}

\citet{lilly09} measured spectroscopic redshifts for galaxies in the COSMOS field and 
made publicly available the spectroscopic redshifts for targets with \MakeUppercase{$I$}~$< 22.5$ (the zCOSMOS bright sample). 
The left panel of Figure~\ref{fig:cosmosonepoint} shows the  most probable photometric redshift $z_p$ based on $B_JV_Jr^+i^+z^+$ photometry 
recalibrated with stellar locus matching for the COSMOS field versus the
 zCOSMOS spectroscopic redshifts $z_s$, for galaxies with BPZ \mbox{ODDS $>$ 0.9}.
The histogram in the upper left corner of the panel shows the level of agreement 
between $z_p$ and $z_s$.
The photometric redshifts have accuracy of \mbox{$\sigma((z_p - z_s)/(1+z_s)) = 0.040$} and bias of \mbox{$\langle(z_p - z_s) / (1 + z_s)\rangle = -0.005$}, 
with \mbox{$|z_p - z_s|/(1+z_s) > 0.1$} outliers (10.5\% of galaxies) excluded.  
Only the galaxies with BPZ \mbox{ODDS $>$ 0.9} are plotted, which includes 6842 of 7959 galaxies.

\subsubsection{COSMOS-30 Photometric Redshifts of Galaxies with \MakeLowercase{$i^+$}~$< 25$}

\citet{ics09} estimated redshifts using the thirty-band photometry available in the COSMOS survey 
for several hundred thousand galaxies to $i^+ < 25$ mag (the COSMOS-30 redshifts). These authors used the {LePhare} code 
(\citealt{arn99}; \citealt{ilb06}) and rest-frame galaxy templates that include narrow line features. 
The accuracy of COSMOS-30 photometric redshifts,  estimated by comparing them to the zCOSMOS spectroscopic redshifts $z_s$ for the faint sample, ranges from 
\mbox{$\sigma((z_{\rm COSMOS-30} - z_s)/(1+z_s))$}=0.007 at $i^{+} < 22.5$ mag to \mbox{$\sigma((z_{\rm COSMOS-30} - z_s)/(1+z_s))$}=0.06 at  
$i^{+} \approx 24$ mag.

The right panel of Figure \ref{fig:cosmosonepoint} compares the most probable photometric redshift $z_p$ we estimate from recalibrated COSMOS $B_JV_Jr^+i^+z^+$ magnitudes 
to the COSMOS-30 photometric redshift.
The photometric redshifts have a standard deviation of \mbox{$\sigma((z_p - z_{\rm COSMOS-30})/(1+z_{\rm COSMOS-30})) = 0.042$} and bias of \mbox{$\langle(z_p - z_{\rm COSMOS-30}) / (1 + z_{\rm COSMOS-30})\rangle = 0.002$}, 
with \mbox{$|(z_p - z_{\rm COSMOS-30})/(1+z_{\rm COSMOS-30})| > 0.1$} outliers (14.2\%) excluded.
Only galaxies with BPZ \mbox{ODDS $>$ 0.9} are plotted, which includes 81,024 of 113,316 galaxies. 

\label{sec:pdz}

Representing the redshift probability distribution $p(z)$ for each galaxy as 
a single redshift (e.g., the most probable redshift $z_p$) neglects 
the sometimes substantial probability that the true galaxy redshift may
have a very different value (e.g., \citealt{cunha09}; \citealt{abrahamse11}; \citealt{sheldon11}). 
Single-point estimates can
lead to inaccurate conclusions, although 
the degree of inaccuracy will vary with the question being addressed and the precise $p(z)$ distribution. 
The $p(z)$ distributions for the majority of COSMOS galaxies with $B_JV_Jr^+i^+z^+$ photometry
to \MakeLowercase{$i^+$}~$\lesssim 25$
cannot be parameterized with a simple function (e.g., the sum of two Gaussian distributions). 

The COSMOS data allow us to assess the accuracy of the
$B_JV_Jr^+i^+z^+$  posterior redshift probability distributions $p(z)$.  
We partition galaxies from the COSMOS-30 catalog into bins
according to most probable redshift $z_p$, with bin widths of 0.2.
For the galaxies in each redshift bin, we calculate the stacked $p(z)$ distribution $\sum_{gal} p(z)$ from our $B_JV_Jr^+i^+z^+$ data. 
In Figure \ref{fig:pdztest}, we compare the 
stacked distribution from $B_{J}V_{J}r^{+}i^{+}z^{+}$ photometry  
to the redshift distribution for thirty-band \citet{ics09} COSMOS-30 photometric 
redshifts for galaxies with $i^+ < 25$ mag and with $z_p$ in four redshift ranges.  
We include all galaxies with BPZ \mbox{ODDS $>$ 0.5}.
The comparisons show agreement between the $p(z)$ distribution we estimate 
and the \citet{ics09} COSMOS-30 estimate, which uses substantially more complete photometry.

\begin{figure*}
\begin{center} $
\begin{array}{cc}
\includegraphics[angle=0,width=3.25in]{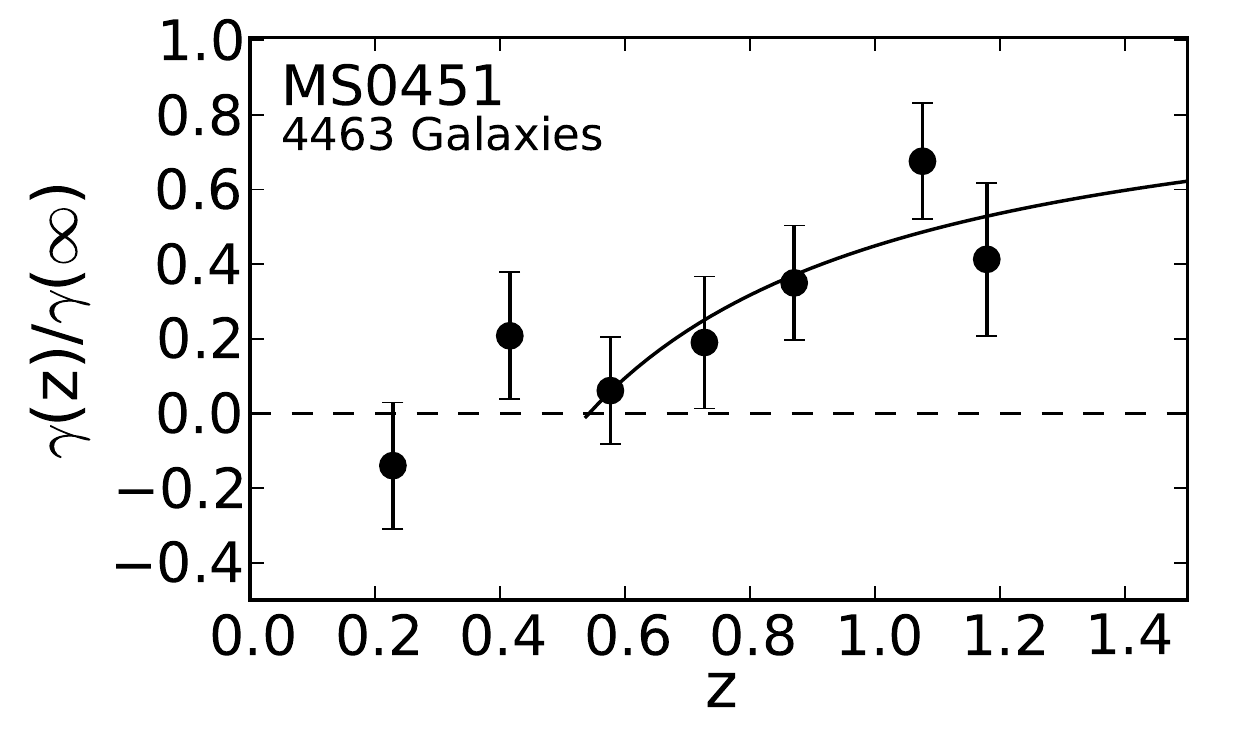} &
\includegraphics[angle=0,width=3.25in]{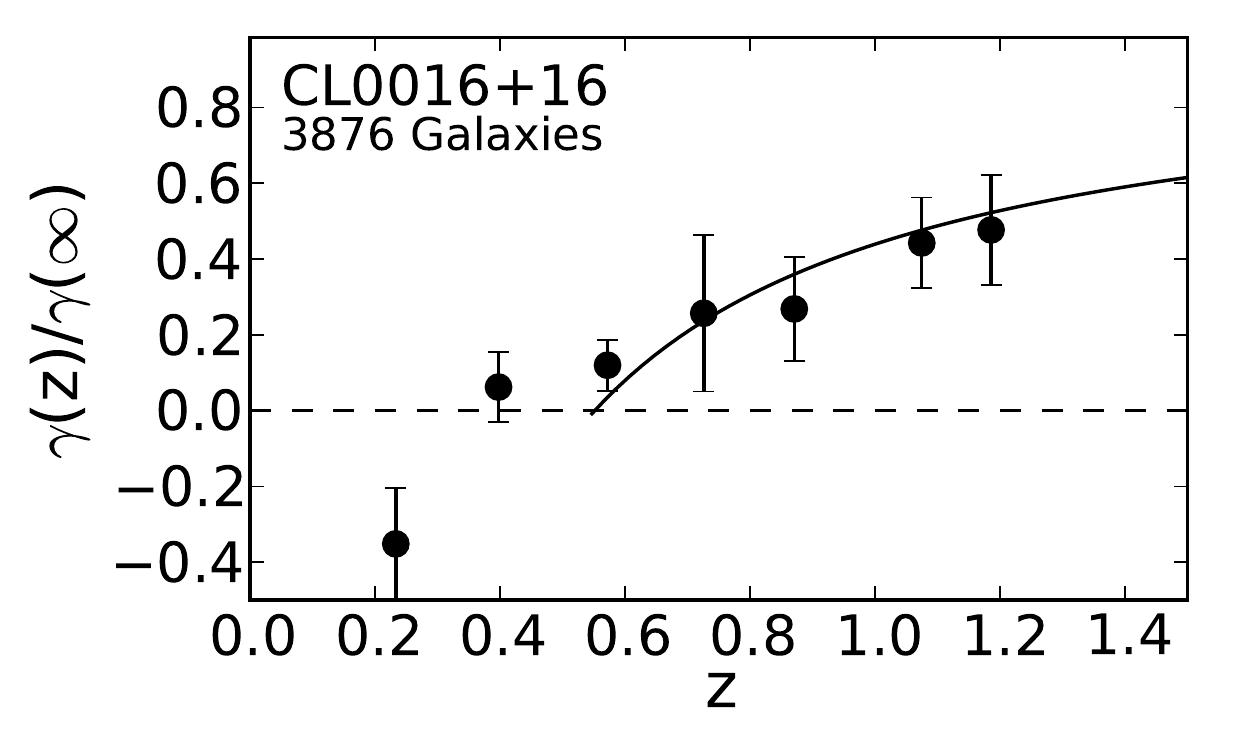} \\
\includegraphics[angle=0,width=3.25in]{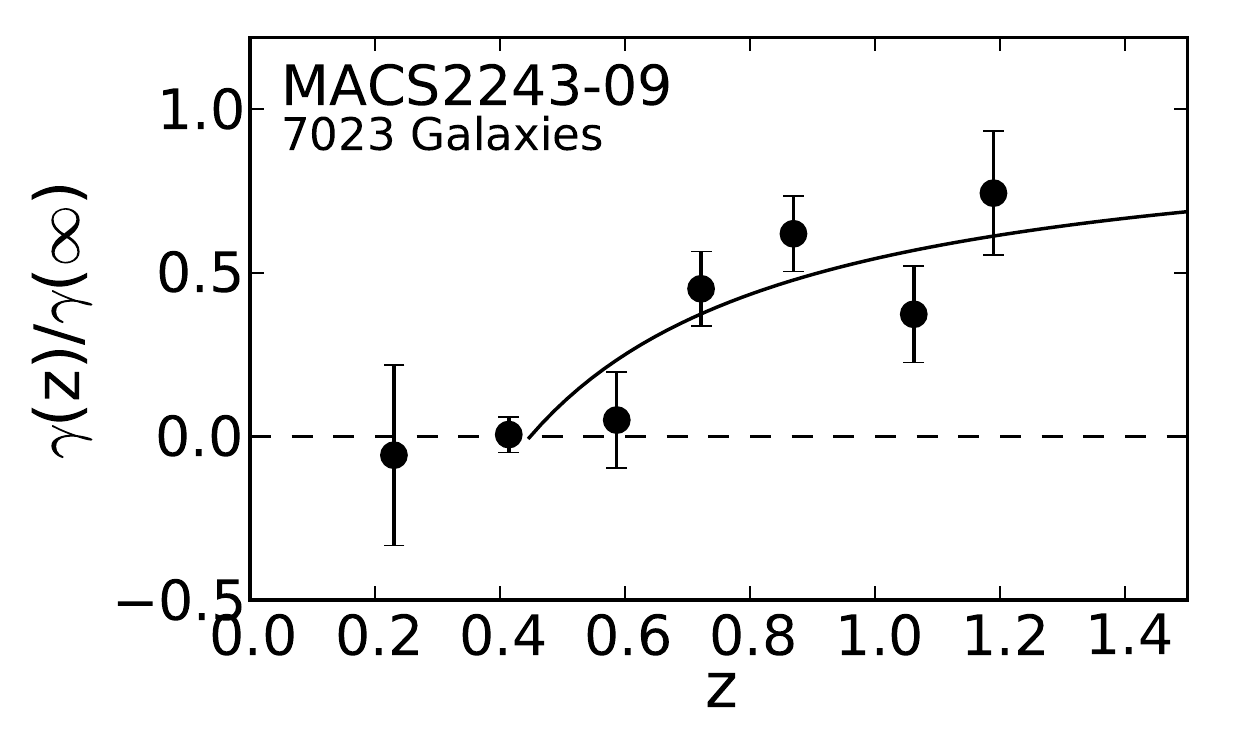} &
\includegraphics[angle=0,width=3.25in]{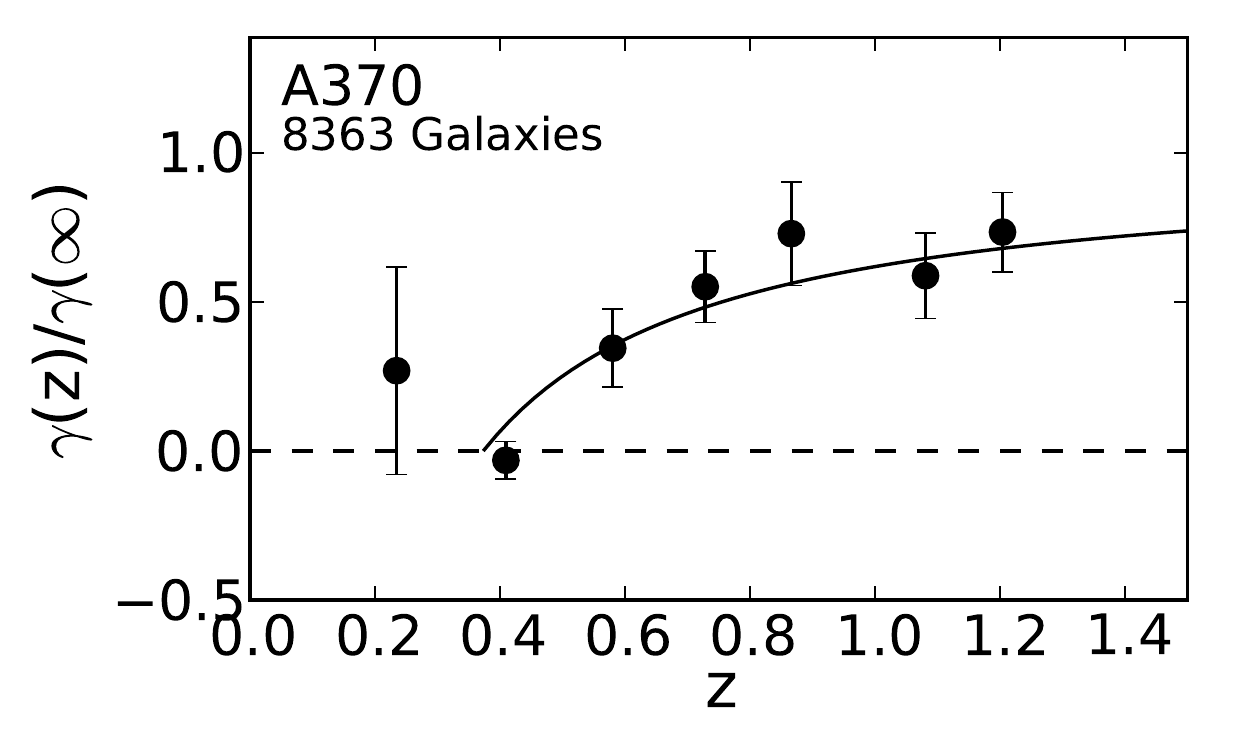} \\
\includegraphics[angle=0,width=3.25in]{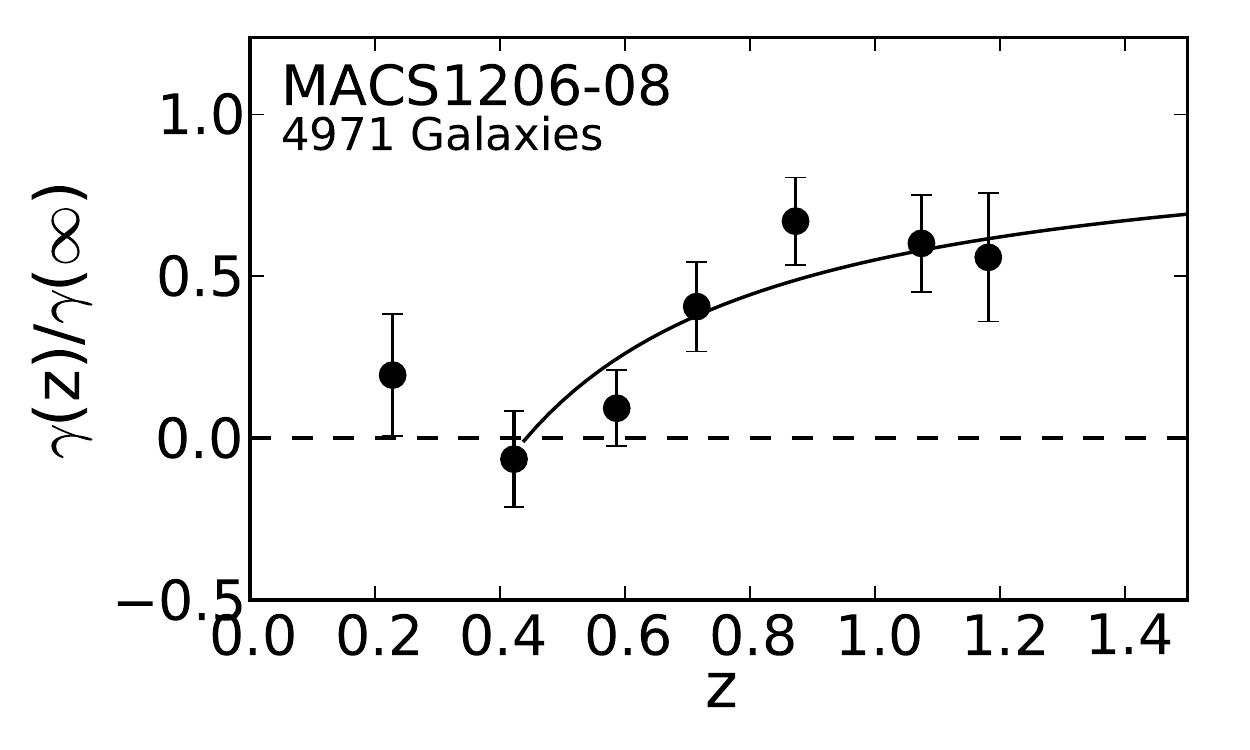} &
\includegraphics[angle=0,width=3.25in]{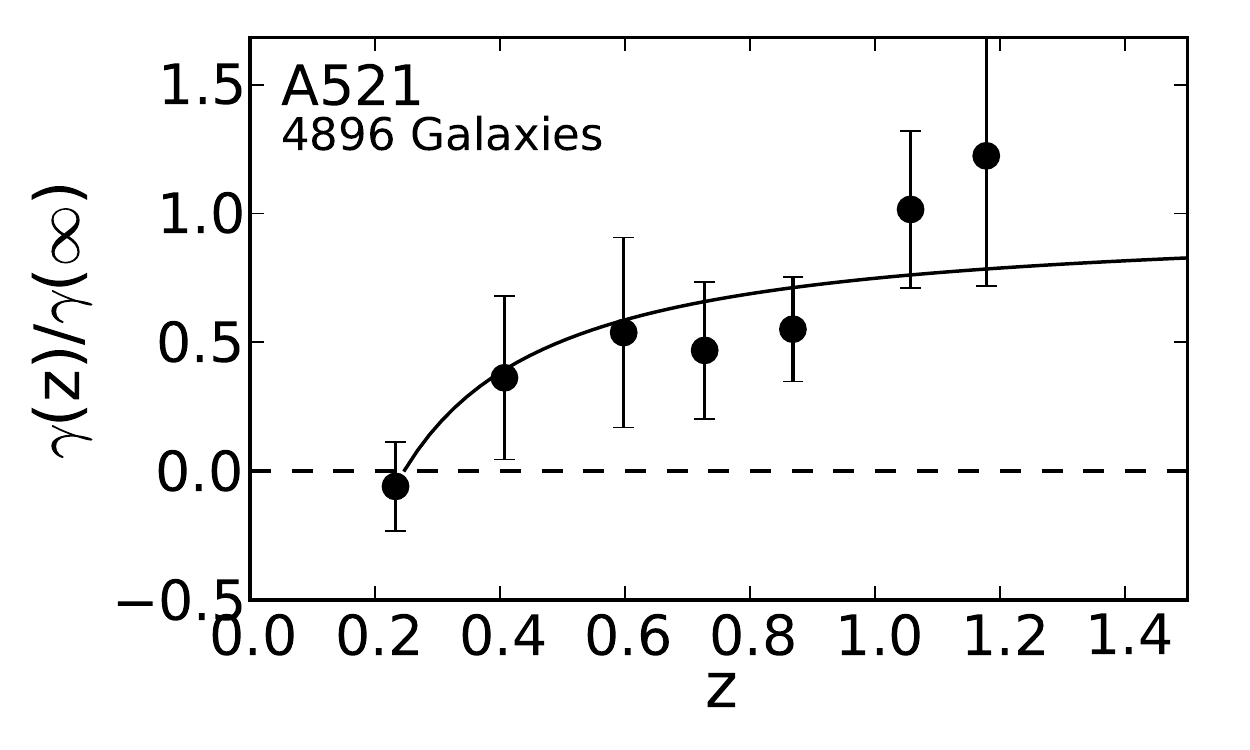} \\
\includegraphics[angle=0,width=3.25in]{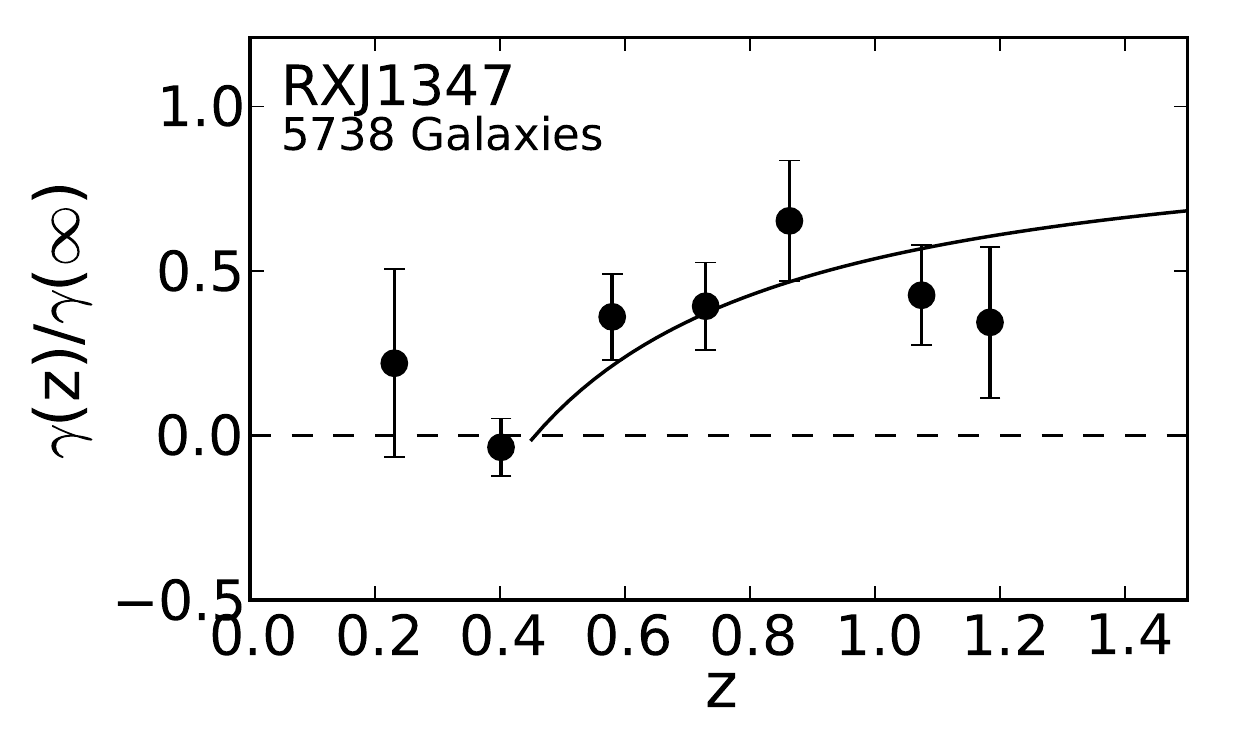} &
\includegraphics[angle=0,width=3.25in]{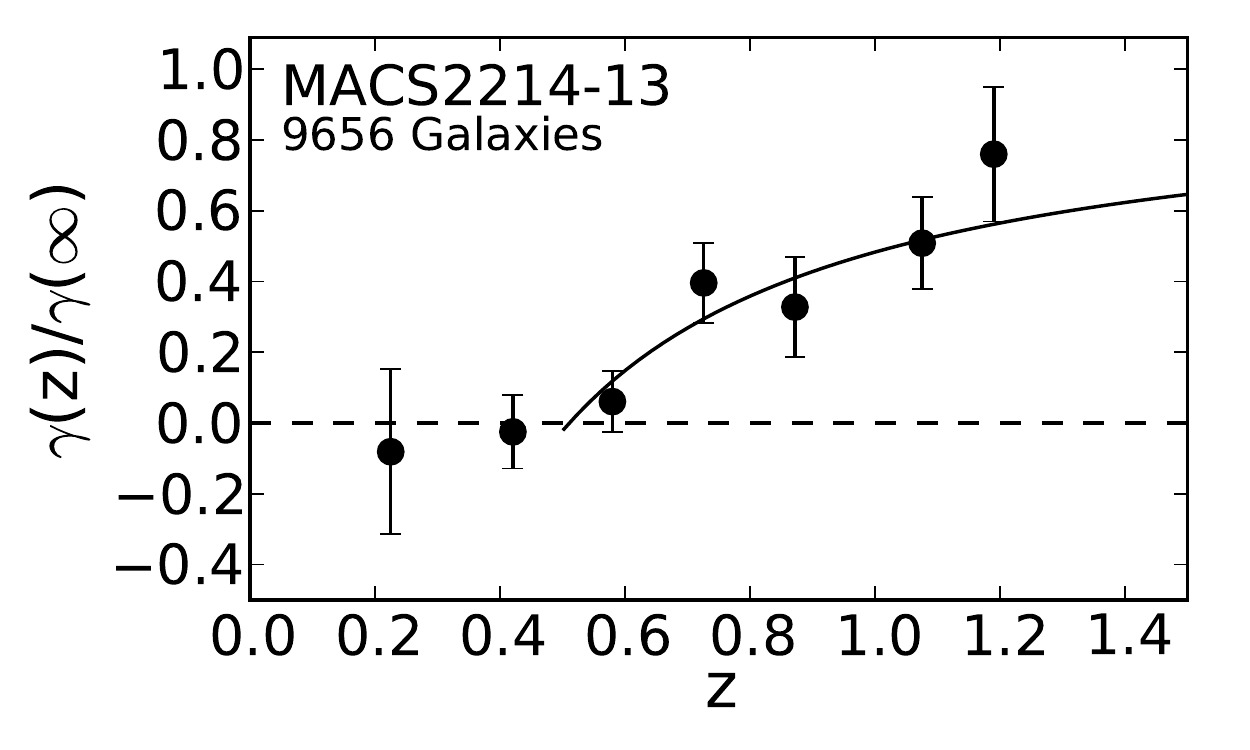}
\end{array} $
\end{center} 
\caption{
The measured weak-lensing shear as a function of most probable redshift $z_p$ for source galaxies
in eight different massive cluster fields, with photometric redshifts measured with at least
five filters.  
The shear $\gamma_t(z)$ is normalized to the expected asymptotic value $\gamma_t(z = \infty)$.
The points show the measured amplitude of the tangential shear profile, assuming a singular isothermal model for the cluster mass-density distributions.
The solid curve in each plot shows the cosmological-distance--redshift relation 
for sources behind the cluster for our reference $\Lambda$-CDM cosmology \cosmology~normalized to fit the shear amplitude. 
The error bars on the measured values are the 68\% uncertainties estimated from bootstrapped galaxy samples.
The measured shear is based on galaxies for which BPZ ODDS $> 0.5$. 
}
\label{fig:shearratio1}
\end{figure*}

\section{Tomographic Weak Lensing}
\label{sec:shearratio}

Correlations in the cosmic shear field hold significant promise as a tool for investigating dark energy (e.g., \citealt{hoekstra06}; \citealt{massey07}; \citealt{fu08}; \citealt{schrabback10}). 
However, accessing the statistical power available from upcoming surveys 
\surveys~to constrain cosmological parameters 
will require a complete characterization of 
the biases of both shear measurements and photometric redshift algorithms.
The massive galaxy clusters in our sample provide ideal proving grounds for these techniques because the deep gravitational potentials of the galaxy clusters should yield strong tomographic shear signals.

The growth in shear with increasing redshift of a galaxy behind a gravitational lens ($z_S$)
is sensitive to cosmological parameters through its 
dependence on the ratios of angular diameter distances (\citealt{jain03}; \citealt{tkb07}).
The tangential shear at a radius $r$ from the cluster center is given by
\begin{equation}
\gamma_t(r,z_S)  =  {\overline{\Sigma}(<r) - \overline{\Sigma}(r) \over  \Sigma_{c}},
\end{equation}
where $\overline{\Sigma}(<r)$ is the mean surface mass density inside radius $r$ and $\overline{\Sigma}(r)$ is the azimuthally averaged surface mass density  at radius $r$. 
$\Sigma_c$ is the critical surface mass density,
\begin{equation}
\Sigma_c \equiv \frac{c^2}{4 \pi G} \frac{D_S}{D_L D_{LS}},
\label{eqn:ratio}
\end{equation}
where $D_{S}$, $D_{L}$, and $D_{LS}$ are the angular diameter distances from the observer to the lensed source galaxy, from the observer to the lens (the cluster), and from the lens to the source galaxy, respectively.
The dependence of the weak gravitational shear on angular diameter distances is
therefore
\begin{equation}
\gamma_{t} \propto \frac{D_{LS}D_L}{D_S}.
\label{eq:shearscale}
\end{equation}
Since $D_L$ is the same for all sources lensed by a given galaxy cluster, 
the dependence of shear on the source redshift is simply
$$\gamma_{t} \propto \frac{D_{LS}}{D_S} \equiv \beta(z).$$

Shape measurements yield an estimate for the reduced shear $g$ rather than the shear $\gamma$; 
the two are related by 
\begin{equation}
g  =  \frac{\beta_{S}\gamma_{\infty}}{1 - \beta_{S}\kappa_{\infty}},
\label{eqn:reducedshear}
\end{equation}
where 
\begin{equation}
\beta_S = \frac{\beta(z_S)}{\beta_{\infty}} .
\end{equation}
$\kappa_{\infty}$ and $\gamma_{\infty}$ are the convergence and shear, respectively, 
of a lensed source at $z = \infty$.

To estimate the actual reduced shear $g$ from the measured shear $\hat g$,
we use a linear shear correction with slope $m(r_h)$ and intercept $c$ measured from STEP2 simulations \citep{mhb07}:
\begin{equation}
g = \frac{\hat g - c}{1 + m(r_h)},
\label{eqn:shearcalibration}
\end{equation}
where $m(r_h)$ is a function of the object size $r_h$ relative to the PSF size of the observation (see
Papers I and III).

\citet{tkb07} projected the constraints on cosmological parameters that would be possible from
a five-band optical survey of galaxy clusters with masses $M \approx10^{15} M_{\odot}$.
For a survey of fifty of the most massive low- to moderate-redshift galaxy clusters with five-band photometry
and a median source galaxy redshift of $z=0.7$, similar to this paper's sample, \citet{tkb07} forecast that the dark-energy equation-of-state parameter $w$ may be constrained to within $\pm$0.6, after combination with a WMAP prior.
This anticipated precision is for models that allow curvature and evolving $w$, and improved constraints will therefore be possible 
for a flat cosmology with a static value for $w$.
For a five-band survey covering 10000 deg$^2$ with median source galaxy redshift of $z=0.7$, \citet{tkb07} 
projected a constraint of $\Delta w = 0.0075$.
\citet{tkb07} made the assumption that the errors of photometric redshifts are Gaussian
and that any magnitude- or size-dependent bias in ground-based shear measurements can be controlled to $z = 1.5$.  

In a number of pioneering efforts, an increase in shear signal has been observed with distance of the source galaxies behind groups or clusters.
\citet{wittman01} plotted the measured shear against photometric redshift for the galaxies lensed by a massive cluster at $z=0.276$ (see also \citealt{wittman03}).
\citet{kht07} combined a tomographic analysis of the A901/2 supercluster, with a cosmic-shear analysis of two randomly selected fields using COMBO-17 photometric redshift estimates. 
\citet{mbu11} showed that the shear signal increases with distance of sources behind the A370, ZwCl0024+17, and RXJ1347-11 galaxy clusters, based on color cuts applied to three-filter Subaru imaging. 
\citet{taylor11} measured the redshift dependence of the weak lensing shear behind groups in the COSMOS field
using HST images for shape measurements and the \citet{ics09} thirty-band photometric redshifts.

\subsection{Tomographic Signal Behind Individual Massive Clusters}
\label{sec:shearratio_indiv}

We first examine how, for individual clusters, the variation of measured shear with source redshift compares with the prediction for a fiducial $\Lambda$-CDM cosmology \cosmology.

To determine the shear signal as a function of redshift in each field, we bin galaxies with BPZ ODDS $> 0.5$ according to their most probable redshift $z_p$.
We fit a singular isothermal shear (SIS) profile 
\begin{equation}
\gamma_t(r, z \in z_i) = \frac{A_i}{r}
\label{eqn:SIS}
\end{equation}
to the tangential shear $\gamma_t$ of the galaxies in the $i$th redshift bin $z_i$,
where $r$ is the distance to the center of the galaxy cluster.
We divide each amplitude $A_i$ by the 
asymptotic value $A_{\infty}$ at $z=\infty$ expected for the fiducial $\Lambda$-CDM cosmology:
\begin{equation}
\frac{\gamma_t(z \in z_i)}{\gamma_t(\infty)} = \frac{A_i}{A_{\infty}}.
\end{equation}
Figure~\ref{fig:shearratio1} shows the variation of $\gamma_t(z \in z_i) / \gamma(\infty)$ versus mean redshift for
eight cluster fields with excellent filter coverage (SuprimeCam $B_JV_JR_CI_Cz^+$ and, in some cases, also MegaPrime $u^*g^*r^*i^*z^*$). 
The error bars are uncertainties we estimate from bootstrapped samples of galaxies.
These comparisons show good agreement between the observed variation in $\gamma_t(z) / \gamma(\infty)$ with redshift and the expectation for the fiducial cosmology.

\subsection{Tomographic Signal From a Stack of 27 Massive Clusters}

For a cosmology with zero spatial curvature (i.e., a `flat' universe),
the angular diameter distance $D(z)$ is equal to the product of the
comoving distance $\omega_z$ and the scale factor $a_z$,
\begin{equation}
D(z) = \omega_z \times a_z.
\end{equation}
\citet{taylor11} used this relation
to stack the measured shear as a function of distance for multiple lenses. 

For a flat geometry, Equation~\ref{eq:shearscale} simplifies to
\begin{equation}
\gamma_t(z) \propto \frac{\omega_L(\omega_S -
  \omega_L)}{\omega_S(1+z_L)},
\label{eqn:flatshear}
\end{equation}
where $\omega_{S}$ and $\omega_L$ are comoving
distances from the observer to the source and lens, respectively, 
and $z_L$ is the redshift of the lens.
Defining the ratio $x$ of the two comoving distances $\omega_{S}$ and $\omega_L$ as
\begin{equation}
x =  \omega_S/\omega_L, 
\label{eqn:x_distance}
\end{equation}
we can express Equation \ref{eqn:flatshear} in terms of $x$:
\begin{equation}
\gamma_t(x) \propto \frac{\omega_L}{1+z_L}(1-\frac{1}{x}).
\end{equation}
Therefore, in a flat universe, we can express the tangential shear ratio $\gamma_t(x) / \gamma_t(x=\infty)$ simply in terms of $x$:
\begin{equation}
\frac{\gamma_t(x)}{\gamma_t(x=\infty)} = 1 - \frac{1}{x}.  
\label{eq:shear_dist_equiv}
\end{equation}
We note that $\gamma_t(x = \infty)$ is not physical for the reference cosmology, because the ratio of comoving distances
$x$ remains finite as the source redshift $z$ approaches infinity.

When estimating $\gamma_t(x) / \gamma_t(x = \infty)$ for each galaxy, we divide the measured shear by the shear expected for a 
source galaxy at $x=\infty$ given the best-fit mass profile for the cluster (Paper III).
We place the galaxies in the 27 cluster fields with photometry through at least the $B_JV_JR_CI_Cz^+$ bands (see Table 1 of Paper I)
into bins according to $x$. 
The value of $\gamma_t(x) / \gamma_t(x=\infty)$ for the sample of galaxies in each redshift bin
is then estimated with a weighted Gaussian-kernel density estimator. 
Only the galaxies that are used to estimate cluster masses in Paper III are included; see that paper for a detailed description of the galaxy selection criteria.
Galaxies are included only if the most probable photometric redshift is
$z_p < 1.25$.

\begin{figure}
\includegraphics[width=\hsize]{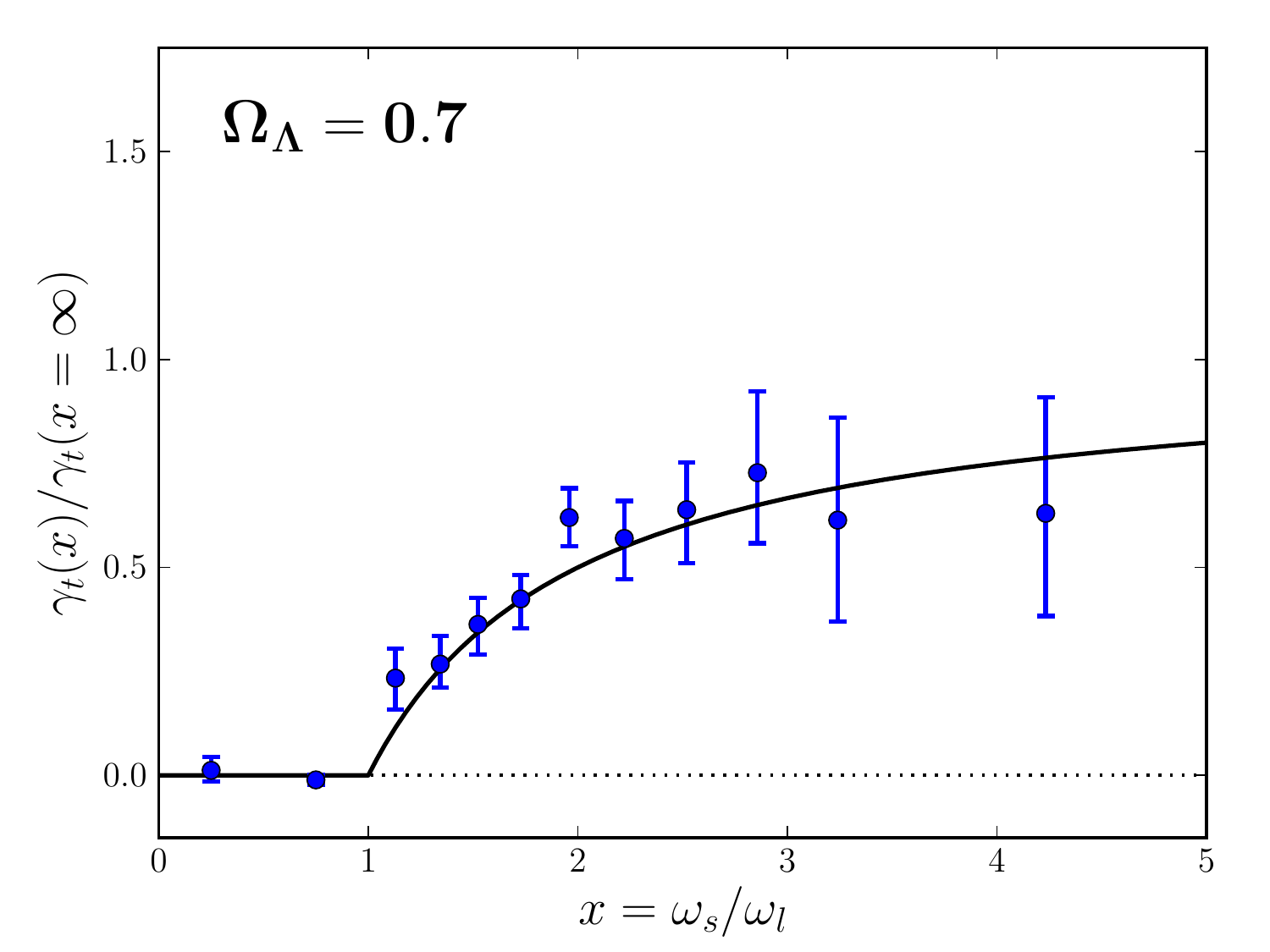}
\caption{The stacked shear ratio for 27 clusters as a function of
  scaled comoving distance.  
  The horizontal axis is the ratio of comoving distances from the observer to the source galaxy and to the lens.
  The vertical axis is the shear signal divided by that for a hypothetical (although physically disallowed) source at $x = \infty$ calculated for the fiducial cosmology and the best-fit cluster mass (from Paper III).
  The  black curve is the predicted shear scaling $1-1/x$ for the reference cosmology. Points are the
  modes (maximum likelihood) measured from a Gaussian-kernel density
  estimator, and the error bars represent the 68\% confidence interval as estimated from bootstrapped samples of galaxies in
  each bin. }
\label{fig:shearratio_stack}
\end{figure}

In Figure~\ref{fig:shearratio_stack}, the points correspond to the stacked shear signal calculated 
in the reference cosmology, and the error bars are estimated from bootstrapped galaxy samples at each redshift.
The data show good qualitative agreement with the variation expected for the reference cosmology.
We defer to a follow-up paper the determination of cosmological constraints from the tomographic lensing signal, using the
entire redshift probability distribution $p(z)$ for each galaxy.

\section{Conclusions}
\label{sec:conclusions}

Converting the signal measured by CCD sensors mounted on a wide-field camera 
to calibrated AB magnitudes with better than several percent accuracy
has typically required time-consuming calibration exposures as well as optimal observing conditions. 
When useful overlap with survey photometry is not available (e.g., limited dynamic range, color transformations from survey magnitudes), 
repeated standard-star observations in photometric conditions are necessary for accurate zeropoint calibration.
Dedicated exposures sequences of dense stellar fields have been considered necessary to measure 
the position-dependent zeropoint error ($\sim$0.04-0.06 mag for SuprimeCam) present after 
correcting by dome or sky-flat calibration exposures.

For the Subaru and CFHT imaging that we analyzed, dedicated calibration data were not always available and conditions were not always photometric.
We have developed 
improved tools that use only the instrumental stellar magnitudes in exposures of science targets to produce accurate star-flat models and color calibrations.
These have sufficient power to enable high-quality photometric redshift estimates without spectroscopic training sets.
We use \speczcluster~galaxy spectra in \speczfields~cluster fields across 
the sky to measure an accuracy of \mbox{$\sigma((z_p - z_s)/(1+z_s)) \approx $~\accuracy}.

We achieve $\lesssim0.01$ - 0.02 mag color calibration by matching the observed stellar locus in color-color space to a model stellar locus.
We developed a spectroscopic model for the dereddened SDSS stellar locus to estimate
more accurately the locus expected for SuprimeCam and MegaPrime filters.
This step is important because nonlinear color transformations exist between our SuprimeCam $B_JV_JR_CI_Cz^+$ magnitudes and the SDSS $u'g'r'i'z'$ filters used to measure the stellar locus. 
We created a simple $\chi^2$ GOF statistic that makes possible a solution for 
all unknown zeropoints simultaneously, without correlated errors. 
These improvements, including the correction of the SDSS stellar locus for Galactic extinction,
were necessary to estimate robust and reliable photometric redshifts from our Subaru and CFHT imaging.
Interested readers may find the Python code as well as a spectroscopic model for the locus at \codeurl.  
Calibration requires only a catalog of stellar magnitudes and the total transmission function for each bandpass filter. 

Summarizing the redshift posterior probability distribution $p(z)$ for each galaxy as 
a single redshift (e.g., the most probable redshift $z_p$) disregards 
any probability that the true galaxy redshift may
have a very different value (e.g., catastrophic outliers). 
The $p(z)$ distribution contains more information
than the most probable redshift, or other single-point estimates. 
We find good agreement between
the stacked $\sum_{gal} p(z)$ and the redshift distribution estimated from thirty COSMOS photometric bands by 
\citet{ics09}. The $p(z)$ distributions will provide a more
powerful, less biased tool for science analysis than single-point estimates for ongoing and future surveys \surveys.

The extreme gravitational potentials of massive galaxy clusters are expected to yield
a simple, strong growth in shear signal with increasing distance of the source galaxy  behind the cluster lens.
The shape of the growth of shear behind a stack of 27 clusters with increasing redshift shows agreement with a fiducial $\Lambda$-CDM cosmology. 
The average shear we measure for foreground galaxies is consistent with 
zero. 

The software we make available can produce state-of-the-art photometric accuracy without
dedicated observations or the need to estimate nonlinear color terms. 
Stellar colors will be a useful tool for checking the accuracy of zeropoints obtained through ubercalibration of surveys (e.g.,  \citealt{ivezic07}; \citealt{padmanabhan08}; \citealt{schlafly12})
or close monitoring of atmospheric transparency (e.g., \citealt{bab10}; \citealt{blake11}). 

\section*{Acknowledgments}

We thank F. William High, Jorg Dietrich, and Hendrik Hildebrandt for their expert advice and assistance on photometric calibration and redshift estimation. We also appreciate helpful discussions with Thomas Erben, C-J Ma, Daniel Coe, and Carlos Cunha.  

This work is supported in part by the U.S. Department of Energy under
contract number DE-AC02-76SF00515. This work was also supported by the
National Science Foundation under Grant Nos. AST-0807458 and AST-1140019. 
MTA and PRB acknowledge the support of NSF grant PHY-0969487. AM acknowledges the
support of NSF grant AST-0838187.  The authors acknowledge support
from programs HST-AR-12654.01-A, HST-GO-12009.02-A, and
HST-GO-11100.02-A provided by NASA through a grant from the Space
Telescope Science Institute, which is operated by the Association of
Universities for Research in Astronomy, Inc., under NASA contract NAS
5-26555. This work is also supported by the National Aeronautics and
Space Administration through Chandra Award Numbers TM1-12010X,
GO0-11149X, GO9-0141X , and GO8-9119X issued by the Chandra X-ray
Observatory Center, which is operated by the Smithsonian Astrophysical
Observatory for and on behalf of the National Aeronautics Space
Administration under contract NAS8-03060.  DEA recognizes the support
of a Hewlett Foundation Stanford Graduate Fellowship.

Based in part on data collected at Subaru Telescope (University of
Tokyo) and obtained from the SMOKA, which is operated by the Astronomy
Data Center, National Astronomical Observatory of Japan.  Based on
observations obtained with MegaPrime/MegaCam, a joint project of CFHT
and CEA/DAPNIA, at the Canada-France-Hawaii Telescope (CFHT), which is
operated by the National Research Council (NRC) of Canada, the
Institute National des Sciences de l'Univers of the Centre National de
la Recherche Scientifique of France, and the University of Hawaii.
This research used the facilities of the Canadian Astronomy Data
Centre operated by the National Research Council of Canada with the
support of the Canadian Space Agency.  This research has made use of
the VizieR catalogue access tool, CDS, Strasbourg, France. Funding for
SDSS-III has been provided by the Alfred P. Sloan Foundation, the
Participating Institutions, the National Science Foundation, and the
U.S. Department of Energy Office of Science. The SDSS-III web site is
http://www.sdss3.org/. This research has made use of the NASA/IPAC
Extragalactic Database (NED), which is operated by the Jet Propulsion
Laboratory, Caltech, under contract with NASA.

\bibliography{paper2}

\appendix

\label{lastpage}

\end{document}